\documentclass[12pt]{article}

\usepackage[T1]{fontenc}
\usepackage[utf8]{inputenc}
\usepackage{lmodern}
\usepackage{amsthm}
\usepackage{amssymb}
\usepackage{amstext}
\usepackage{bm}
\usepackage{graphicx}
\usepackage{epstopdf}
\usepackage[export]{adjustbox}
\usepackage{subcaption}
\usepackage[breaklinks=true]{hyperref}
\usepackage{amsmath}
\usepackage{setspace}
\usepackage{morefloats} 
\usepackage[capitalize]{cleveref}

\crefalias{subequation}{equation}
\crefformat{pluraleq}{Eqs.~(#2#1#3)}

\renewcommand{\baselinestretch}{1.0} 
\makeatletter
\newcounter{parentsubcaption}
\newenvironment{subsubcaption}
 {
 
 \refstepcounter{sub\@captype}%
  \protected@edef\theparentsubcaption{\@nameuse{thesub\@captype}}%
  \setcounter{parentsubcaption}{\value{sub\@captype}}%
  \setcounter{sub\@captype}{0}%
  \@namedef{thesub\@captype}{\alph{sub\@captype}.\theparentsubcaption}%
  \ignorespaces
}{%
  \setcounter{sub\@captype}{\value{parentsubcaption}}%
  \ignorespacesafterend
}
\makeatother

\hypersetup{colorlinks=true,
  pdftitle={Nonclassical Particle Transport in 1-D Random Periodic Media},
  pdfauthor={Richard Vasques and Kai Krycki and Rachel Slaybaugh}
}

\newcommand{\bl}{\big<}
\newcommand{\bg}{\big>}

\newcommand{\eps}{\varepsilon}

\newcommand{\ux}{{\bm x}}

\newcommand{\uomega}{{\bf \Omega}}
\newcommand{\unabla}{{\bf \nabla}}

\newcommand{\seta}{\mathcal{A}}
\newcommand{\setb}{\mathcal{B}}
 \newcommand{\Keywords}[1]{\vspace{12pt}\par\noindent
{\small{\bf Keywords\/}: #1}}

\begin{document}

\title{Nonclassical Particle Transport in 1-D Random Periodic Media}
\author{{\bf R.\ Vasques $^{\dagger,}$}\footnote{Email: \texttt{richard.vasques@fulbrightmail.org}} , {\bf K.\ Krycki $^\ddagger$}, {\bf R.\ N.\ Slaybaugh $^\dagger$}\\ \\
\em {\bf $^\dagger$}University of California, Berkeley\\
\em Department of Nuclear Engineering\\
\em 4155 Etcheverry Hall, Berkeley, CA 94720-1730\\
\and \\
\em {\bf $^\ddagger$}Aachen Institute for Nuclear Training GmbH \\
\em Jesuitenstraße 4, 52062 Aachen, Germany}
\date{}
\maketitle

\begin{abstract}

We investigate the accuracy of the recently proposed nonclassical transport equation.
This equation contains an extra independent variable compared to the classical transport equation (the path-length $s$), and models particle transport taking place in homogenized random media in which a particle's distance-to-collision is not exponentially distributed.
To solve the nonclassical equation one needs to know the $s$-dependent ensemble-averaged total cross section, $\Sigma_t(\mu,s)$, or its corresponding path-length distribution function, $p(\mu,s)$.
We consider a 1-D spatially periodic system consisting of alternating solid and void layers, randomly placed in the $x$-axis.
We obtain an analytical expression for $p(\mu,s)$ and use this result to compute the corresponding $\Sigma_t(\mu,s)$.
Then, we proceed to numerically solve the nonclassical equation for different test problems in rod geometry; that is, particles can move only in the directions $\mu=\pm 1$.
To assess the accuracy of these solutions, we produce ``benchmark" results obtained by (i) generating a large number of physical realizations of the system, (ii) numerically solving the transport equation in each realization, and (iii) ensemble-averaging the solutions over all physical realizations.
We show that the numerical results validate the nonclassical model; the solutions obtained with the nonclassical equation accurately estimate the ensemble-averaged scalar flux in this 1-D random periodic system, greatly outperforming the widely-used atomic mix model in most problems. 

\end{abstract}

\pagebreak

\doublespacing

\section{Introduction}

The classical theory of linear particle transport defines the total cross section $\Sigma_t$ as independent of the path-length $s$ (the distance traveled by the particle since its previous interaction) and of the direction of flight $\uomega$.
This definition leads to an exponential probability density function for a particle's distance-to-collision:
\begin{align}\label{eq1}
p(s) = \Sigma_t e^{-\Sigma_t s}.
\end{align}

However, a nonexponential attenuation law for the particle flux arises in certain inhomogeneous media in which the scattering centers are spatially correlated.
This ``nonclassical" behavior occurs in certain important applications, such as neutron transport in Pebble Bed Reactors (in which a nonexponential $p(s)$ arises due to the pebble arrangement within the core) and photon transport in atmospheric clouds (in which the locations of the water droplets in the cloud seem to be correlated in ways that measurably affect the radiative transfer within the cloud).

An approach to this type of nonclassical transport problem was recently proposed \cite{lar07,larvas11}, with the assumption that the positions of the scattering centers are correlated but independent of direction $\uomega$.
Existence and uniqueness of solutions are rigorously discussed in \cite{fra10}.
This nonclassical theory was extended in \cite{vaslar14a} to include angular-dependent path-length distributions in order to investigate anisotropic diffusion of neutrons in 3-D PBR cores.

A similar kinetic equation with path-length as an independent variable has been rigorously derived for the periodic Lorentz gas in a series of papers by Golse et al.~(cf.~\cite{gol12} for a review), and by Marklof \& Str\"ombergsson (cf.~\cite{mar11,mar15}).
Furthermore, related work has been performed by Grosjean in \cite{gro51}; it considers a generalization of neutron transport that includes arbitrary path-length distributions, and presents a derivation of diffusion solutions for infinite isotropic point and plane source problems.
 
Assuming monoenergetic transport and isotropic scattering, the nonclassical linear Boltzmann equation with angular-dependent path-length distributions and isotropic source is writen as
\begin{align}\label{eq2}
\frac{\partial\psi}{\partial s}(\ux,\uomega,s) + &\uomega\cdot\unabla \psi(\ux,\uomega,s) + \Sigma_t(\uomega,s)\psi(\ux,\uomega,s) 
\\&= \frac{\delta(s)}{4\pi}\left[ c\int_{4\pi}\int_0^\infty \Sigma_t(\uomega',s')\psi(\ux,\uomega',s')ds' d\Omega' + Q(\ux) \right]\,, \nonumber
\end{align}
where $\ux = (x,y,z)=$ position, $\uomega = (\Omega_x,\Omega_y,\Omega_z)=$ direction of flight (with $|\uomega|=1$), $\psi$ is the nonclassical angular flux, $c$ is the scattering ratio (such that the scattering cross section $\Sigma_s= c\Sigma_t$), and $Q$ is the source.
Here, the nonclassical angular-dependent ensemble-averaged total cross section $\Sigma_t(\uomega,s)$ is defined as
\begin{align}\label{eq3}
\Sigma_t(\uomega,s)ds =  \begin{array}{l}
\text{ the probability (ensemble-averaged over all physical}\vspace{-10pt}\\
\text{ realizations) that a particle, scattered or born at any}\vspace{-10pt}\\
\text{ point $\ux$ and traveling in the direction $\uomega$, will experience}\vspace{-10pt}\\
\text{ a collision between $\ux + s\uomega$ and $\ux + (s+ds)\uomega$.}
\end{array}
 \end{align}
The underlying path-length distribution and the above nonclassical cross section are related \cite{vaslar14a} by
\begin{align}\label{eq4}
	p(\uomega,s) = \Sigma_t(\uomega,s)\exp\left( -\int_0^s\Sigma_t(\uomega,s')ds'\right).
\end{align}
It has been shown that, if $p(s)$ is independent of $\uomega$, \cref{eq2} can be converted to an integral equation for the scalar flux that is identical to the integral equation that can be constructed for certain diffusion-based approximations \cite{siap15,vas16}.

Moreover, if the path-length distribution function is an exponential as given in \cref{eq1}, \cref{eq2} reduces to the classical linear Boltzmann equation
\begin{subequations}\label[pluraleq]{eq5}
\begin{align}\label{eq5a}
\uomega\cdot{\unabla} \Psi(\ux,\uomega) + \Sigma_t\Psi(\ux,\uomega) = \frac{1}{4\pi}\left[\int_{4\pi}\Sigma_s\Psi(\ux,\uomega')d\Omega' + Q(\ux) \right]\,
\end{align}
for the classical angular flux 
\begin{align}
\Psi(\ux,\uomega) = \int_0^\infty \psi(\ux,\uomega,s)ds.
\end{align} 
\end{subequations}

Numerical results have been provided for the asymptotic diffusion limit of this nonclassical theory \cite{larvas11,vaslar09,vas13,vaslar14b}, and for moment models of the nonclassical equation in the diffusive regime \cite{kry13}.
However, very few results have been presented for the nonclassical \textit{transport} equation.
This is because one must know $\Sigma_t(\uomega,s)$, or $\Sigma_t(s)$ in the case of angular-independent path lengths, in order to solve \cref{eq2}. 

In this paper we investigate the accuracy of the 1-D nonclassical transport equation.
We consider a 1-D random periodic system: a spatially periodic system consisting of alternating layers, randomly placed on the $x$-axis.
This means that we only know which material is present at any given point $x$ in a probabilistic sense.
The 1-D version of \cref{eq2} is written as
\begin{align}\label{eq6}
\frac{\partial\psi}{\partial s}(x,\mu,s) + \mu\frac{\partial \psi}{\partial x}(x,\mu,s) &+ \Sigma_t(\mu,s)\psi(x,\mu,s) 
\\& = \frac{\delta(s)}{2}\left[ c\int_{-1}^1\int_0^\infty \Sigma_t(\mu',s')\psi(x,\mu',s')ds' d\mu' + Q(x) \right]\,. \nonumber
\end{align}
This system was chosen because we can obtain an analytical expression for the distribution function $p(\mu,s)$ of a particle's distance-to-collision in the direction $\mu$.
Then, using the identity \cite{vaslar14a}
\begin{align}\label{eq7}
\Sigma_t(\mu,s)=\frac{p(\mu,s)}{1-\int_0^sp(\mu,s')ds'},
\end{align}
one can obtain a solution for \cref{eq6}.

The numerical results presented in this paper consider transport in {\em rod geometry}, in which particles can only move in the directions $\mu = \pm 1$.
Solutions are given for a total of 72 solid-void test problems.
To analyze the accuracy of these results, we compare them against ``benchmark" numerical results, obtained by ensemble-averaging the solutions of the transport equation over a large number of physical realizations of the random system.
Furthermore, we compare the performance of the nonclassical model against the widely-used atomic mix model.

This paper is an expanded version of a recent conference paper \cite{mc15}.
The remainder of this paper is organized as follows.
In \cref{sec2} we sketch the 1-D random periodic system under consideration.
In \cref{sec3} we analytically derive the path-length distribution function for the periodic random system; explicit expressions for solid-void media are given in \cref{sec3A}.
In \cref{sec4} we define the parameters of the test problems and describe the benchmark, atomic mix, and nonclassical approaches to solve them.
In \cref{sec5} we examine the numerical results that confirm the accuracy of the nonclassical model.
We conclude with a discussion in \cref{sec6}.

\section{The 1-D Random Periodic System}\label{sec2}

Let us consider a 1-D physical system similar to the one introduced in \cite{zuc94}, consisting of alternating layers of two distinct materials (labeled 1 and 2) periodically arranged.
The period is given by $\ell = \ell_1 + \ell_2$, where $\ell_i$ represents the length of each layer of material $i \in \{1,2\}$.
A sketch of the periodic system is given in \cref{fig1}.

This periodic system is {\em randomly placed} in the infinite line $-\infty < x < \infty$, such that the probability $P_i$ of finding material $i$ in a given point $x$ is $\ell_i/\ell$.
Therefore, the cross sections and source are stochastic functions of space; that is, if $x$ is in material $i$, then
\begin{subequations}\label[pluraleq]{eq8}
\begin{align}
\Sigma_t(x) &= \Sigma_{ti}\, ,\\
\Sigma_s(x) &= c_i \Sigma_{ti}\, ,\\
Q(x) &= Q_i(x) \, ,
\end{align}
\end{subequations}
where $\Sigma_{ti}$, $c_i$, and $Q_i$ represent the total cross section, scattering ratio, and source in material $i$. 

\section{The Path-length Distribution Function}\label{sec3}

Given a physical realization of the 1-D system described in \cref{sec2}, let us examine a particle that is born (or scatters) at a point $x$ in a layer of material $i \in \{1,2\}$ with direction of flight $\mu\neq 0$.
We define $x_0$ to be the horizontal distance between $x$ (the point in which the collision or birth event took place) and the next intersection between layers in the direction $\mu$.
We also define:
\begin{subequations}\label[pluraleq]{eq9}
\begin{align}
p_{A_i}(x_0,\mu,s) &= \begin{array}{l}
\text{ the probability that a particle born or scattered in}\vspace{-10pt}\\
\text{ material $i$, at a horizontal distance $x_0$ of the next}\vspace{-10pt}\\
\text{ intersection, with direction of flight $\mu$, will travel a}\vspace{-10pt}\\
\text{ distance $s$ without colliding;}
\end{array}
\\
p_{B_i}(x_0,\mu,s)ds &= \begin{array}{l}
\text{ the probability that a particle born or scattered in}\vspace{-10pt}\\
\text{ material $i$, at a horizontal distance $x_0$ of the next}\vspace{-10pt}\\
\text{ intersection, with direction of flight $\mu$, will experience}\vspace{-10pt}\\
\text{  a collision between $s$ and $s+ds$.}
\end{array}
\end{align}
\end{subequations}
For $\mu\neq 0$, we can write
\begin{subequations}\label[pluraleq]{eq10}
\begin{align}
p_{A_i}(x_0,\mu,s) &=  \left\{
\begin{array}{ll}
e^{-\Sigma_{ti}s}, & \text{if } 0\leq s|\mu|\leq x_0\\
(e^{-\Sigma_{ti}x_0/|\mu|})(e^{-\Sigma_{tj}(s-x_0/|\mu|)}), &\text{if } x_0< s|\mu|\leq x_0+\ell_j\\
(e^{-\Sigma_{ti}(s-\ell_j/|\mu|)})(e^{-\Sigma_{tj}\ell_j/|\mu|}), &\text{if } x_0+\ell_j< s|\mu|\leq x_0+\ell\\
\,\,\,\vdots & 
\end{array}
\right.
\end{align}
and
\begin{align}
p_{B_i}(x_0,\mu,s) &=  \left\{
\begin{array}{ll}
\Sigma_{ti}, & \text{if } 0\leq s|\mu|\leq x_0\\
\Sigma_{tj}, &\text{if } x_0< s|\mu|\leq x_0+\ell_j\\
\Sigma_{ti}, &\text{if } x_0+\ell_j< s|\mu|\leq x_0+\ell\\
\,\,\,\vdots & 
\end{array}
\right.
\,\,,
\end{align}
\end{subequations}
such that
\begin{subequations}\label[pluraleq]{eq11}
\begin{align}
p_{A_i}(x_0,\mu,s) &=  \left\{
\begin{array}{ll}
e^{-\Sigma_{ti}s}, & \text{if } 0\leq s|\mu|\leq x_0  \\
e^{-\Sigma_{tj}s-(\Sigma_{ti}-\Sigma_{tj})(x_0+n\ell_i)/|\mu|}, & \text{if } x_0+n\ell< s|\mu|\leq  x_0+n\ell+\ell_j \\
e^{-\Sigma_{ti}s-(\Sigma_{tj}-\Sigma_{ti})(n+1)\ell_j/|\mu|}, & \text{if } x_0+n\ell+\ell_j< s|\mu|\leq x_0+(n+1)\ell
\end{array}
\right.
\end{align}
and
\begin{align}
p_{B_i}(x_0,\mu,s) &=  \left\{
\begin{array}{ll}
\Sigma_{ti}, & \text{if } 0\leq s|\mu|\leq x_0\\
\Sigma_{tj}, &\text{if } x_0+n\ell< s|\mu|\leq  x_0+n\ell+\ell_j\\
\Sigma_{ti}, &\text{if } x_0+n\ell+\ell_j< s|\mu|\leq x_0+(n+1)\ell
\end{array}
\right.
\,\,.
\end{align}
\end{subequations}
Here, $n=0, 1, 2, ...$; $i,j \in\{1,2\}$; $i\neq j$; and $\ell = \ell_i+\ell_j$.
It is clear that
\begin{align}\label{eq12}
p_{C_i}(x_0,\mu,s)ds &= \begin{array}{l}
\text{ the probability that a particle born or scattered in}\vspace{-10pt}\\
\text{ material $i$, at a horizontal distance $x_0$ of the next}\vspace{-10pt}\\
\text{ intersection, with direction of flight $\mu$, will experience}\vspace{-10pt}\\
\text{ its \textit{first collision} while traveling a distance between $s$}\vspace{-10pt}\\
\text{  and $s+ds$}
\end{array}
\\&= \,\,\,p_{A_i}(x_0,\mu,s)\times p_{B_i}(x_0,\mu,s)ds, \nonumber
\end{align}
and the \textit{ensemble-averaged} path-length distribution function of particles born or scattered in material $i$ with direction of flight $\mu$ is given by
\begin{align}\label{eq13}
p_i(\mu,s) &= \frac{1}{\ell_i}\int_0^{\ell_i} p_{C_i}(x_0,\mu,s) dx_0.
\end{align}
Finally, the ensemble-averaged path-length distribution function for particles born {\em anywhere} in the 1-D random periodic system with direction of flight $\mu$ is given by the weighted average
\begin{align}\label{eq14}
p(\mu,s) &= \lambda_1 p_1(\mu,s) + \lambda_2p_2(\mu,s),
\end{align}
where $\lambda_i$ is the probability that any given birth or scattering event takes place in material $i$.
It is easy to see that if $\Sigma_{t1}=\Sigma_{t2}$, \cref{eq11} to \labelcref{eq14} yield the exponential
\begin{align}\label{eq15}
p(\mu,s)=p(s)=\Sigma_{t1}e^{-\Sigma_{t1}s},
\end{align}
as given in \cref{eq1}. 

\subsection{Solid-Void Medium}
\label{sec3A}

The numerical results included in this paper are for solid-void systems.
We define material 2 as the void, such that $\lambda_2=\Sigma_{t2}=Q_2=0$, $\lambda_1=1$, and $p(\mu,s) = p_1(\mu,s)$.
Depending on the lengths $\ell_i$ of the material layers, \cref{eq13} yields the following expressions for $p(\mu,s)$:
\begin{subequations}\label[pluraleq]{eq16}
\begin{itemize}
\item Case 1: $\ell_1<\ell_2$
\end{itemize}
\begin{align}
p(\mu,s) = \left\{
\begin{array}{ll}
\frac{\Sigma_{t1}}{\ell_1}(n\ell +\ell_1-s|\mu|)e^{-\Sigma_{t1}(s-n\ell_2/|\mu|)}, & \text{if } n\ell\leq s|\mu| \leq n\ell+\ell_1\\
0, & \text{if } n\ell+\ell_1 \leq s|\mu| \leq n\ell+\ell_2\\
\frac{\Sigma_{t1}}{\ell_1}(s|\mu|-n\ell-\ell_2)e^{-\Sigma_{t1}[s-(n+1)\ell_2/|\mu|]}, & \text{if } n\ell+\ell_2 \leq s|\mu| \leq (n+1)\ell\\
\end{array}
\right.
\end{align}
\begin{itemize}
\item Case 2: $\ell_1=\ell_2$
\end{itemize}
\begin{align}
p(\mu,s) = \left\{
\begin{array}{ll}
\frac{\Sigma_{t1}}{\ell_1}(n\ell +\ell_1-s|\mu|)e^{-\Sigma_{t1}(s-n\ell_2/|\mu|)}, & \text{if } n\ell\leq s|\mu| \leq n\ell+\ell_1\\
\frac{\Sigma_{t1}}{\ell_1}(s|\mu|-n\ell-\ell_2)e^{-\Sigma_{t1}[s-(n+1)\ell_2/|\mu|]}, & \text{if } n\ell+\ell_2 \leq s|\mu| \leq (n+1)\ell\\
\end{array}
\right.
\end{align}
\begin{itemize}
\item Case 3: $\ell_1>\ell_2$
\end{itemize}
\begin{align}
p(\mu,s) = \left\{
\begin{array}{ll}
\frac{\Sigma_{t1}}{\ell_1}(n\ell +\ell_1-s|\mu|)e^{-\Sigma_{t1}(s-n\ell_2/|\mu|)}, & \\
\hspace{6cm}\text{if } n\ell\leq s|\mu| \leq n\ell+\ell_2 & \\
\frac{\Sigma_{t1}}{\ell_1}[(n\ell +\ell_2-s|\mu|)(1-e^{\Sigma_{t1}\ell_2/|\mu|}) +\ell_1-\ell_2]e^{-\Sigma_{t1}(s-n\ell_2/|\mu|)}, & \\
\hspace{6cm}\text{if } n\ell+\ell_2\leq s|\mu| \leq n\ell+\ell_1 & \\
\frac{\Sigma_{t1}}{\ell_1}(s|\mu|-n\ell-\ell_2)e^{-\Sigma_{t1}[s-(n+1)\ell_2/|\mu|]}, & \\
\hspace{6cm}\text{if } n\ell+\ell_1 \leq s|\mu| \leq (n+1)\ell & 
\end{array}
\right.
\end{align}
\end{subequations}
where $n=0, 1, 2, ...$ .
The first and second moments of $p(\mu,s)$ in \cref{eq16} are given by
\begin{subequations}\label[pluraleq]{eq17}
\begin{align}
\overline{s} &= \int_0^\infty sp(\mu,s)ds = \frac{\ell_1+\ell_2}{\Sigma_{t1}\ell_1}\, ,\\
\overline{s^2}(\mu) &= \int_0^\infty s^2p(\mu,s)ds = \frac{2\ell_1+4\ell_2}{\Sigma_{t1}^2\ell_1}+\frac{\ell_2^2}{\Sigma_{t1}\ell_1|\mu|}\left(\frac{e^{\Sigma_{t1}\ell_1/|\mu|}+1}{e^{\Sigma_{t1}\ell_1/|\mu|}-1}\right)\, .
\end{align}
\end{subequations}
We point out that the {\em mean free path} $\overline{s}$ does not depend on the direction $\mu$ and it is equivalent to the inverse of the volume-averaged total cross section.
On the other hand, the {\em mean square free path} $\overline{s^2}$ is a function of $|\mu|$.

\Cref{fig2} depicts examples of path-length distributions and nonclassical cross sections assuming $\Sigma_{t1}=1$ and direction of flight $\mu=\pm 1$.
\Cref{fig2a,fig2c,fig2e} show a comparison between numerically obtained (through Monte Carlo) $p(s)$ and the analytical expressions given in \cref{eq16}. 
\Cref{fig2b,fig2d,fig2f} show the corresponding $\Sigma_t(s)$ obtained with \cref{eq7}. 
The ``saw-tooth" behavior of $\Sigma_t(s)$ is consistent with the physical process and can be easily understood.
For instance, in the case of $\ell_1=\ell_2=1$ (Case 2):
\begin{itemize}
\item[\textbf{1.}] A particle is born or scatters in material 1. The path-length $s$ is set to 0, and $\Sigma_t(0) = \Sigma_{t1} = 1$\vspace{-8pt}
\item[\textbf{2.}] At $s=1$, the $x$-coordinate {\em must be} in material 2. Thus, $\Sigma_t(1) = \Sigma_{t2} = 0$\vspace{-8pt}
\item[\textbf{3.}] At $s=2$, the $x$-coordinate {\em must be} back in material 1. Thus, $\Sigma_t(2) = \Sigma_{t1} = 1$\vspace{-8pt}
\end{itemize}
The exceptions would be particles born exactly at {\em interface points}, which form a set of measure zero.

\section{Test Problems and Models}\label{sec4}

The test problems simulated in this paper consider only {\em rod geometry} transport (particles can only travel in the directions $\mu = \pm1$) taking place in a finite 1-D random periodic system with vacuum boundaries.
The classical transport equation is written as
\begin{subequations}\label[pluraleq]{eq18}
\begin{align}
&\pm \frac{\partial \Psi^{\pm}}{\partial x}(x) + \Sigma_t(x)\Psi^{\pm}(x) 
= \frac{\Sigma_s(x)}{2}\left[\Psi^{+}(x)+\Psi^{-}(x)\right]+ \frac{Q(x)}{2}\,,
\,\,\,-X\leq x\leq X,\\
&\Psi^+(-X) = \Psi^-(X) = 0\,,
\end{align}
\end{subequations}
where $\Psi^{\pm}(x) = \Psi(x,\mu=\pm 1)$ and the stochastic parameters $\Sigma_t(x)$, $\Sigma_s(x)$, and $Q(x)$ are given by \cref{eq8}.

We are interested in how accurately the nonclassical model predicts the ensemble-averaged scalar flux $\bl\Phi\bg$ (over all physical realizations).
To this end, we compare the nonclassical results against ``benchmark" results obtained by averaging the solutions of the transport equation over a large number of physical realizations of the random system.
Finally, we compare the performance of the nonclassical model against the widely-known atomic mix model.

We consider 2 sets of problems ({$\seta$ and $\setb$), each divided in 3 subsets according to the choices of the lengths $\ell_i$ of the material layers.
For each subset we present results for 12 different choices of scattering ratios ranging from purely absorbing to diffusive; namely $c_1 \in$ \{0.0; 0.1; 0.2; 0.3; 0.4; 0.5; 0.6; 0.7; 0.8; 0.9; 0.95; 0.99\}.
We assume vacuum boundaries at $x = \pm 10$.
Material 2 is defined as void, and the parameters of material 1 are given in \cref{tab1}.
The source $Q_1(x)$ is defined as 
\begin{align}\label{eq19}
Q_1(x) = \left\{
\begin{array}{cl}
q_1, & \text{if} -0.5\leq x\leq 0.5\\
0, &\text{otherwise}\\
\end{array}
\right .\, ; 
\end{align}
that is, particles are born {\em near the center} of the random system.
The reason for this choice of source region can be visualized in \cref{fig3}, in which the ``wavy" pattern that arises from the periodic structure can be seen in \cref{fig3a}.
If we allow $Q_1=1$ for $-X\leq x\leq X$, the solution is smoother, and the pattern is harder to identify (\cref{fig3b}).    

\subsection{The Benchmark Model}

The random quality of the 1-D system arises from its random placement in the $x$-axis.
To obtain a single physical realization one can simply choose a continuous segment of two full layers (one of each material) and randomly place the coordinate $x=0$ in this segment, which also defines the boundaries $\pm X$.

Given this fixed realization of the system, the cross sections and source in \cref{eq18} are now deterministic functions of space.
We use the diamond spatial differencing scheme with mesh interval $\triangle x=2^{-7}$ to solve for the angular flux $\Psi$, obtaining the scalar flux $\Phi(x) = \Psi^+(x)+\Psi^-(x)$ (see \cref{fig4}).
This procedure is repeated for different realizations of the random system.
Finally, we calculate the ensemble-averaged {\em benchmark} scalar flux $\bl\Phi_B\bg(x)$ by averaging the resulting scalar fluxes over all physical realizations (as shown in \cref{fig3a}). 

Clearly, the number of different realizations that can be computed is limited by the spatial discretization, with the maximum number of different realizations being $\ell/\triangle x$.
For all test problems in this paper, differences in the numerical results for $\bl\Phi_B\bg(x)$ were negligible when increasing the number of mesh intervals and realizations.
Thus, we have concluded that these benchmark results are adequately accurate for the scope of this work.

\subsection{The Atomic Mix Model}

The {\em atomic mix model} \cite{pom91,dum00} consists of replacing in the classical transport equation the stochastic parameters (cross sections and source) by their volume-averages.
This model is known to be accurate in 1-D geometry when the material layers are optically thin.
The atomic mix equation in rod geometry for the test problems in this paper is given by
\begin{subequations}\label[pluraleq]{eq20}
\begin{align}\label{eq19a}
&\pm \frac{\partial \bl\Psi^{\pm}\bg}{\partial x}(x) + \bl\Sigma_t\bg\bl\Psi^{\pm}\bg(x) 
= \frac{\bl\Sigma_s\bg}{2}\left[\bl\Psi^{+}\bg(x)+\bl\Psi^{-}\bg(x)\right]+ \frac{\bl Q\bg(x)}{2}\,,\\
& \hspace{12cm} -X\leq x\leq X,\nonumber\\
&\bl\Psi^+\bg(-X) = \bl\Psi^-\bg(X) = 0\,,
\end{align}
where 
\begin{align}
\bl\Sigma_t\bg &= P_1\Sigma_{t1} + P_2\Sigma_{t2} = \frac{\ell_1}{\ell}\Sigma_{t1},\\
\bl\Sigma_s\bg &= P_1c_1\Sigma_{t1} + P_2c_2\Sigma_{t2} = \frac{\ell_1}{\ell}c_1\Sigma_{t1},\\
\bl Q\bg(x) &= P_1 Q_1(x) + P_2 Q_2(x) = \frac{\ell_1}{\ell}Q_1(x). \label{eq20e}
\end{align}
\end{subequations}
We solve \cref{eq20} for the ensembled-averaged angular flux $\bl\Psi\bg$ using a diamond spatial differencing scheme with mesh interval $\triangle x=2^{-7}$.
The ensemble-averaged {\em atomic mix} scalar flux is given by $\bl\Phi_{AM}\bg(x) = \bl\Psi^+\bg(x)+\bl\Psi^-\bg(x)$.
An example is depicted in \cref{fig5}.

\subsection{The Nonclassical Model}

For the rod geometry test problems included in this work, we rewrite the nonclassical \cref{eq6} in an initial value form (cf.\ \cite{vaslar14a}) as
\begin{subequations}\label[pluraleq]{eq21}
\begin{align}
&\frac{\partial\psi^{\pm}}{\partial s}(x,s) \pm \frac{\partial \psi^{\pm}}{\partial x}(x,s) + \Sigma_t(s)\psi^{\pm}(x,s)  = 0,\,\,\, -X\leq x\leq X,\,\, s>0 \label{eq21a}\\
& \psi^{\pm}(x,0)= \frac{c}{2} \int_0^\infty \Sigma_t(s')[\psi^{+}(x,s')+\psi^{-}(x,s')]ds' + \frac{\bl Q\bg(x)}{2}, \,\,\, -X\leq x\leq X,\label{eq21b}\\
& \psi^+(-X,s) = \psi^-(X,s) = 0\,,\,\,\, s\geq 0\,,
\end{align}
\end{subequations}
where $\psi^{\pm}(x,s) = \psi(x,\mu=\pm 1, s)$, $\bl Q\bg(x)$ is given by \cref{eq20e}, and the nonclassical cross section $\Sigma_t(s)=\Sigma(\mu=\pm 1,s)$ is given by \cref{eq7,eq16} (see \cref{fig2}).

For the numerical solution of this system, we can interpret the path-length $s$ as a pseudo-time variable.
We then solve \cref{eq21} using a finite volume method with explicit pseudo-time discretization according to \cite{hll83}.
Specifically, we adapt the scheme introduced in \cite{kry13} for moment models of the nonclassical transport equation.

This method is of first order in the pseudo-time variable $s$ and in the spatial variable $x$.
We choose a uniform grid $(x_m,s^n)$, where $x_{m+1} = x_{m} + \Delta x$ for all $m\in\mathbb{Z}$, and $s^{n+1} = s^n + \Delta s$ for all $n\in\mathbb{N}_0$.
Furthermore, we define $\psi_{m}^{n,\pm}= \psi^\pm(x_{m},s^n)$, $Q_m = \bl Q\bg (x_m)$, and $\Sigma_t^n=\Sigma_t(s^n)$.
The fully discretized system reads
\begin{subequations}\label[pluraleq]{eq22}
\begin{align}
	& \frac{\psi^{n+1,\pm}_m-\psi^{n,\pm}_m}{\Delta s} \pm \frac{\psi^{n,\pm}_{m+1}-\psi_{m-1}^{n,\pm}}{2\Delta x} - 
			\frac{\psi^{n,\pm}_{m+1}-2\psi_m^{n,\pm}+\psi_{m-1}^{n,\pm}}{2\Delta x}   + \Sigma_t^n \psi^{n,\pm}_m = 0, \label{eq22a}\\
	&  \psi_m^{0,\pm} = \frac{c}{2}\sum\limits_{n=0}^\infty \omega_n \Sigma_t^n\left( \psi_m^{n,+} + \psi^{n,-}_m\right)  +  \frac{Q_m}{2},\label{eq22b}	
\end{align}
\end{subequations}
for some infinite quadrature rule given by the weights $\omega_n$.
The second order central differences arise as a numerical diffusion term, which is typical for HLL finite volume schemes.

In our calculations we cut off the integration at $s_{\text{max}}=4X=40$ and use the trapezoidal rule.
We use the same mesh interval $\triangle x=2^{-7}$ as for the previous models, and a CFL number $0.5$ (that is, $\triangle s = 2^{-8}$). 
Because of the coupling of the initial value to the full solution in \cref{eq21}, this system is solved in a source-iteration manner, where we iterate between \cref{eq22a,eq22b}.
Finally, the ensemble-averaged {\em nonclassical} scalar flux is given by $\bl\Phi_{NC}\bg(x) = \int_0^{40}[\psi^+(x,s)+\psi^-(x,s)]ds$.
An example is depicted in \cref{fig6}.

It was shown in \cite{kry13} that the contraction rate for the source iteration is given by the scattering ratio $c$.
The maximum number of source iterations to converge the solution in problem set $\seta$ was 417 (problem $\seta_3$ with $c_1=0.99$); and in problem set $\setb$ was 251 (problem $\setb_3$ with $c_1=0.99$).

\section{Numerical Results}\label{sec5}

The atomic mix model inherently approximates the path-length distribution function by the exponential $p(s) = \bl\Sigma_t\bg e^{-\bl\Sigma_t\bg s}$.
The nonclassical model uses the correct $p(\mu,s)$ that was analytically obtained in \cref{eq16}.
In this section we compare the accuracy of these two models in predicting the benchmark solutions obtained for the test problem sets $\seta$ and $\setb$. 

For a better analysis of these results, we define the relative errors of the models with respect to the benchmark solutions as
\begin{subequations}\label[pluraleq]{eq23}
\begin{align}
 Err_{AM}&= \frac{\bl\Phi_{AM}\bg(x)-\bl\Phi_B\bg(x)}{\bl\Phi_B\bg(x)}=\text{Atomic Mix Relative Error},\\
 Err_{NC}&= \frac{\bl\Phi_{NC}\bg(x)-\bl\Phi_B\bg(x)}{\bl\Phi_B\bg(x)}=\text{Nonclassical Relative Error}.
\end{align}
\end{subequations}

\subsection{Problem Set $\seta$}

The lengths of the material 1 layers in this set are the same order as a mean free path; that is, $\ell_1\Sigma_{t1} = O(1)$.
It has been shown \cite{larvas05} that, in the diffusive asymptotic limit, the diffusion coefficient of such problems is correctly estimated by the atomic mix model.
For the rod geometry problems in set $\seta$, this diffusion coefficient is given by
\begin{align}\label{eq24}
D = \frac{\ell_1+\ell_2}{\Sigma_{t1}\ell_1} = \frac{1}{\bl\Sigma_t\bg} =
\left\{
\begin{array}{cl}
3.0 & \text{for set $\seta_1$}\\
2.0 & \text{for set $\seta_2$}\\
1.5 & \text{for set $\seta_3$}\\
\end{array}
\right . \, .
\end{align}  
Therefore, we expect the atomic mix predictions of the ensemble-averaged scalar flux to {\em improve} as the scattering ratio increases and the system becomes more diffusive.

On the other hand, the diffusion coefficient obtained by applying the same asymptotic analysis to the the nonclassical equation (see \cref{appa}) is given by 
\begin{align}\label{eq25}
D_{NC} = \frac{1}{2}\frac{\overline{s^2}}{\overline{s}} \approx 
\left\{
\begin{array}{cl}
3.0277 & \text{for set $\seta_1$}\\
2.0410 & \text{for set $\seta_2$}\\
1.5137 & \text{for set $\seta_3$}\\
\end{array}
\right . \, ,
\end{align}  
where $\overline{s}$ and $\overline{s^2}$ are defined in \cref{eq17}.
The solution of the nonclassical transport equation has been shown to converge to the solution of the nonclassical diffusion equation in the diffusive asymptotic limit \cite{ans16}.
Thus, we expect the nonclassical predictions of the ensemble-averaged scalar flux to {\em deteriorate} as the system becomes diffusive, underestimating the correct solution.

\Cref{fig7} depicts the ensemble-averaged scalar fluxes obtained with each model for the purely absorbing case (\cref{fig7a,fig7c,fig7e}) and for the diffusive case $c_1=0.99$ (\cref{fig7b,fig7d,fig7f}).
The benchmark solutions present a sinuous shape due to the periodic structure of the random systems.
This pattern becomes less noticeable as the solid/void ratio increases, and as the system becomes more diffusive.
It is important to point out that the nonclassical model is able to capture this sinuous behavior, while the atomic mix model yields a smooth curve.

It is easier to analyze the accuracy of these models by examining the relative errors to the benchmark solution.
\Cref{figerrA1,figerrA2,figerrB1,figerrB2,figerrC1,figerrC2} show the (absolute) percentage error of the nonclassical and atomic mix predictions of the ensemble-averaged scalar flux with respect to the benchmark solutions.
The error plots confirm the theoretical predictions; atomic mix becomes more accurate as the system becomes more diffusive, while the accuracy of the nonclassical model decreases.

The nonclassical model clearly outperforms atomic mix for all the problems in $\seta_1$ and for most of the problems in sets $\seta_2$ and $\seta_3$.
The exceptions take place for the cases $c_1=0.95$ and $c_1=0.99$, in which the accuracy of the atomic mix model overtakes that of the nonclassical.
\Cref{tab2,tab3,tab4} show that the nonclassical model tends to
underestimate the scalar flux, while atomic mix overestimates the solution.
The nonclassical model never reaches an error larger than 3.7\% in estimating the solutions' peak (at $x=0$).
On the other hand, the atomic mix estimate exceeds 5\% error in several problems, reaching a maximum of 8.24\%.

It can also be seen from the results at the boundaries that the atomic mix model generates a solution with a large tail and it greatly overestimates the outgoing flux, in some problems by several orders of magnitude.
The nonclassical model, however, never reaches an error larger than 4.7\%.

\subsection{Problem Set $\setb$}

Following the work presented in \cref{sec3A}, \cref{fig14} shows the path-length distributions and nonclassical cross sections of problem set $\setb$.
We have chosen the parameters of this set such that:
\begin{itemize}
\item[i.] The optical thickness of each layer of material 1 is one order of magnitude larger than a mean free path: $\ell_1\Sigma_{t1} = 10$;
\vspace{-4pt}
\item[ii.] The volume-averaged parameters remain the same in all problems in the set: $\bl \Sigma_t \bg = \bl q_1 \bg = 0.5$.
\end{itemize}
The large optical thickness implies that the problems in this set are {\em not} the type of problems for which the atomic mix model is known to yield the correct aymptotic diffusive limit.
By fixing the volume-averaged parameters, the atomic mix model will yield exactly the same ensemble-averaged scalar flux for all problems in set $\setb$ (which is the same as in $\seta_2$).
The goal is to investigate whether the nonclassical model will outperform atomic mix for the diffusive cases.

\Cref{fig15} depicts the ensemble-averaged scalar fluxes obtained with each model for the purely absorbing case (\cref{fig15a,fig15c,fig15e}) and for the diffusive case $c_1=0.99$ (\cref{fig15b,fig15d,fig15f}).
The sinuous pattern of the benchmark solution is easier to notice in set $\setb_3$, with the largest solid/void ratio.
As in the case in set $\seta$, the nonclassical model is able to capture the sinuous behavior.
The atomic mix model generates the same smooth solution for each choice of $c_1$, unable to capture the differences in the scalar flux caused by the different choices of $\ell_i$, $\Sigma_{ti}$, and $q_i$.

\Cref{figerrD1,figerrD2,figerrE1,figerrE2,figerrF1,figerrF2} show the percentage error of the nonclassical and atomic mix predictions of the ensemble-averaged scalar flux with respect to the benchmark solutions {\em in logarithmic scale}.
The changes in the accuracy of both models have a different pattern than in problem set $\seta$.
The atomic mix solutions tend to grossly overestimate the ensemble-averaged scalar flux in most of the system, with errors at $x=0$ reaching 36\% as seen in \cref{tab5,tab6,tab7}.
Once $x$ approaches the boundaries, the atomic mix model systematically underestimates the solution, with errors in the outgoing flux exceeding 50\% in most test problems and reaching over 80\% in the least diffusive systems.

Once again, the nonclassical model underestimates the solution in diffusive systems.
For most problems the nonclassical error in estimating the ensemble-averaged scalar flux at $x=0$ is less than 4\%.
The exceptions are the most diffusive problems, with scattering ratios $c_1 = 0.95$ and $c_1 = 0.99$.
Nevertheless, even in these diffusive cases the nonclassical model greatly outperforms the atomic mix approach.

\section{Conclusion}\label{sec6}

This work presents an investigation of the accuracy of the nonclassical transport theory in estimating the ensemble-averaged scalar flux in 1-D random periodic media.
The analytical portion of the paper considers transport in a {\em slab} consisting of alternating layers of any 2 materials.
The following simplifying assumptions are made for the numerical simulations: (i) the 1-D system is a periodic arrangement of {\em solid and void} layers randomly placed in the $x$-axis; and (ii) particle transport takes place in {\em rod geometry}.
This paper is an expanded version of a recent conference paper \cite{mc15}, in which numerical solutions for the nonclassical transport equation were provided for the first time. 

A total of 72 test problems are analyzed.
We show that the nonclassical theory greatly outperforms the atomic mix model in estimating the ensemble-averaged scalar flux for most problems and that it qualitatively preserves the sinuous shape of the solution.
The few cases in which atomic mix is more accurate are part of a class of diffusive problems in which the atomic mix model is known to converge to the correct diffusive limit (diffusive problems in set $\seta$).
In this small subset of problems the nonclassical model converges to a diffusion solution with an unphysically large diffusion coefficient, causing the nonclassical solution to underestimate the ensemble-averaged scalar flux.
However, for diffusive problems that are {\em not} in the atomic mix limit (set $\setb$), the nonclassical model is clearly superior to the atomic mix approach.

This gain in accuracy comes at a cost: the path-length distribution function $p(s)$ (and its corresponding $\Sigma_t(s)$) must be known in order to solve the nonclassical transport equation.
Despite the extra work, it is our expectation that the gain in accuracy will prove the effort worthwhile in the important nuclear system where nonclassical transport takes place, such as in Pebble Bed and Boiling Water reactor cores.
In particular, the nonclassical theory represents an alternative to current methods that might yield more accurate estimates of the eigenvalue and eigenfunction in a criticality calculation.

Future work includes (i) performing a thorough numerical investigation of the nonclassical theory in slab geometry to further validate our analytical results; (ii) comparing the gain in accuracy against other models and experimental data; and (iii) dropping the periodic assumption to investigate results in more realistic random media.
We point out that step (iii) cannot be performed with the analytical approach to obtain the path-lengths presented in this paper.
It requires either a numerical approach to estimate $p(\mu,s)$, or a (much) more complex mathematical theory.

\section*{Acknowledgments}

This paper was prepared by Richard Vasques and Rachel Slaybaugh under award number NRC-HQ-84-14-G-0052 from the Nuclear Regulatory Commission.
The statements, findings, conclusions, and recommendations are those of the authors and do not necessarily reflect the view of the U.S. Nuclear Regulatory Commission.

\appendix

\begin{center}
\textbf{APPENDIX}
\end{center}

\section{1-D Asymptotic Analysis}\label{appa}

Following \cite{vaslar14a},  we scale the parameters of \cref{eq6} such that $\Sigma_t = O(1)$, $ 1-c = O(\varepsilon^2) $, $Q=O(\varepsilon^2)$,  $\partial \psi / \partial s = O(1)$, and $\mu \partial \psi/\partial x = O(\eps)$, with $\varepsilon \ll 1$.
In this scaling, \cref{eq6} becomes
  \begin{align}
    &\frac{\partial \psi}{\partial s}  (x, \mu,s) 
      + \eps\mu\frac{\partial\psi}{\partial x}(x, \mu, s)
       + \Sigma_t(\mu,s) \psi( x, \mu, s)   = \label{eqap1}\\
   & \quad\quad= \frac{\delta(s)}{2}\int_{-1}^1\int_0^{\infty}[1-\eps^2(1-c)]\Sigma_t(\mu',s') 
      \psi(x, \mu', s') \, ds' d\mu' + + \varepsilon^2 \delta(s)\frac{Q(x)}{2}\nonumber \, .
  \end{align}
Let us define $\hat\psi( x, \mu, s)$ such that
  \begin{align}\label{eqap2}
   \psi( x, \mu, s) &\equiv 
          \hat\psi(x, \mu, s) \frac{e^{-\int_0^s \Sigma_t(\mu,s') ds'}}{\overline{s}}\,,
  \end{align}
where $\overline{s} = \frac{1}{2}\int_{-1}^1\int_0^\infty s p(\mu,s)dsd\mu$.
Then, using \cref{eq4}, \cref{eqap2} becomes the following equation for $\hat\psi(x, \mu, s)$:
  \begin{align}\label{eqap3}
    &\frac{\partial \hat\psi}{\partial s} (x, \mu,s) 
      + \varepsilon \mu\frac{\partial\hat\psi}{\partial x}(x,\mu, s)  = \\
   & \quad= \frac{\delta(s)}{2} \int_{-1}^1 \int_0^{\infty} [1-\eps^2(1-c)] p(\mu',s')
      \hat\psi(x, \mu', s') \, ds' d\mu'       + \varepsilon^2 \delta(s) \overline{s} \frac{Q(x)}{2} \,.\nonumber
  \end{align}
This equation is mathematically equivalent to:
   \begin{subequations}\label[pluraleq]{eqap4}
   \begin{align}
      &\frac{\partial \hat\psi}{\partial s} (x, \mu,s) 
         + \varepsilon \mu\frac{\partial\hat\psi}{\partial x}(x, \mu, s) = 0 \,, \quad s > 0 \,,\label{eqap4a}\\
       &\hat\psi(x, \mu, 0)  =\frac{1}{2}\int_{-1}^1 [1-\eps^2(1-c)] \int_0^{\infty} p(\mu',s')
\hat\psi(x, \mu', s') ds' d\mu' + \varepsilon^2 \overline{s} \frac{Q(x)}{2} \,,
   \end{align}
   \end{subequations}
where $\hat\psi(x,\mu,0) = \hat\psi(x,\mu,0^+)$. Integrating \cref{eqap4a} over $0 < s' < s$ we obtain:
   \begin{align}
      \hat\psi( x, \mu, s)  &= \hat\psi(x,\mu, 0) - \varepsilon \mu\frac{\partial}{\partial x} \int_0^s \hat\psi(x, \mu, s') \, ds' \\
      & = \frac{1}{2}\int_{-1}^1 [1-\eps^2(1-c)] \int_0^{\infty} p(\mu',s')
\hat\psi(x, \mu', s') ds' d\mu' + \nonumber\\
         &\hspace{3.5cm} + \varepsilon^2 \overline{s} \frac{Q(x)}{2} - \varepsilon \mu\frac{\partial}{\partial x} \int_0^s \hat\psi(x, \mu, s') \, ds' \,.\nonumber
   \end{align}
Introducing into this equation the ansatz 
   \begin{equation}
 \hat\psi(x, \mu, s) =  \sum_{n=0}^{\infty} \varepsilon^n
       \hat\psi_n(x, \mu, s) 
       \end{equation}
and equating the coefficients of different powers of $\varepsilon$, we obtain for $n \ge 0$:
   \begin{align}\label{eqap7}
\hat\psi_n(x,\mu, s) &= \frac{1}{2}\int_{-1}^1 \int_0^{\infty} p(\mu',s')
\hat\psi_n(x, \mu', s') ds' d\mu'  - \mu\frac{\partial}{\partial x} \int_0^s \hat\psi_{n-1}(x, \mu, s') \, ds' \\
& \quad\quad -\frac{1-c}{2}\int_{-1}^1 \int_0^{\infty} p(\mu',s')
\hat\psi_{n-2}(x, \mu', s') ds' d\mu' + \delta_{n,2} \overline{s} \frac{Q( x)}{2} \,, 
   \nonumber
   \end{align}
with $\hat\psi_{-1}=\hat\psi_{-2}=0$.
\Cref{eqap7} with $n=0$ has the general solution
   \begin{equation}
      \hat\psi_0(x, \mu, s) = \frac{\hat\phi_0(x)}{2} \,,
   \end{equation}
where $\hat\phi_0(x)$ is undetermined at this point.
For $n=1$, \cref{eqap7} has a particular solution of the form:
    \begin{equation}
      \hat\psi^{part}_1(x, \mu, s) = - \frac{s\mu}{2}\frac{d \hat\phi_0}{d x}(x) \,,
   \end{equation}   
and its general solution is given by 
   \begin{equation}
      \hat\psi_1( x, \mu, s) =  \frac{1}{2}\left[\hat\phi_1( x) - s\mu\frac{d \hat\phi_0}{d x}(x)\right] \,,
  \end{equation}  
where $\hat\phi_1(x)$ is undetermined.

 \Cref{eqap7} with $n=2$ has a solvability condition, which is obtained by operating on it by $\int_{-1}^1\int_0^{\infty} p(\mu,s) ( \cdot ) ds d\mu$; the solvability condition yields
   \begin{align}\label{eqap11}
    0 = \frac{1}{2}\int_{-1}^1\int_0^{\infty}p(\mu,s)&\left(\frac{(s\mu)^2}{2} \frac{d^2\hat\phi_0}{dx^2}(x)\right)ds d\mu \\
    & - \frac{1-c}{2} \int_{-1}^1\int_0^{\infty} p(\mu,s) \hat\phi_0( x) \, ds d\mu + \overline{s} Q( x)\,.\nonumber 
   \end{align}
Thus, using the fact that $\int_{0}^\infty p(\mu,s)ds =1$, we can rewrite \cref{eqap11} as:
\begin{subequations}\label{eqap12}
\begin{align}
      -D_{NC}\frac{d^2\hat\phi_0}{dx^2}(x) + \frac{1-c}{\overline{s} } \hat\phi_0(x) = Q(x)\,,\label{eqap12a}
      \end{align}
      where $D_{NC}$ is the nonclassical diffusion coefficient given by
      \begin{align}\label{eqap12b}
      D_{NC} = \frac{1}{4\overline{s}}\int_{-1}^1\mu^2\int_{0}^\infty s^2p(\mu,s) dsd\mu \, .
   \end{align}
   \end{subequations}
Therefore, the solution $\psi(x, \mu, s)$ of \cref{eqap3} satisfies
   \begin{equation}\label{eqap13}
      \psi(x, \mu, s) = \frac{\hat\phi_0(x)}{2} \frac{e^{- \int_0^s \Sigma_t(\mu, s') ds'}} 
         {\overline{s}} + O(\varepsilon) \,,
   \end{equation} 
where $\hat\phi_0(x)$ satisfies \cref{eqap12}.
The classical angular flux can be obtained to leading order by integrating \cref{eqap13} over $0 < s < \infty$.
For transport in {\em rod geometry}, \cref{eqap12b} yields
\begin{align}
D_{NC} = \frac{1}{2}\frac{\overline{s^2}}{\overline{s}},
\end{align} 
where $\overline{s^2} = \int_0^\infty s^2 p(s)ds$.

\begin{singlespace}
\bibliography{references}
\bibliographystyle{ans}
\end{singlespace}

\pagebreak
\begin{table}[hp]
\centering
\caption{Parameters of test problems}
\label{tab1} 
\begin{tabular}{||c|c|c|c|c||c|c|c|c|c||} \hline \hline
\textbf{Set}  & $\ell_1$ & $\ell_2$ & $\Sigma_{t1}$ &$q_1$ & \textbf{Set}  & $\ell_1$ & $\ell_2$ & $\Sigma_{t1}$ &$q_1$ \\ \hline\hline
$\seta_1$ & 0.5 & 1.0 & 1.0 & 1.0 & $\setb_1$ & 20/3 & 40/3 & 1.5 & 1.5\\
\hline
$\seta_2$ & 1.0 & 1.0 & 1.0 & 1.0 & $\setb_2$ & 10 & 10 & 1.0 & 1.0\\
\hline
$\seta_3$ & 1.0 & 0.5 & 1.0 & 1.0 & $\setb_3$ & 40/3 & 20/3 & 0.75 & 0.75\\
 \hline\hline  
  \end{tabular}
\end{table}

\pagebreak
\begin{table}[hp]
\centering
\caption{Ensemble-averaged scalar fluxes for problem set $\seta_1$}
\label{tab2} 
\begin{tabular}{||c|c||c|c|c||c|c||} \hline \hline
  & $c$ & $\bl\phi_B\bg$ & $\bl\phi_{AM}\bg$ &$\bl\phi_{NC}\bg$ & $Err_{AM}$ & $ Err_{NC}$\\ \hline\hline
& 0.0 & 0.1420 & 0.1537 & 0.1421 & 0.0824 & 0.0006 \\
\cline{2-7}
& 0.1 & 0.1509 & 0.1628 & 0.1509 & 0.0787 & 0.0002 \\
\cline{2-7}
& 0.2 & 0.1614 & 0.1734 & 0.1613 & 0.0747 & -0.0001 \\
\cline{2-7}
& 0.3 & 0.1740 & 0.1862 & 0.1738 & 0.0706 & -0.0006 \\
\cline{2-7}
& 0.4 & 0.1895 & 0.2021 & 0.1893 & 0.0662 & -0.0012 \\
\cline{2-7}
$x=0$ & 0.5 & 0.2094 & 0.2223 & 0.2091 & 0.0616 &  -0.0019 \\
\cline{2-7}
& 0.6 & 0.2360 & 0.2493 & 0.2353 & 0.0567 & -0.0026 \\
\cline{2-7}
& 0.7 & 0.2735 & 0.2876 & 0.2725 & 0.0515 & -0.0036 \\
\cline{2-7}
& 0.8 & 0.3316 & 0.3469 & 0.3300 & 0.0462 & -0.0048 \\
\cline{2-7}
& 0.9 & 0.4360 & 0.4541 & 0.4333 & 0.0413 & -0.0063 \\
\cline{2-7}
& 0.95 & 0.5287 & 0.5496 & 0.5249 & 0.0397 & -0.0072 \\
\cline{2-7}
& 0.99 & 0.6472 & 0.6728 & 0.6421 & 0.0395 & - 0.0079  \\
\hline\hline
& 0.0 & 0.0063 & 0.0071 & 0.0061 & 0.1294 & -0.0326 \\
\cline{2-7}
& 0.1 & 0.0076 & 0.0085 & 0.0074 & 0.1128 & -0.0313 \\
\cline{2-7}
& 0.2 & 0.0093 & 0.0103 & 0.0091 & 0.0972 & -0.0301 \\
\cline{2-7}
& 0.3 & 0.0116 & 0.0126 & 0.0113 & 0.0826& -0.0289 \\
\cline{2-7}
& 0.4 & 0.0148 & 0.0158 & 0.0143 & 0.0693 & -0.0278 \\
\cline{2-7}
$x=10$ & 0.5 & 0.0191 & 0.0202 & 0.0186 & 0.0571 & -0.0267 \\
\cline{2-7}
& 0.6 & 0.0255 & 0.0267 & 0.0248 & 0.0464 & -0.0256 \\
\cline{2-7}
& 0.7 & 0.0354 & 0.0367 & 0.0345 & 0.0371 & -0.0244 \\
\cline{2-7}
& 0.8 & 0.0520 & 0.0535 & 0.0508 & 0.0297 & -0.0231 \\
\cline{2-7}
& 0.9 & 0.0841 & 0.0863 & 0.0823 & 0.0251 & -0.0216 \\
\cline{2-7}
& 0.95 & 0.1141 & 0.1169 & 0.1117 & 0.0246 & -0.0206 \\
\cline{2-7}
& 0.99 & 0.1533 & 0.1573 & 0.1503 & 0.0259 & -0.0196 \\
\hline\hline
  \end{tabular}
\end{table}

\pagebreak
\begin{table}[hp]
\centering
\caption{Ensemble-averaged scalar fluxes for problem set $\seta_2$}
\label{tab3} 
\begin{tabular}{||c|c||c|c|c||c|c||} \hline \hline
  & $c$ & $\bl\phi_B\bg$ & $\bl\phi_{AM}\bg$ &$\bl\phi_{NC}\bg$ & $Err_{AM}$ & $ Err_{NC}$\\ \hline\hline
&0.0 & 0.2049 & 0.2213 & 0.2048 & 0.0798 & -0.0006 \\
\cline{2-7}
& 0.1 & 0.2181 & 0.2347 & 0.2179 & 0.0760 & -0.0009 \\
\cline{2-7}
& 0.2 & 0.2337 & 0.2506 & 0.2334 & 0.0720 & -0.0013 \\
\cline{2-7}
& 0.3 & 0.2527 & 0.2698 & 0.2522 & 0.0677 & -0.0019 \\
\cline{2-7}
& 0.4 & 0.2762 & 0.2936 & 0.2755 & 0.0631 & -0.0026 \\
\cline{2-7}
$x=0$ & 0.5 & 0.3065 & 0.3243 & 0.3054 & 0.0582 & -0.0035 \\
\cline{2-7}
& 0.6 & 0.3475 & 0.3658 & 0.3458 & 0.0527 & -0.0049\\
\cline{2-7}
& 0.7 & 0.4072 & 0.4263 & 0.4045 & 0.0467 & -0.0069 \\
\cline{2-7}
& 0.8 & 0.5054 & 0.5255 & 0.5003 & 0.0398 & -0.0100\\
\cline{2-7}
& 0.9 & 0.7067 & 0.7291 & 0.6950 & 0.0316 & -0.0165\\
\cline{2-7}
& 0.95 & 0.9254 & 0.9502 & 0.9035 & 0.0267 & -0.0237 \\
\cline{2-7}
& 0.99 & 1.2915 & 1.3204 & 1.2451 & 0.0223 & -0.0359 \\
\hline\hline
& 0.0 & 0.0017 & 0.0033 & 0.0018 & 0.9112 & 0.0057 \\
\cline{2-7}
& 0.1 & 0.0023 & 0.0040 & 0.0023 & 0.7419 & 0.0058 \\
\cline{2-7}
& 0.2 & 0.0031 & 0.0049 & 0.0031 & 0.5953 & 0.0063 \\
\cline{2-7}
& 0.3 & 0.0043 & 0.0063 & 0.0043 & 0.4695 & 0.0070 \\
\cline{2-7}
& 0.4 & 0.0060 & 0.0082 & 0.0060 & 0.3628 & 0.0075 \\
\cline{2-7}
$x=10$ & 0.5 & 0.0087 & 0.0111 & 0.0088 & 0.2733 & 0.0077 \\
\cline{2-7}
& 0.6 & 0.0132 & 0.0158 & 0.0133 & 0.1993 & 0.0072 \\
\cline{2-7}
& 0.7 & 0.0211 & 0.0241 & 0.0212 & 0.1390 & 0.0055 \\
\cline{2-7}
& 0.8 & 0.0371 & 0.0405 & 0.0372 & 0.0910 & 0.0016 \\
\cline{2-7}
& 0.9 & 0.0769 & 0.0811 & 0.0764 & 0.0536 & -0.0070 \\
\cline{2-7}
& 0.95 & 0.1257 & 0.1305 & 0.1236 & 0.0384 & -0.0162 \\
\cline{2-7}
& 0.99 & 0.2126 & 0.2185 & 0.2061 & 0.0278 & -0.0305 \\
\hline\hline
  \end{tabular}
\end{table}

\pagebreak
\begin{table}[hp]
\centering
\caption{Ensemble-averaged scalar fluxes for problem set $\seta_3$}
\label{tab4} 
\begin{tabular}{||c|c||c|c|c||c|c||} \hline \hline
  & $c$ & $\bl\phi_B\bg$ & $\bl\phi_{AM}\bg$ &$\bl\phi_{NC}\bg$ & $Err_{AM}$ & $ Err_{NC}$\\ \hline\hline
& 0.0 & 0.2732 & 0.2835 & 0.2730 & 0.0376 & -0.0007 \\
\cline{2-7}
& 0.1 & 0.2908 & 0.3012 & 0.2905 & 0.0359 & -0.0010 \\
\cline{2-7}
& 0.2 & 0.3117 & 0.3223 & 0.3112 & 0.0341 & -0.0013  \\
\cline{2-7}
& 0.3 & 0.3369 & 0.3477 & 0.3363 & 0.321 & -0.0018 \\
\cline{2-7}
& 0.4 & 0.3683 & 0.3793 & 0.3674 & 0.0300 & -0.0023 \\
\cline{2-7}
$x=0$ & 0.5 & 0.4087 & 0.4201 & 0.4075 & 0.0277 & -0.0031 \\
\cline{2-7}
& 0.6 & 0.4637 & 0.4754 & 0.4618 & 0.0252 & -0.0041 \\
\cline{2-7}
& 0.7 & 0.5442 & 0.5564 & 0.5412 & 0.0224 & -0.0056 \\
\cline{2-7}
& 0.8 & 0.6788 & 0.6919 & 0.6733 & 0.0192 & -0.0081\\
\cline{2-7}
& 0.9 & 0.9715 & 0.9868 & 0.9582 & 0.0157 & -0.0137 \\
\cline{2-7}
& 0.95 & 1.3295 & 1.3481 & 1.3018 & 0.0140 & -0.0209 \\
\cline{2-7}
& 0.99 & 2.0777 & 2.1055 & 2.0011 & 0.0134 & -0.0369 \\ 
\hline\hline
& 0.0 & 0.0004 & 0.0026 & 0.0004 & 4.8188 &  -0.0070 \\
\cline{2-7}
& 0.1 & 0.0006 & 0.0030 & 0.0006 & 3.6072 & -0.0073 \\
\cline{2-7}
& 0.2 & 0.0009 & 0.0034 & 0.0009 & 2.6478 & -0.0073 \\
\cline{2-7}
& 0.3 & 0.0014 & 0.0041 & 0.0014 & 1.8989 & -0.0071\\
\cline{2-7}
& 0.4 & 0.0022 & 0.0052 & 0.0022 & 1.3238 & -0.0070 \\
\cline{2-7}
$x=10$ & 0.5 & 0.0036 & 0.0068 & 0.0036 & 0.8906 & -0.0070 \\
\cline{2-7}
& 0.6 & 0.0062 & 0.0097 & 0.0061 & 0.5718 & -0.0076 \\
\cline{2-7}
& 0.7 & 0.0114 & 0.0153 & 0.0113 & 0.3438 & -0.0090 \\
\cline{2-7}
& 0.8 & 0.0237 & 0.0282 & 0.0235 & 0.1868 & -0.0121 \\
\cline{2-7}
& 0.9 & 0.0618 & 0.0670 & 0.0605 & 0.0847 & -0.0196 \\
\cline{2-7}
& 0.95 & 0.1198 & 0.1259 & 0.1164 & 0.0502 & -0.0287 \\
\cline{2-7}
& 0.99 & 0.2570 & 0.2647 & 0.2449 & 0.0302 & -0.0469 \\
\hline\hline
  \end{tabular}
\end{table}

\pagebreak
\begin{table}[hp]
\centering
\caption{Ensemble-averaged scalar fluxes for problem set $\setb_1$}
\label{tab5} 
\begin{tabular}{||c|c||c|c|c||c|c||} \hline \hline
  & $c$ & $\bl\phi_B\bg$ & $\bl\phi_{AM}\bg$ &$\bl\phi_{NC}\bg$ & $Err_{AM}$ & $ Err_{NC}$\\ \hline\hline
& 0.0 & 0.1776 & 0.2213 & 0.1768 & 0.2459 & -0.0045 \\
\cline{2-7}
& 0.1 & 0.1896 & 0.2347 & 0.1892 & 0.2379 & -0.0018 \\
\cline{2-7}
& 0.2 & 0.2037 & 0.2506 & 0.2039 & 0.2302 & 0.0009 \\
\cline{2-7}
& 0.3 & 0.2206 & 0.2698 & 0.2214 & 0.2229 & 0.0036 \\
\cline{2-7}
& 0.4 & 0.2414 & 0.2936 & 0.2329 & 0.2163 & 0.0063 \\
\cline{2-7}
$x=0$ & 0.5 & 0.2678 & 0.3243 & 0.2700 & 0.2111 & 0.0085 \\
\cline{2-7}
& 0.6 & 0.3027 & 0.3658 & 0.3056 & 0.2085 & 0.0098 \\
\cline{2-7}
& 0.7 & 0.3520 & 0.4263 & 0.3550 & 0.2108 & 0.0084 \\
\cline{2-7}
& 0.8 & 0.4294 & 0.5255 & 0.4293 & 0.2237 & -0.0002\\
\cline{2-7}
& 0.9 & 0.5772 & 0.7291 & 0.5585 & 0.2630 & -0.0325 \\
\cline{2-7}
& 0.95 & 0.7271 & 0.9502 & 0.6710 & 0.3067 & -0.0771\\
\cline{2-7}
& 0.99 & 0.9650 & 1.3204 & 0.8160 & 0.3683 & -0.1545 \\
\hline\hline
& 0.0 & 0.0250 & 0.0033 & 0.0248 & -0.8667 & -0.0075 \\
\cline{2-7}
& 0.1 & 0.0270 & 0.0040 & 0.0273 & -0.8515 & 0.0079 \\
\cline{2-7}
& 0.2 & 0.0295 & 0.0049 & 0.0302 & -0.8324 & 0.0248 \\
\cline{2-7}
& 0.3 & 0.0325 & 0.0063 & 0.0339 & -0.8078 & 0.0436 \\
\cline{2-7}
& 0.4 & 0.0364 & 0.0082 & 0.0387 & -0.7756 & 0.0643 \\
\cline{2-7}
$x=10$ & 0.5 & 0.0414 & 0.0111 & 0.0450 & -0.7325 & 0.0871 \\
\cline{2-7}
& 0.6 & 0.0483 & 0.0158 & 0.0537 & -0.6734 & 0.1114 \\
\cline{2-7}
& 0.7 & 0.0587 & 0.0241 & 0.0666 & -0.5899 & 0.1355 \\
\cline{2-7}
& 0.8 & 0.0760 & 0.0405 & 0.0877 & -0.4677 & 0.1532 \\
\cline{2-7}
& 0.9 & 0.1126 & 0.0811 & 0.1284 & -0.2799 & 0.1408 \\
\cline{2-7}
& 0.95 & 0.1529 & 0.1305 & 0.1676 & -0.1463 & 0.0966 \\
\cline{2-7}
& 0.99 & 0.2209 & 0.2185 & 0.2226 & -0.0106 & 0.0077\\
\hline\hline
  \end{tabular}
\end{table}

\pagebreak
\begin{table}[hp]
\centering
\caption{Ensemble-averaged scalar fluxes for problem set $\setb_2$}
\label{tab6} 
\begin{tabular}{||c|c||c|c|c||c|c||} \hline \hline
  & $c$ & $\bl\phi_B\bg$ & $\bl\phi_{AM}\bg$ &$\bl\phi_{NC}\bg$ & $Err_{AM}$ & $ Err_{NC}$\\ \hline\hline
  & 0.0 & 0.1975 & 0.2213 & 0.1972 & 0.1200 & -0.0018 \\
  \cline{2-7}
  & 0.1 & 0.2100 & 0.2347 & 0.2101 & 0.1177 & 0.0004 \\
  \cline{2-7}
  & 0.2 & 0.2245 & 0.2506 & 0.2252 & 0.1159 & 0.0028 \\
  \cline{2-7}
  & 0.3 & 0.2420 & 0.2698 & 0.2433 & 0.1148 & 0.0054 \\
  \cline{2-7}
  & 0.4 & 0.2634 & 0.2936 & 0.2655 & 0.1148 & 0.0080 \\
  \cline{2-7}
  $x=0$ & 0.5 & 0.2904 & 0.3243 & 0.2934 & 0.1168 & 0.0105 \\
  \cline{2-7}
  & 0.6 & 0.3261 & 0.3658 & 0.3301 & 0.1218 & 0.0125 \\
  \cline{2-7}
  & 0.7 & 0.3763 & 0.4263 & 0.3812 & 0.1327 & 0.0129 \\
  \cline{2-7}
  & 0.8 & 0.4548 & 0.5255 & 0.4588 & 0.1553 & 0.0088 \\
  \cline{2-7}
  & 0.9 & 0.6042 & 0.7291 & 0.5972 & 0.2066 & -0.0116 \\
  \cline{2-7}
  & 0.95 & 0.7553 & 0.9502 & 0.7233 & 0.2581 & -0.0423 \\
\cline{2-7}  
  & 0.99 & 0.9946 & 1.3204 & 0.8961 & 0.3276 & -0.0990 \\
\hline\hline
& 0.0 & 0.0246 & 0.0033 & 0.0243 & -0.8646 & -0.0106 \\
\cline{2-7}  
&0.1 & 0.0267 & 0.0040 & 0.0265 & -0.8494 & -0.0048 \\
\cline{2-7}  
&0.2 & 0.0291 & 0.0049 & 0.0292 & -0.8302 & 0.0020 \\
\cline{2-7}
& 0.3 & 0.0322 & 0.0063 & 0.0325 & -0.8055 & 0.0098 \\
\cline{2-7}
& 0.4 & 0.0360 & 0.0082 & 0.0367 & -0.7733 & 0.0188 \\
\cline{2-7}
$x=10$ & 0.5 & 0.0410 & 0.0111 & 0.0422 & -0.7302 & 0.0293 \\
\cline{2-7}
& 0.6 & 0.0480 & 0.0158 & 0.0500 & -0.6711 & 0.0413 \\
\cline{2-7}
& 0.7 & 0.0584 & 0.0241 & 0.0615 & -0.5877 & 0.0545 \\
\cline{2-7}
& 0.8 & 0.0758 & 0.0405 & 0.0808 & -0.4658 & 0.0670 \\
\cline{2-7}
& 0.9 & 0.1124 & 0.0811 & 0.1201 & -0.2786 & 0.0685 \\
\cline{2-7}
& 0.95 & 0.1527 & 0.1305 & 0.1604 & -0.1455 & 0.0500 \\
\cline{2-7}
& 0.99 & 0.2208 & 0.2185 & 0.2211 & -0.0105 & 0.0010 \\
\hline\hline
  \end{tabular}
\end{table}

\pagebreak
\begin{table}[hp]
\centering
\caption{Ensemble-averaged scalar fluxes for problem set $\setb_3$}
\label{tab7} 
\begin{tabular}{||c|c||c|c|c||c|c||} \hline \hline
  & $c$ & $\bl\phi_B\bg$ & $\bl\phi_{AM}\bg$ &$\bl\phi_{NC}\bg$ & $Err_{AM}$ & $ Err_{NC}$\\ \hline\hline
  & 0.0 & 0.2089 & 0.2213 & 0.2088 & 0.0593 & -0.0002 \\
  \cline{2-7}
  & 0.1 & 0.2221 & 0.2347 & 0.2220 & 0.0566 & -0.0004 \\
  \cline{2-7}
  &0.2 & 0.2378 & 0.2506 & 0.2375 & 0.0539 & -0.0009 \\
  \cline{2-7}
  & 0.3 & 0.2566 & 0.2698 & 0.2562 & 0.0512 & -0.0017 \\
  \cline{2-7}
  & 0.4 & 0.2800 & 0.2936 & 0.2791 & 0.0487 & -0.0031 \\
  \cline{2-7}
  $x=0$ & 0.5 & 0.3098 & 0.3243 & 0.3082 & 0.0467 & -0.0053 \\
  \cline{2-7}
  & 0.6 &0.3498 & 0.3658 & 0.3468 & 0.0457 & -0.0086 \\
  \cline{2-7}
  & 0.7 & 0.4071 & 0.4263 & 0.4015 & 0.0471 & -0.0137 \\
  \cline{2-7}
  & 0.8 & 0.4985 & 0.5255 & 0.4875 & 0.0542 & -0.0220 \\
  \cline{2-7}
  & 0.9 & 0.6770 & 0.7291 & 0.6511 & 0.0769 & -0.0382 \\
  \cline{2-7}
  & 0.95 & 0.8612 & 0.9502 & 0.8132 & 0.1033 & -0.0558 \\
  \cline{2-7}
  & 0.99 & 1.1570 & 1.3204 & 1.0563 & 0.1412 & -0.0870 \\
\hline\hline
& 0.0 & 0.0073 & 0.0033 & 0.0073 & -0.5438 & 0.0027 \\
 \cline{2-7}
& 0.1 & 0.0086 & 0.0040 & 0.0086 & -0.5355 & -0.0034 \\
 \cline{2-7}
& 0.2 & 0.0104 & 0.0049 & 0.0103 & -0.5225 & -0.0085 \\
 \cline{2-7}
& 0.3 & 0.0126 & 0.0063 & 0.0124 & -0.5037 & -0.0121 \\
 \cline{2-7}
& 0.4 & 0.0156 & 0.0082 & 0.0154 & -0.4776 & -0.0139 \\
 \cline{2-7}
$x=10$ & 0.5 &0.0198 & 0.0111 & 0.0196 & -0.4422 & -0.0133 \\ 
 \cline{2-7}
&0.6 & 0.0261 & 0.0158 & 0.0258 & -0.3947 & -0.0095 \\
 \cline{2-7}
& 0.7 & 0.0360 & 0.0241 & 0.0359 & -0.3313 & -0.0016 \\
 \cline{2-7}
& 0.8 & 0.0537 & 0.0405 & 0.0543 & -0.2463 & 0.0108 \\
 \cline{2-7}
& 0.9 & 0.0933 & 0.0811 & 0.0955 & -0.1314 & 0.0229 \\
 \cline{2-7}
& 0.95 & 0.1386 & 0.1305 & 0.1413 & -0.0586 & 0.0191 \\
 \cline{2-7}
& 0.99 & 0.2166 & 0.2185 & 0.2151 & 0.0089 & -0.0070\\
\hline\hline
  \end{tabular}
\end{table}

\pagebreak
\begin{figure}[p]
  \centering
  \includegraphics[width=\textwidth]{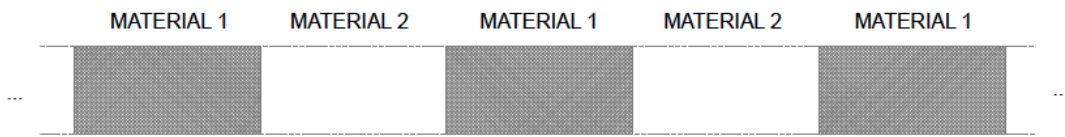}
  \caption{A sketch of the periodic medium}
  \label{fig1}
\end{figure}

\pagebreak
\begin{figure}[p]
    \centering
    \begin{subsubcaption}
    \begin{subfigure}{0.495\textwidth}
        \centering
        \includegraphics[width=\textwidth]{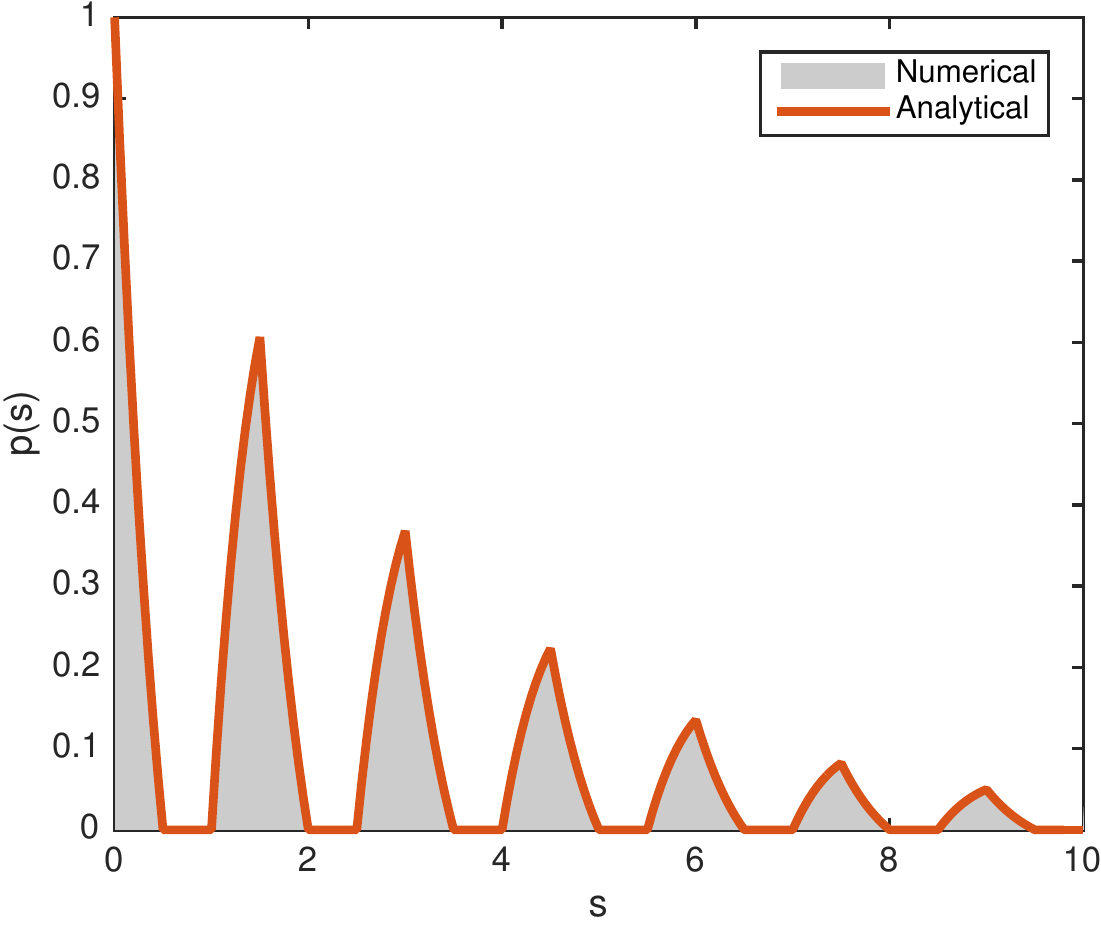}
        \caption{Case 1: $\ell_1=0.5$, $\ell_2=1.0$}
        \label{fig2a}
    \end{subfigure}
    \hfill
    \begin{subfigure}{0.495\textwidth}
        \centering
        \includegraphics[width=\textwidth]{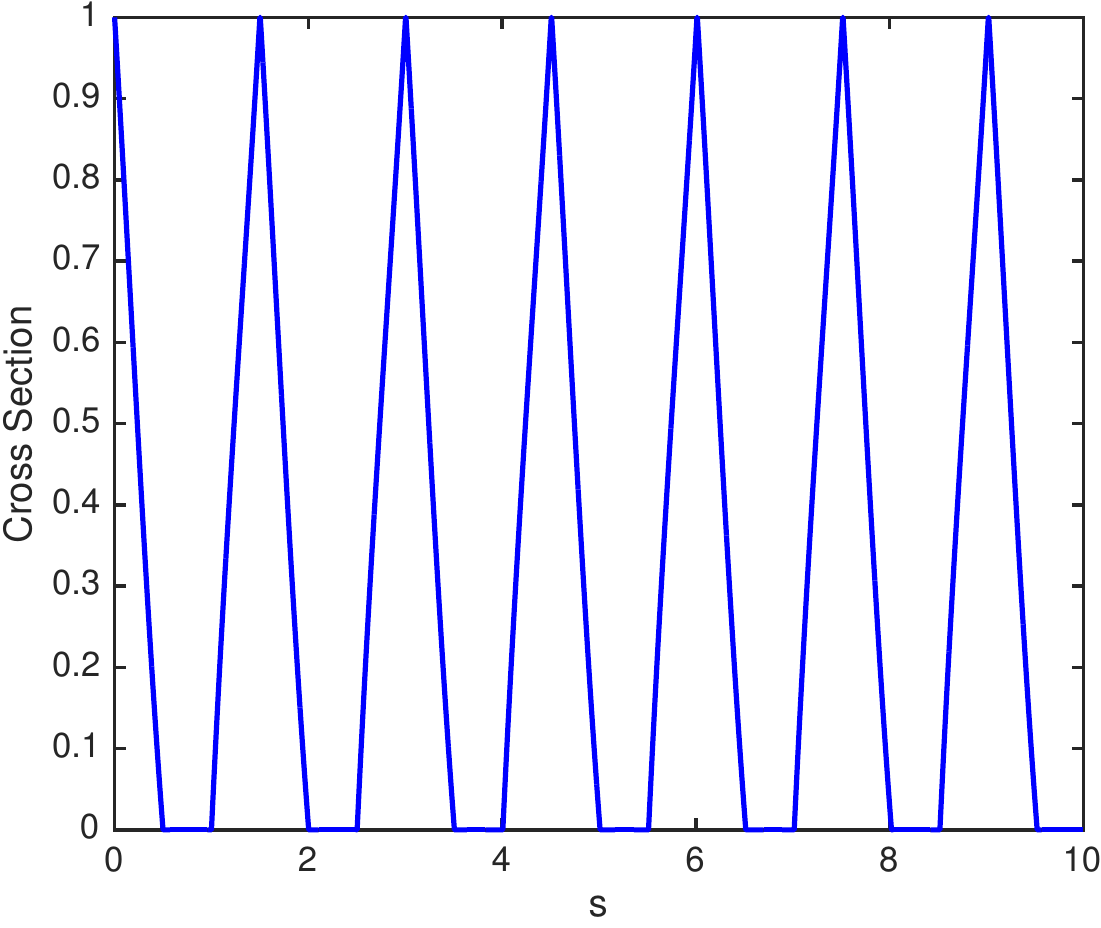}
        \caption{Case 1: $\ell_1=0.5$, $\ell_2=1.0$}
        \label{fig2b}
    \end{subfigure}
    \end{subsubcaption}
    \\
    \begin{subsubcaption}
    \centering
    \begin{subfigure}{0.495\textwidth}
        \centering
        \includegraphics[width=\textwidth]{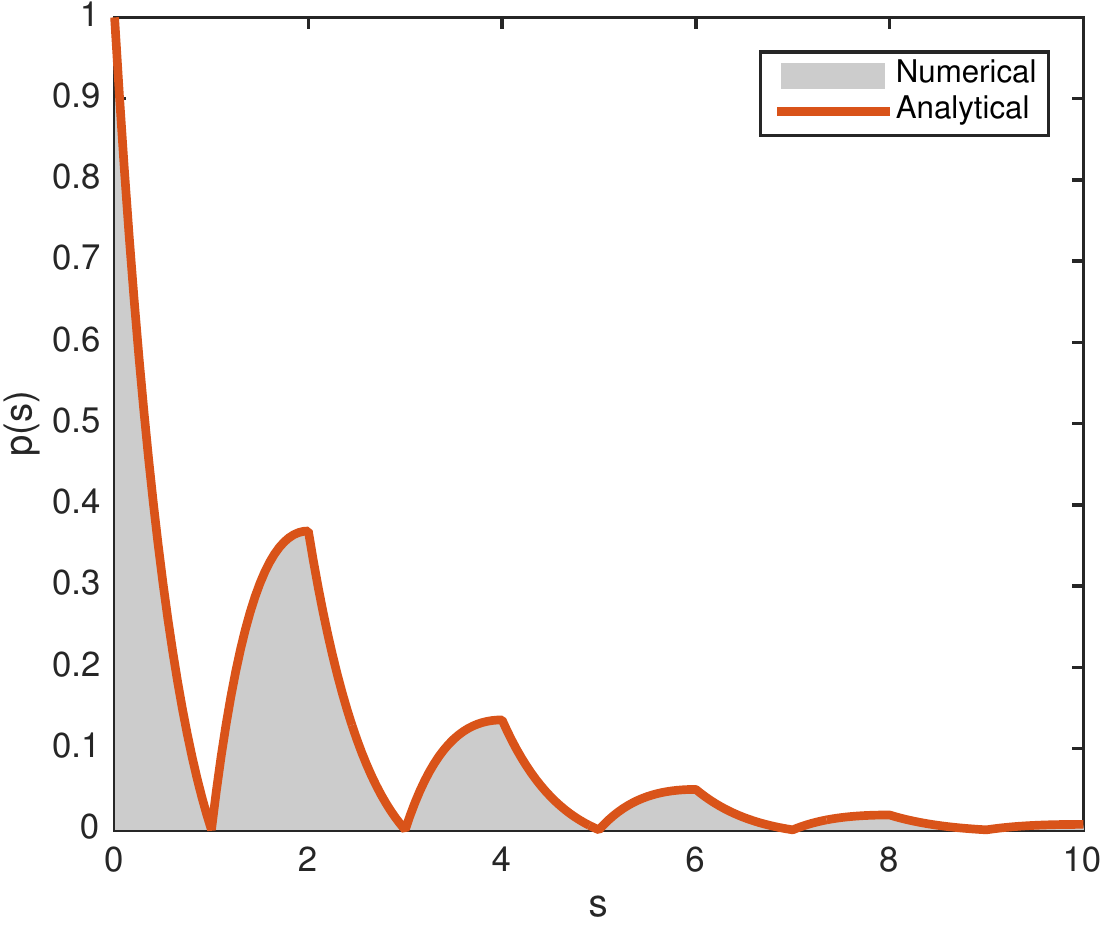}
        \caption{Case 2: $\ell_1=\ell_2=1.0$}
        \label{fig2c}
    \end{subfigure}
    \hfill
    \begin{subfigure}{0.495\textwidth}
        \centering
        \includegraphics[width=\textwidth]{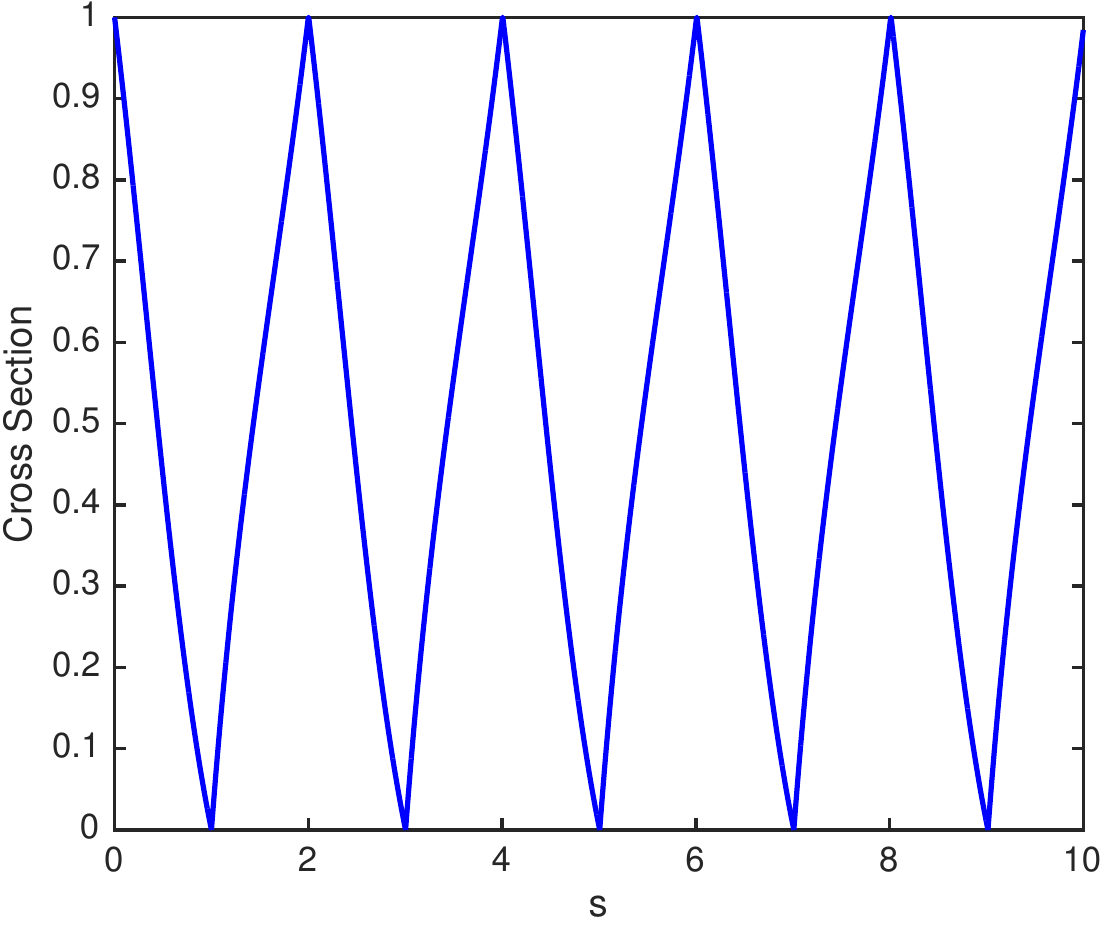}
        \caption{Case 2: $\ell_1=\ell_2=1.0$}
        \label{fig2d}
    \end{subfigure}
    \end{subsubcaption}
    \\
    \begin{subsubcaption}
    \centering
    \begin{subfigure}{0.495\textwidth}
        \centering
        \includegraphics[width=\textwidth]{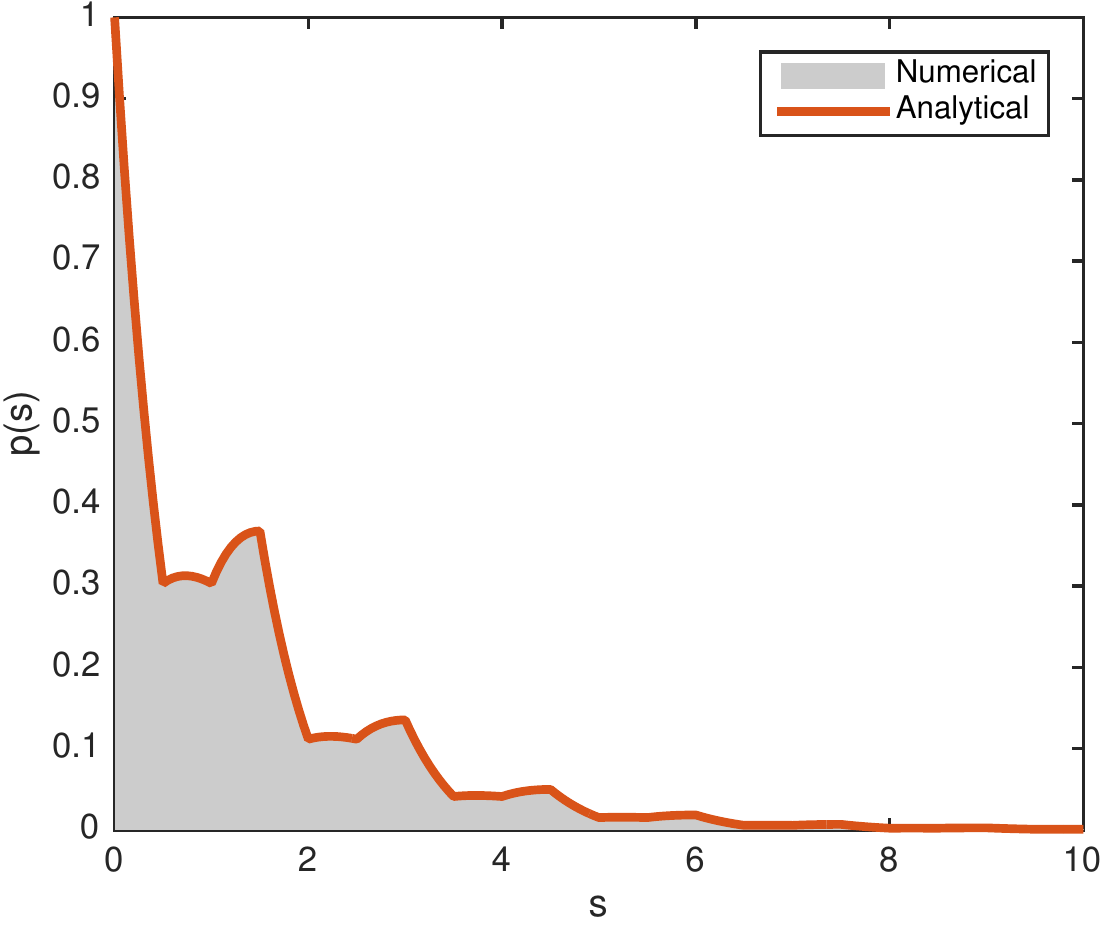}
        \caption{Case 3: $\ell_1=1.0$, $\ell_2=0.5$}
        \label{fig2e}
    \end{subfigure}
    \hfill
    \begin{subfigure}{0.495\textwidth}
        \centering
        \includegraphics[width=\textwidth]{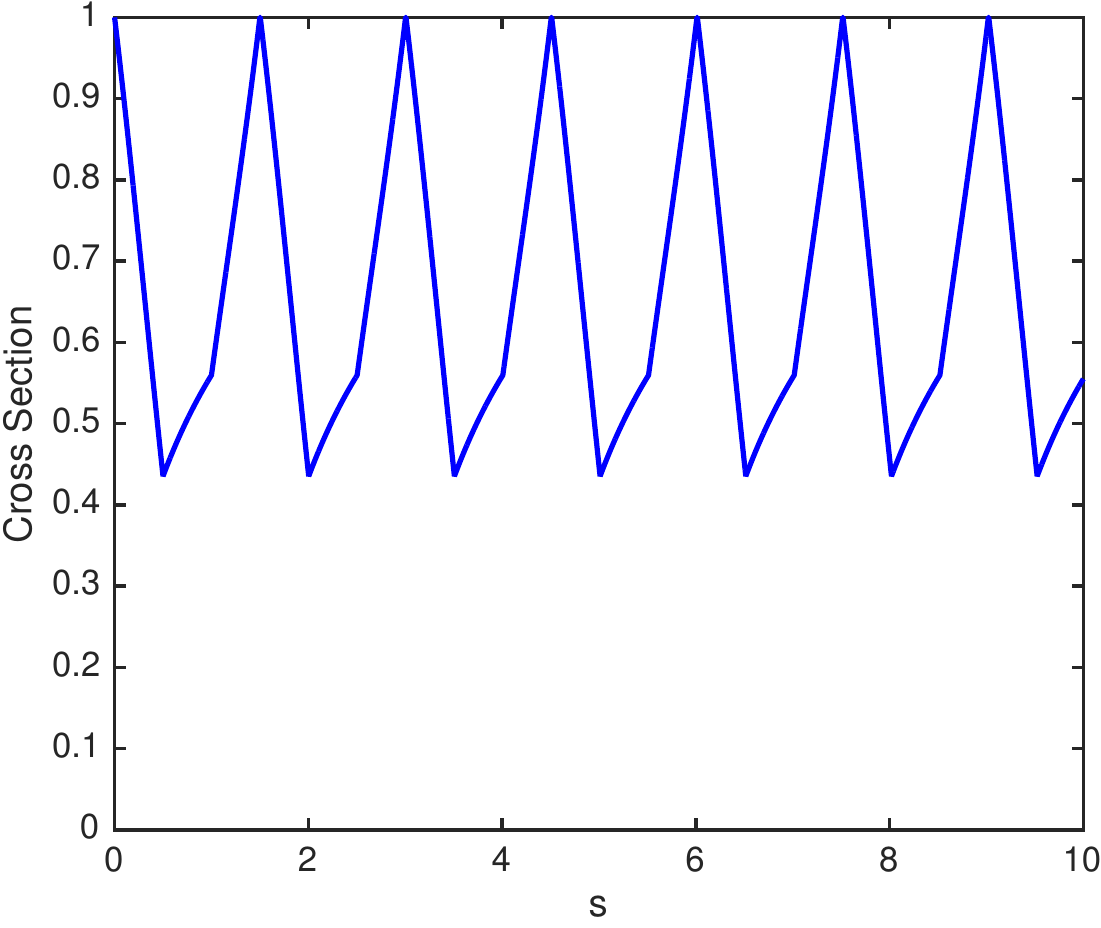}
        \caption{Case 3: $\ell_1=1.0$, $\ell_2=0.5$}
        \label{fig2f}
    \end{subfigure}
    \end{subsubcaption}
    \caption{Path-length distribution functions and corresponding nonclassical cross sections (assuming $\mu=\pm 1$ and $\Sigma_{t1}=1.0$) }
    \label{fig2}
\end{figure}

\pagebreak
\begin{figure}[p]
    \centering
    \begin{subfigure}{0.495\textwidth}
        \centering
        \includegraphics[width=\textwidth]{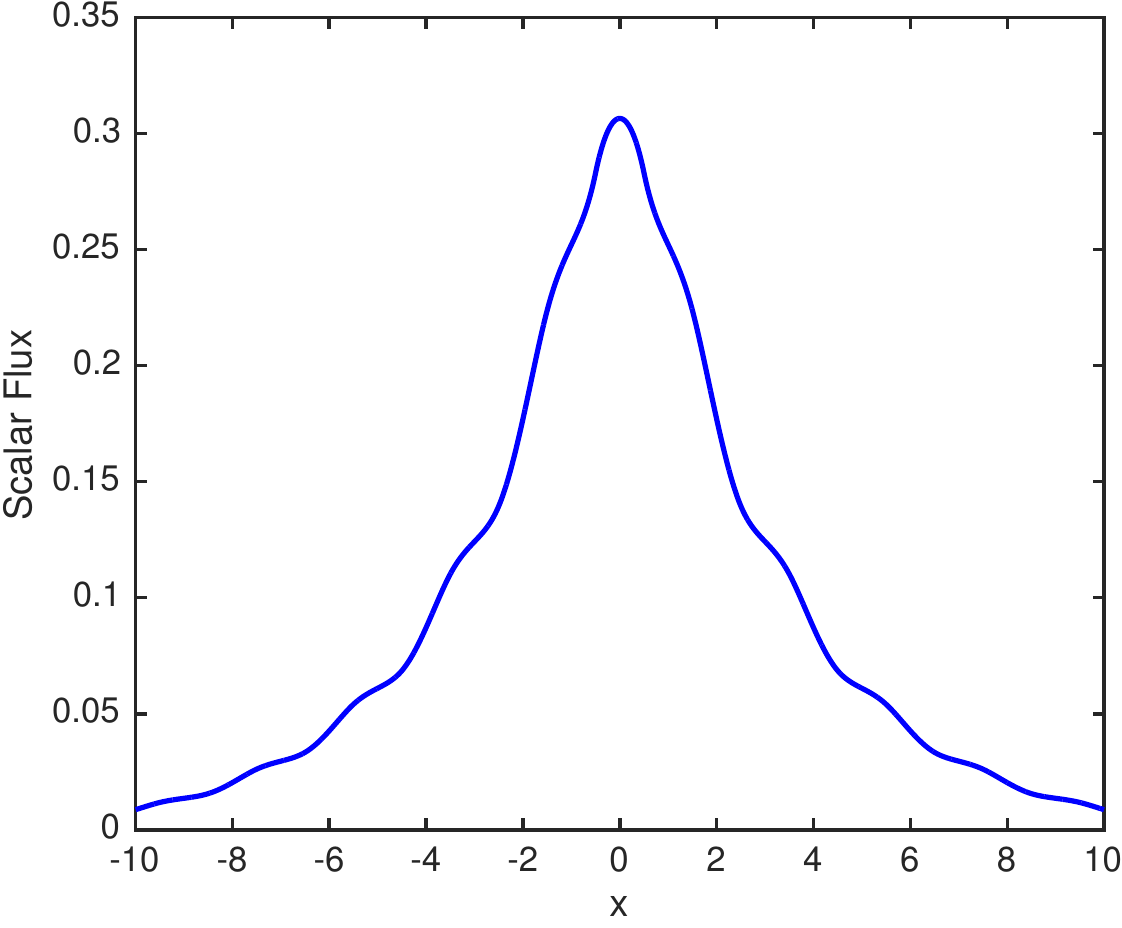}
        \caption{Source $Q_1$ given by \cref{eq19}}
        \label{fig3a}
    \end{subfigure}
    \hfill
    \begin{subfigure}{0.495\textwidth}
        \centering
        \includegraphics[width=\textwidth]{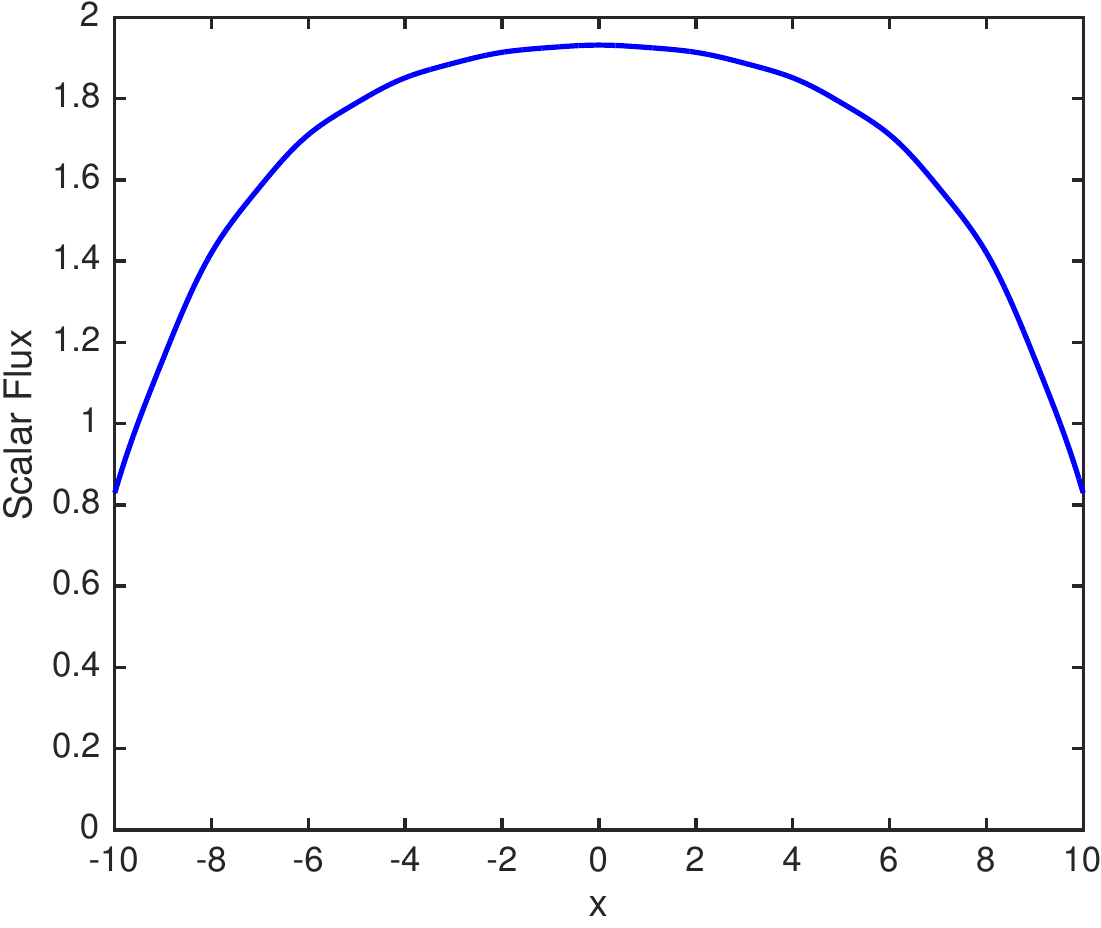}
        \caption{Source $Q_1=1$ for $-10\leq x\leq 10$}
        \label{fig3b}
    \end{subfigure}
    \caption{Ensemble-averaged scalar flux for problem set {$\seta_2$} with $c_1=0.5$}
    \label{fig3}
\end{figure}

\pagebreak
\begin{figure}[p]
  \centering
  \includegraphics[scale=1]{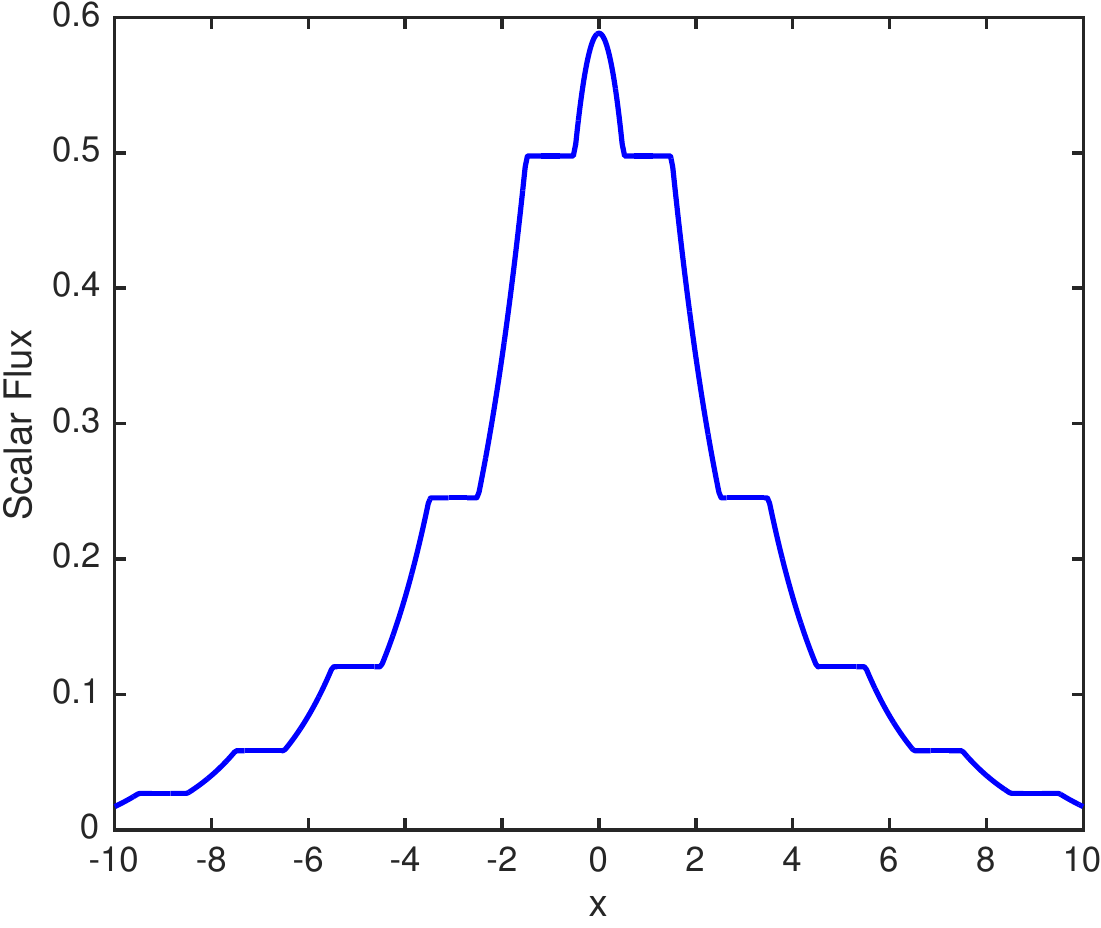}
  \caption{Scalar flux in a fixed realization of problem set {$\seta_2$} with $c_1=0.5$}
  \label{fig4}
\end{figure}

\pagebreak
\begin{figure}[p]
  \centering
  \includegraphics[scale=1]{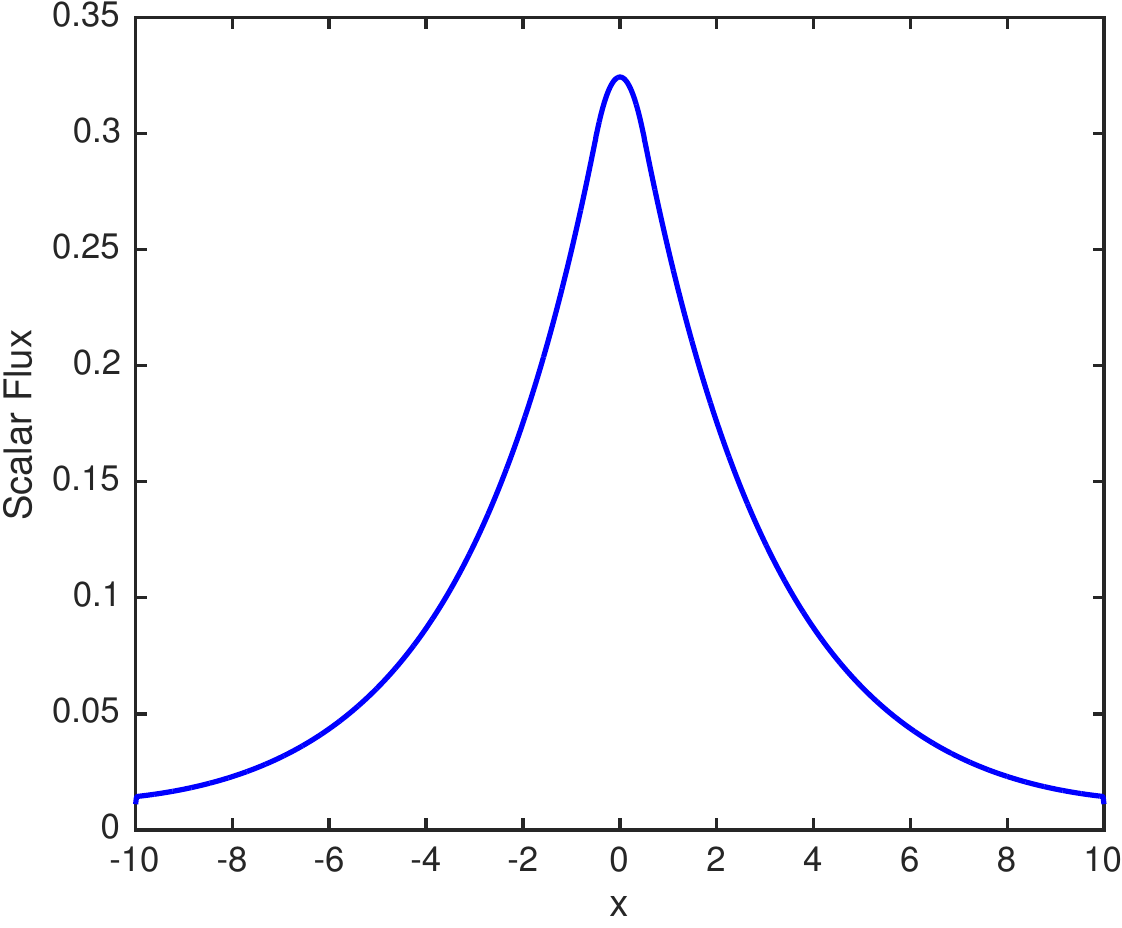}
  \caption{Atomic mix scalar flux for problem set {$\seta_2$} with $c_1=0.5$}
  \label{fig5}
\end{figure}

\pagebreak
\begin{figure}[htb]
  \centering
  \includegraphics[scale=1]{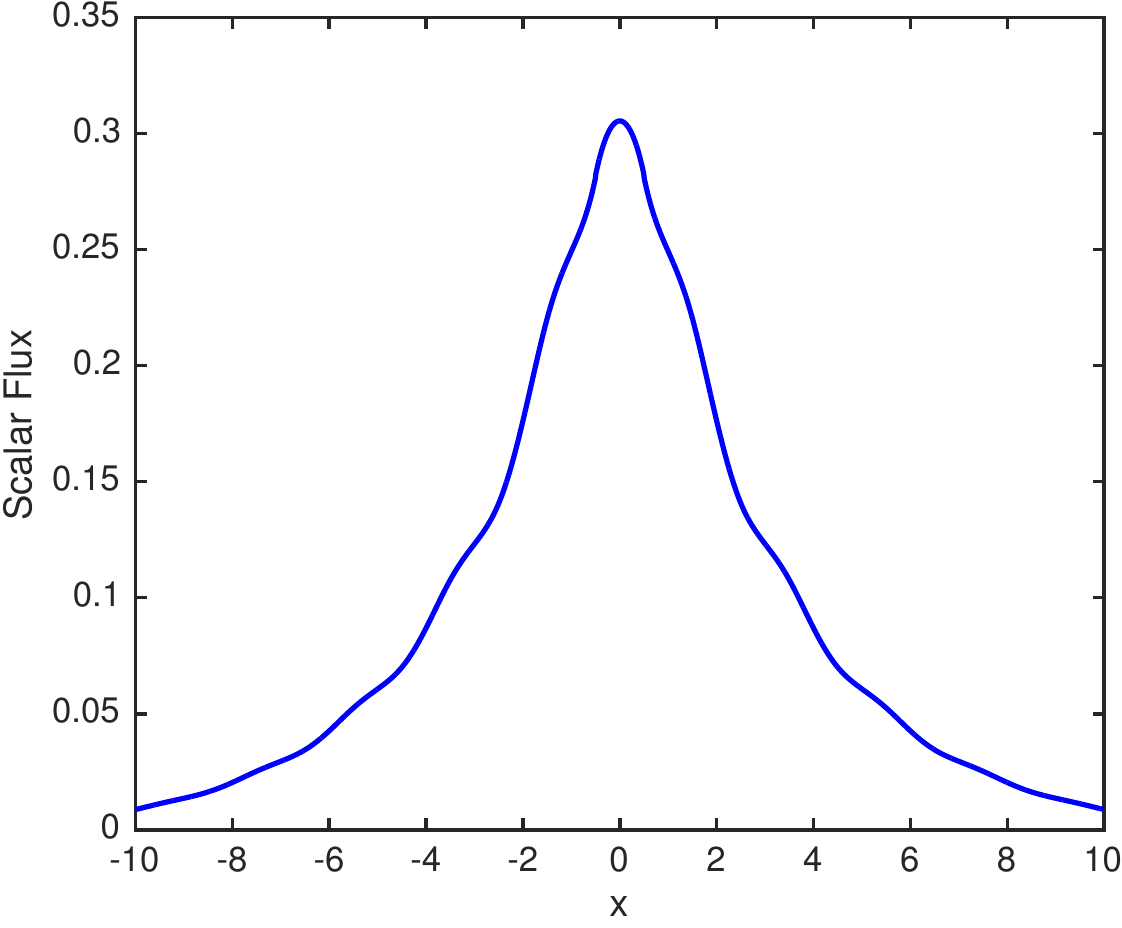}
  \caption{Nonclassical scalar flux for problem set $\seta_2$ with $c_1=0.5$}
  \label{fig6}
\end{figure}

\pagebreak
\setcounter{subfigure}{0}
\begin{figure}[p]
    \centering
    \begin{subsubcaption}
    \begin{subfigure}{0.495\textwidth}
        \centering
        \includegraphics[width=\textwidth]{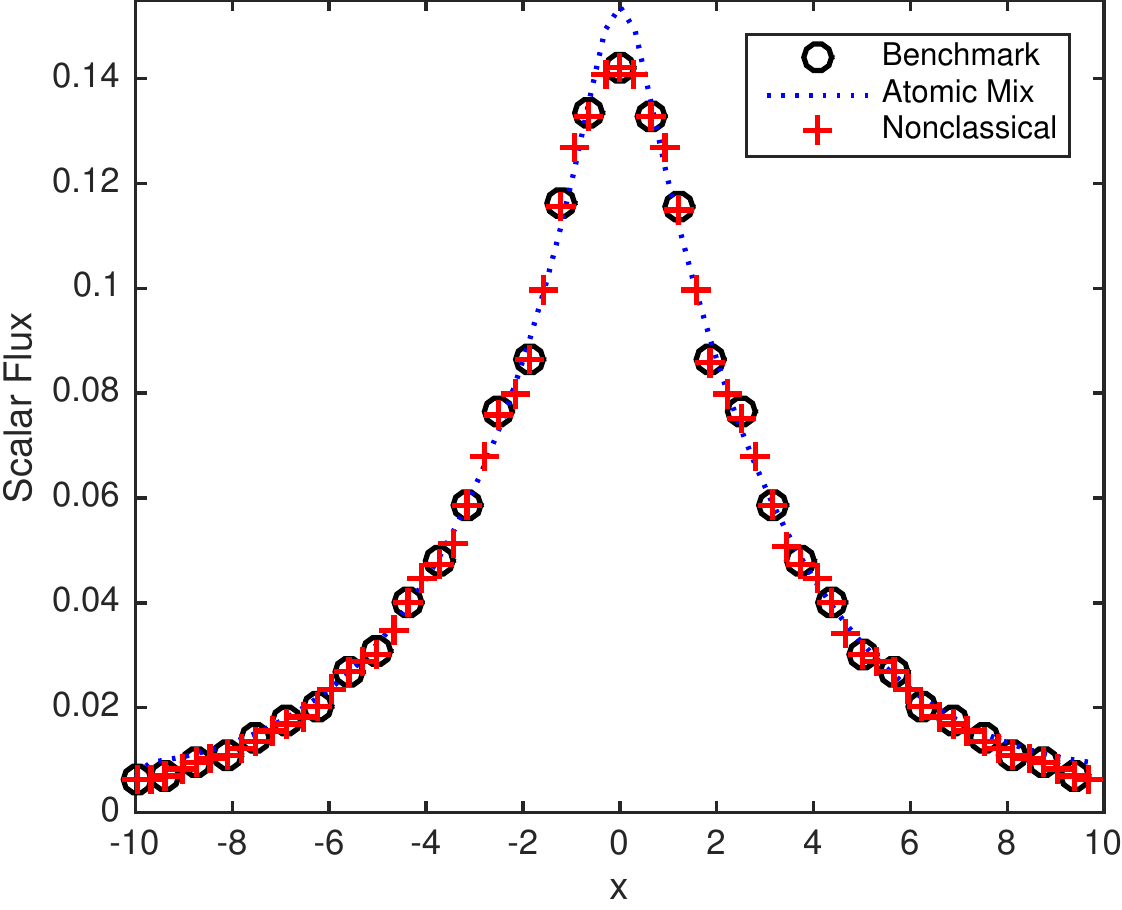}
        \caption{Problem set $\seta_1$ with $c_1=00$}
        \label{fig7a}
    \end{subfigure}
    \hfill
    \begin{subfigure}{0.495\textwidth}
        \centering
        \includegraphics[width=\textwidth]{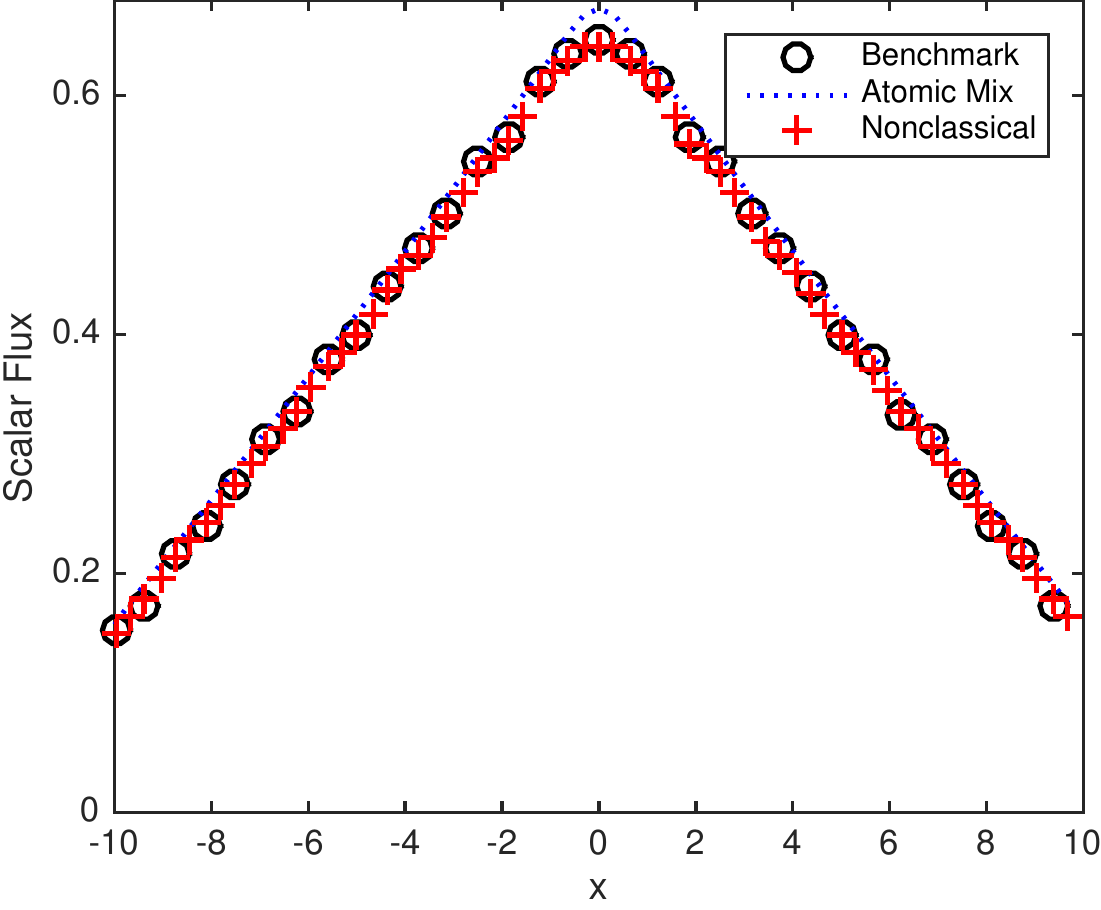}
        \caption{Problem set $\seta_1$ with $c_1=0.99$}
        \label{fig7b}
    \end{subfigure}
        \end{subsubcaption}
    \\
    \begin{subsubcaption}
    \centering
    \begin{subfigure}{0.495\textwidth}
        \includegraphics[width=\textwidth]{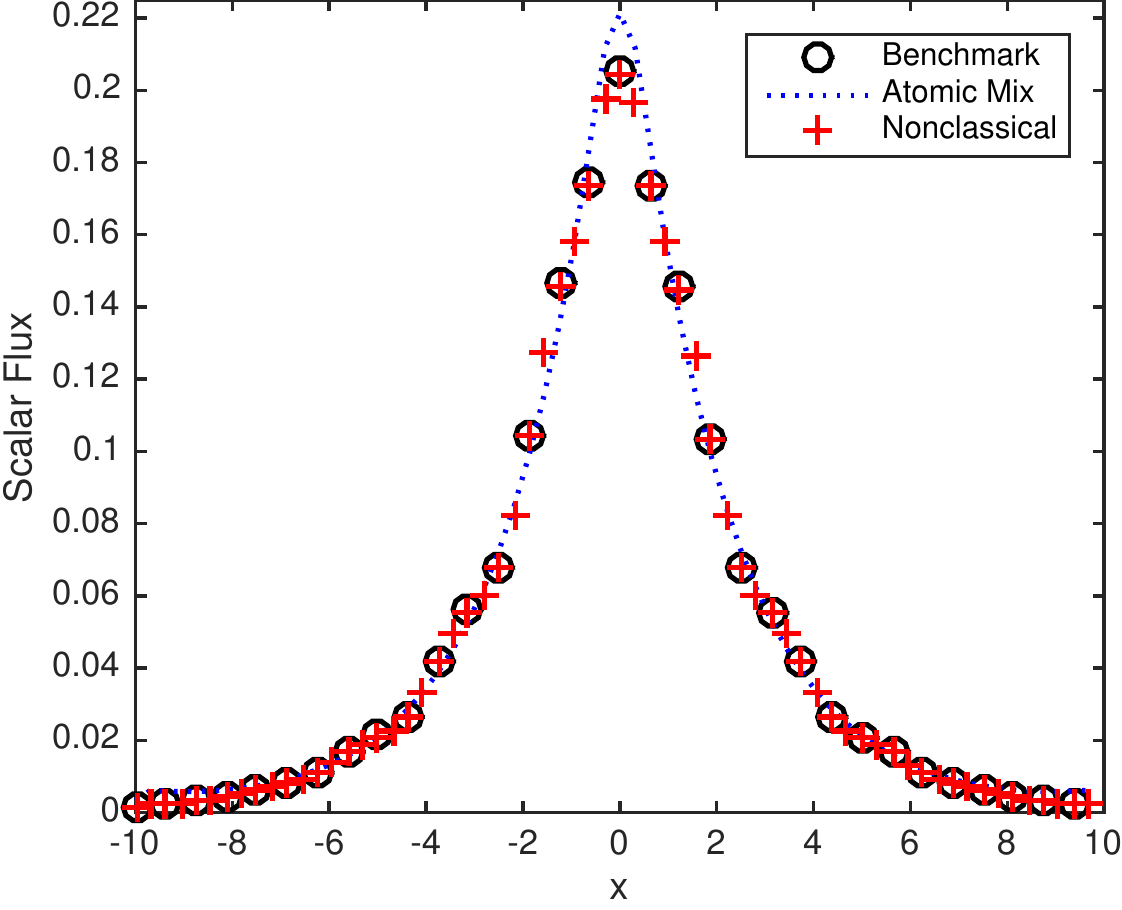}
        \caption{Problem set $\seta_2$ with $c_1=00$}
        \label{fig7c}
    \end{subfigure}
    \hfill
    \begin{subfigure}{0.495\textwidth}
        \centering
        \includegraphics[width=\textwidth]{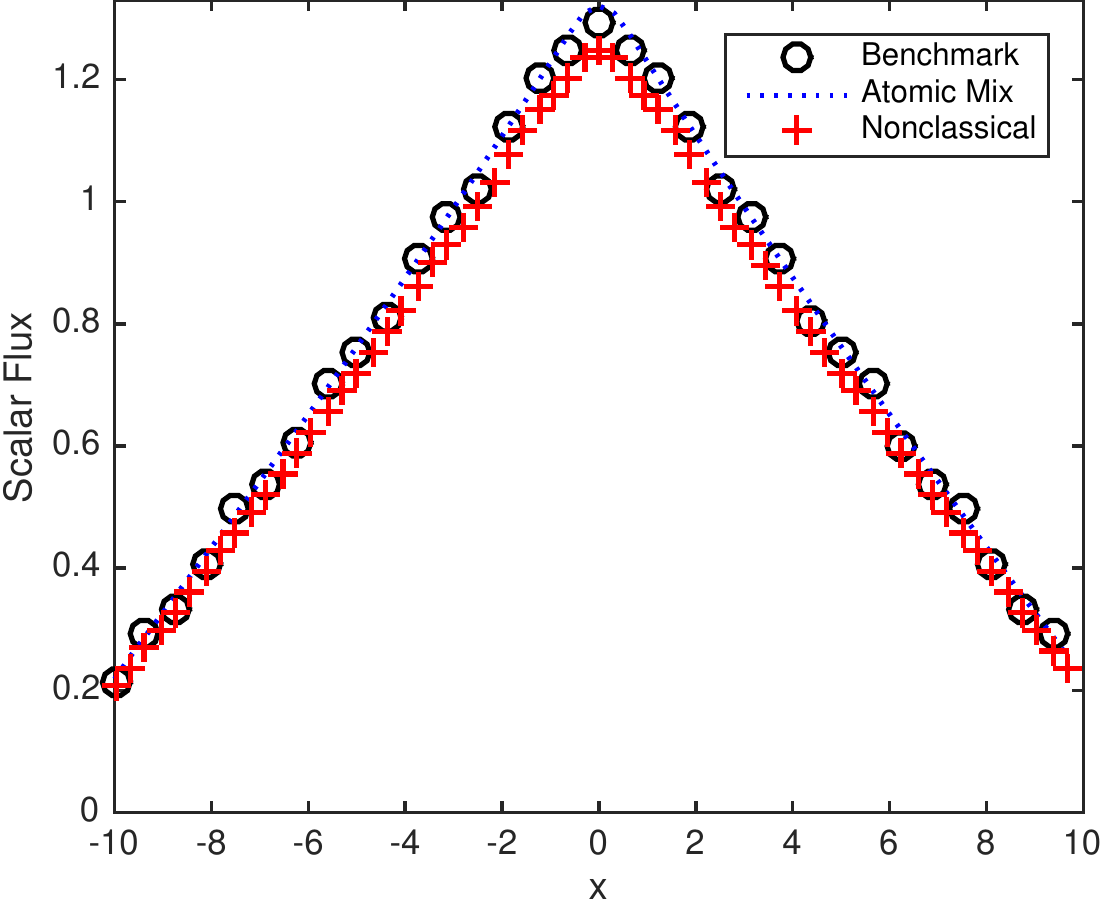}
        \caption{Problem set $\seta_2$ with $c_1=0.99$}
        \label{fig7d}
    \end{subfigure}
        \end{subsubcaption}
    \\
    \begin{subsubcaption}
    \centering
    \begin{subfigure}{0.495\textwidth}
        \includegraphics[width=\textwidth]{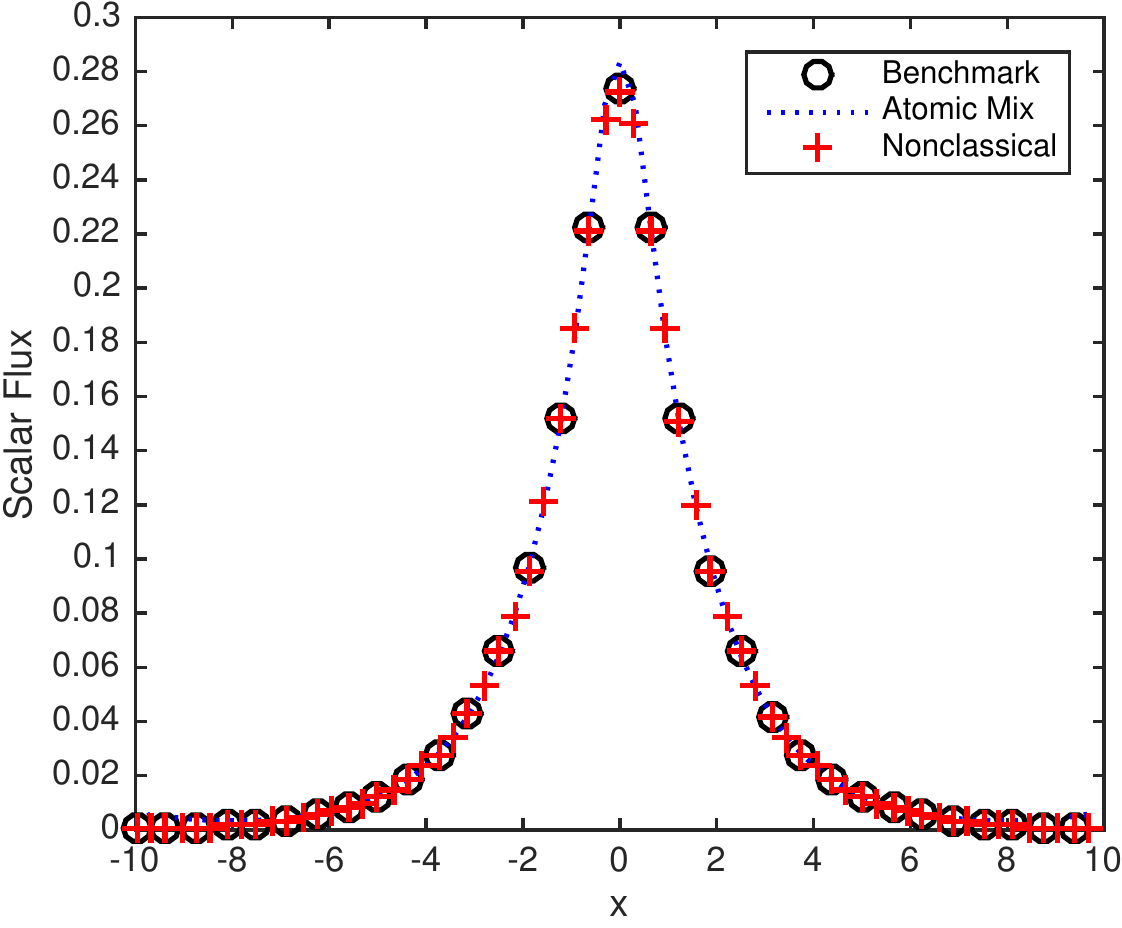}
        \caption{Problem set $\seta_3$ with $c_1=00$}
        \label{fig7e}
    \end{subfigure}
    \hfill
    \begin{subfigure}{0.495\textwidth}
        \centering
        \includegraphics[width=\textwidth]{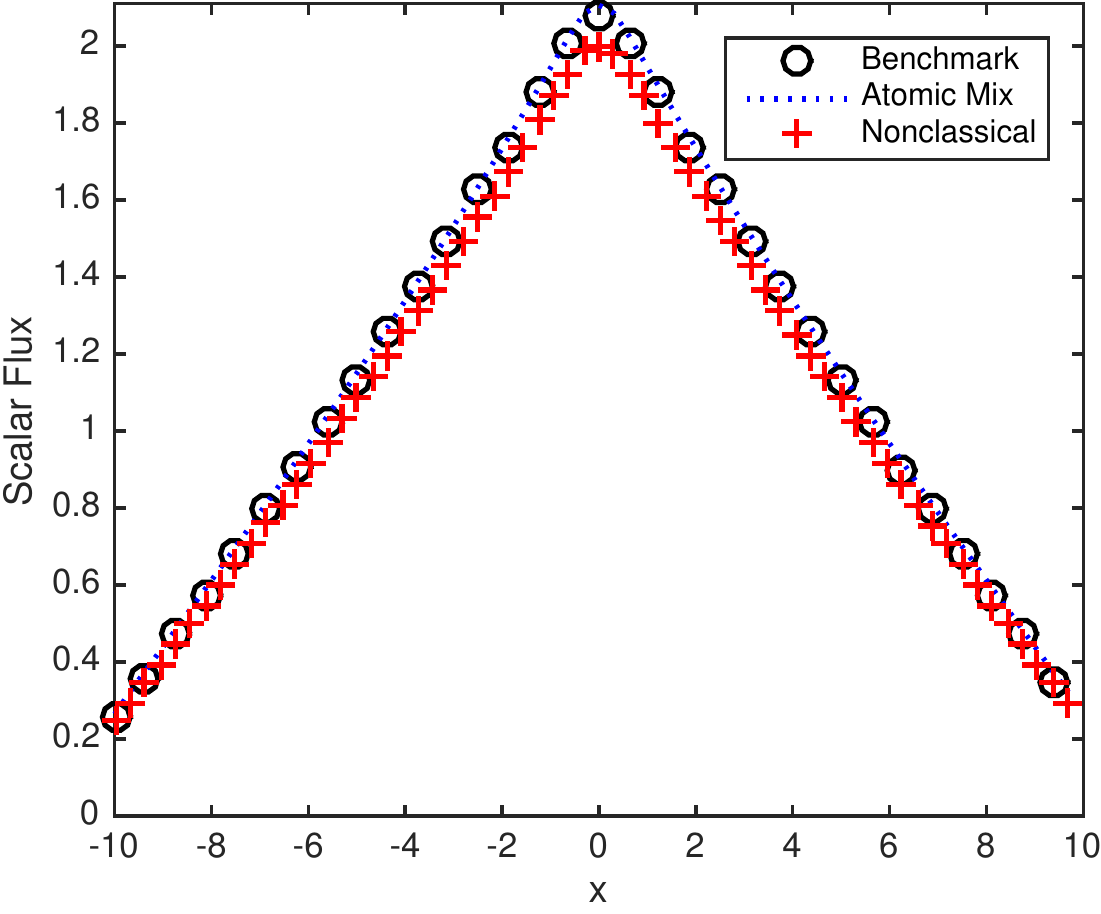}
        \caption{Problem set $\seta_3$ with $c_1=0.99$}
        \label{fig7f}
    \end{subfigure}
    \end{subsubcaption}
    \\
        \caption{Ensemble-averaged scalar fluxes for problem set $\seta$}
    \label{fig7}
\end{figure}

\pagebreak
\begin{figure}[p]
    \centering
    \begin{subfigure}{0.495\textwidth}
        \centering
        \includegraphics[width=\textwidth]{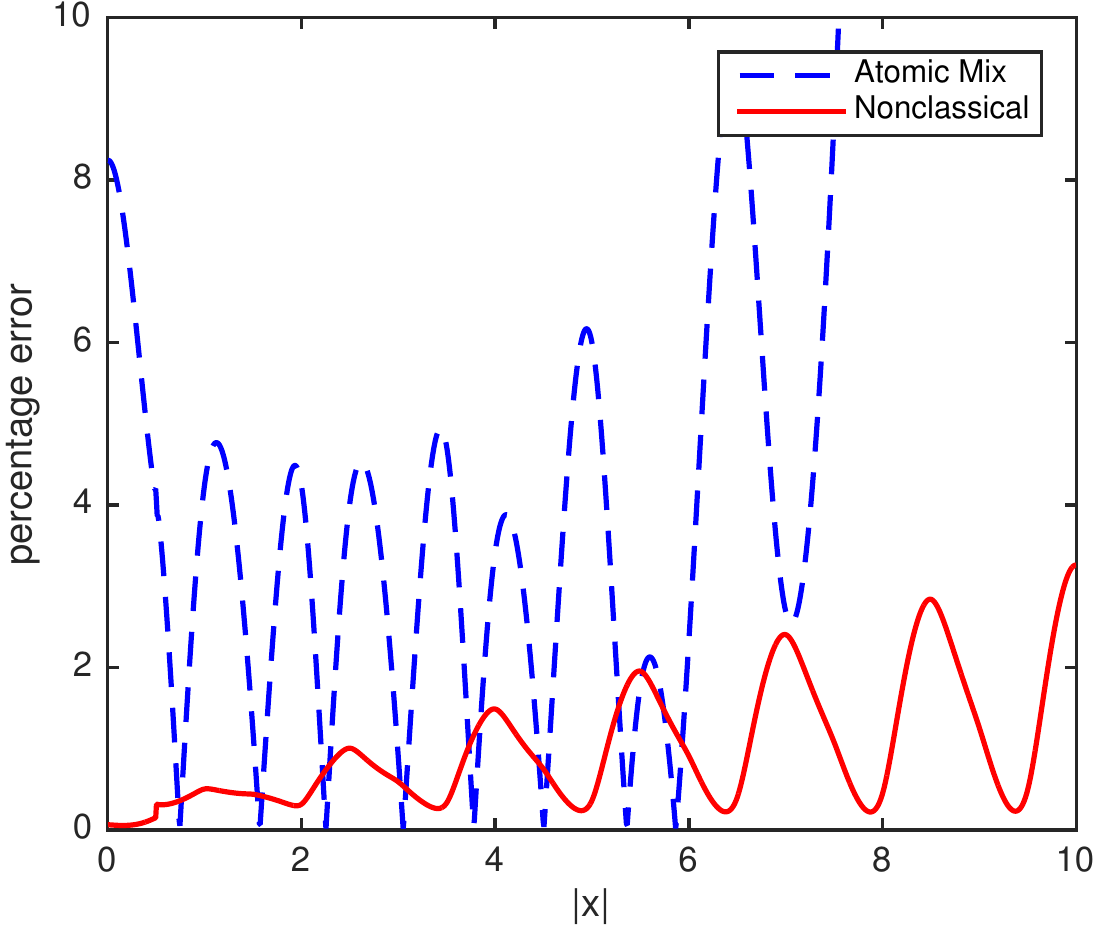}
        \caption{$c_1 = 0.0$}
        \label{figerrA00}
    \end{subfigure}
    \hfill
    \begin{subfigure}{0.495\textwidth}
        \centering
        \includegraphics[width=\textwidth]{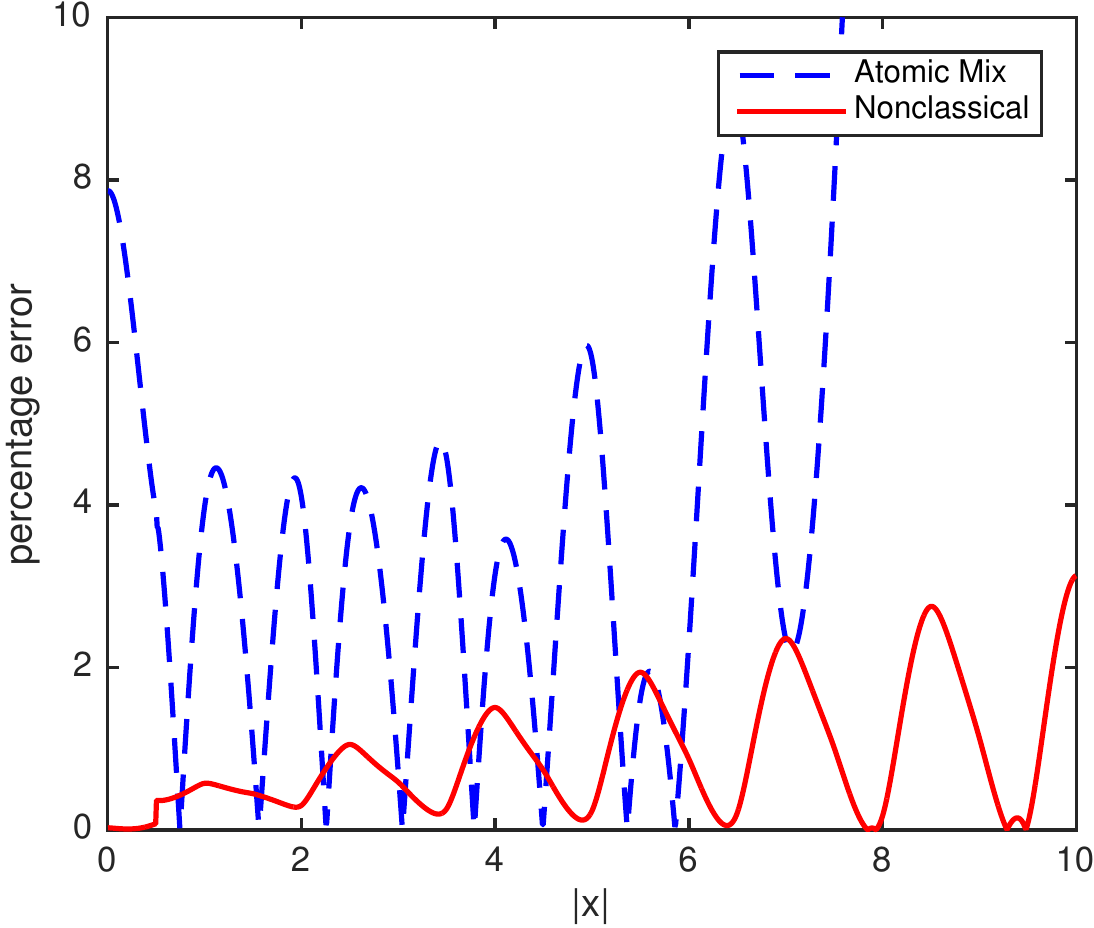}
        \caption{$c_1 = 0.1$}
        \label{figerrA10}
    \end{subfigure}
    \\
    \centering
    \begin{subfigure}{0.495\textwidth}
        \centering
        \includegraphics[width=\textwidth]{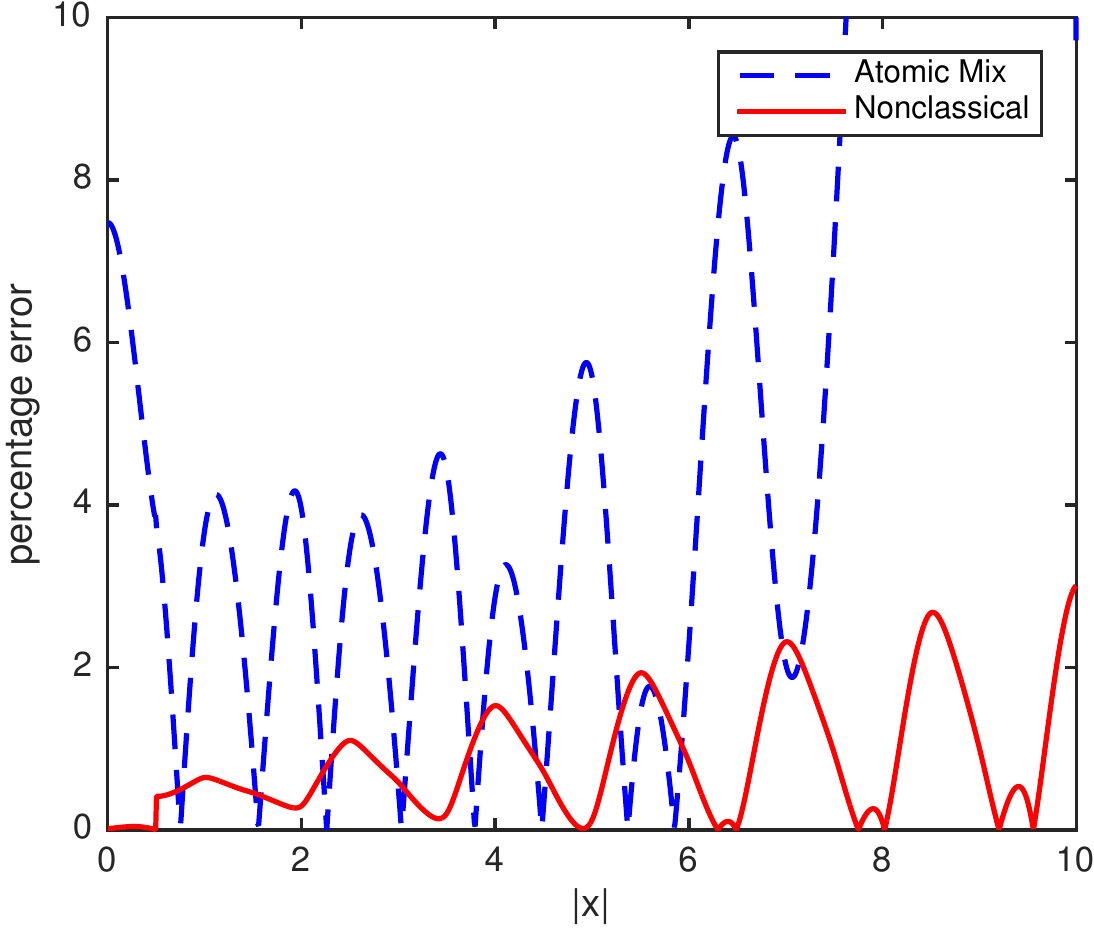}
        \caption{$c_1 = 0.2$}
        \label{figerrA20}
    \end{subfigure}
    \hfill
    \begin{subfigure}{0.495\textwidth}
        \centering
        \includegraphics[width=\textwidth]{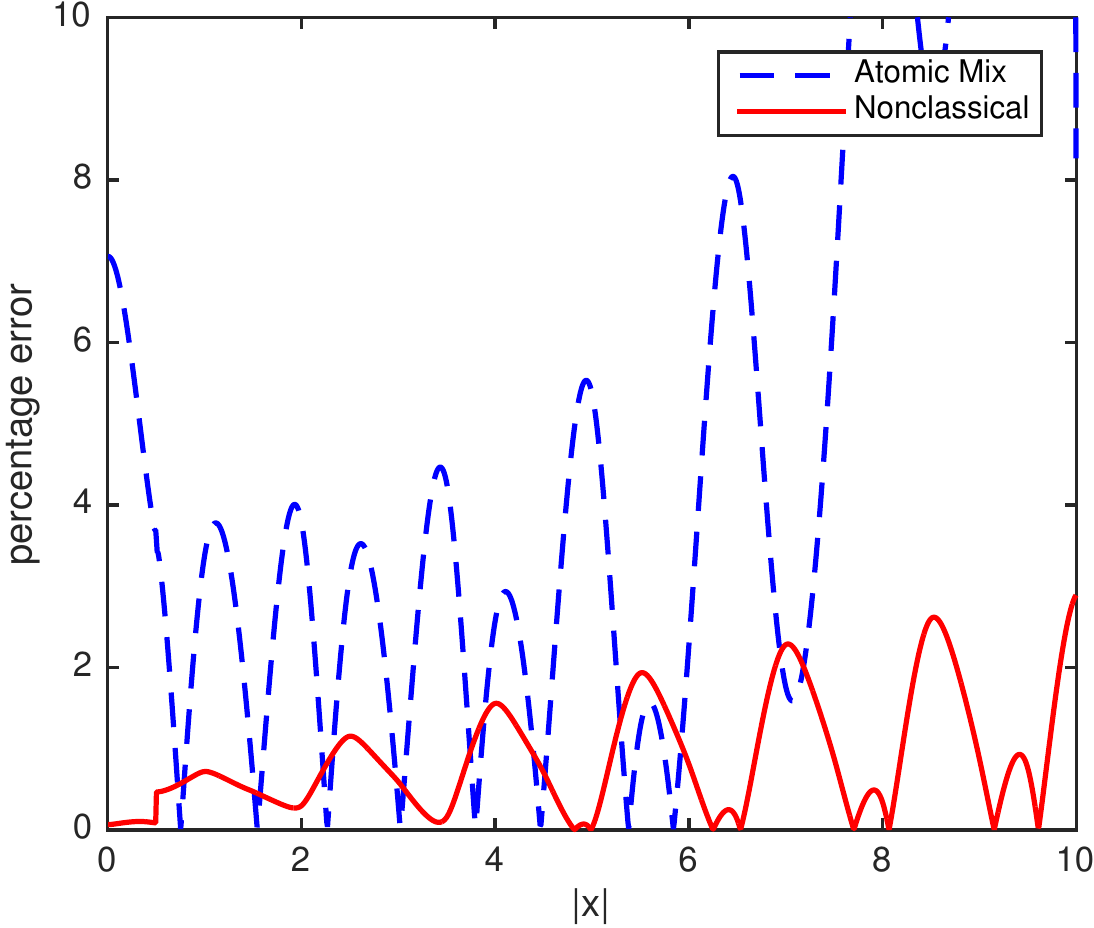}
        \caption{$c_1 = 0.3$}
        \label{figerrA30}
    \end{subfigure}
    \\
    \centering
    \begin{subfigure}{0.495\textwidth}
        \centering
        \includegraphics[width=\textwidth]{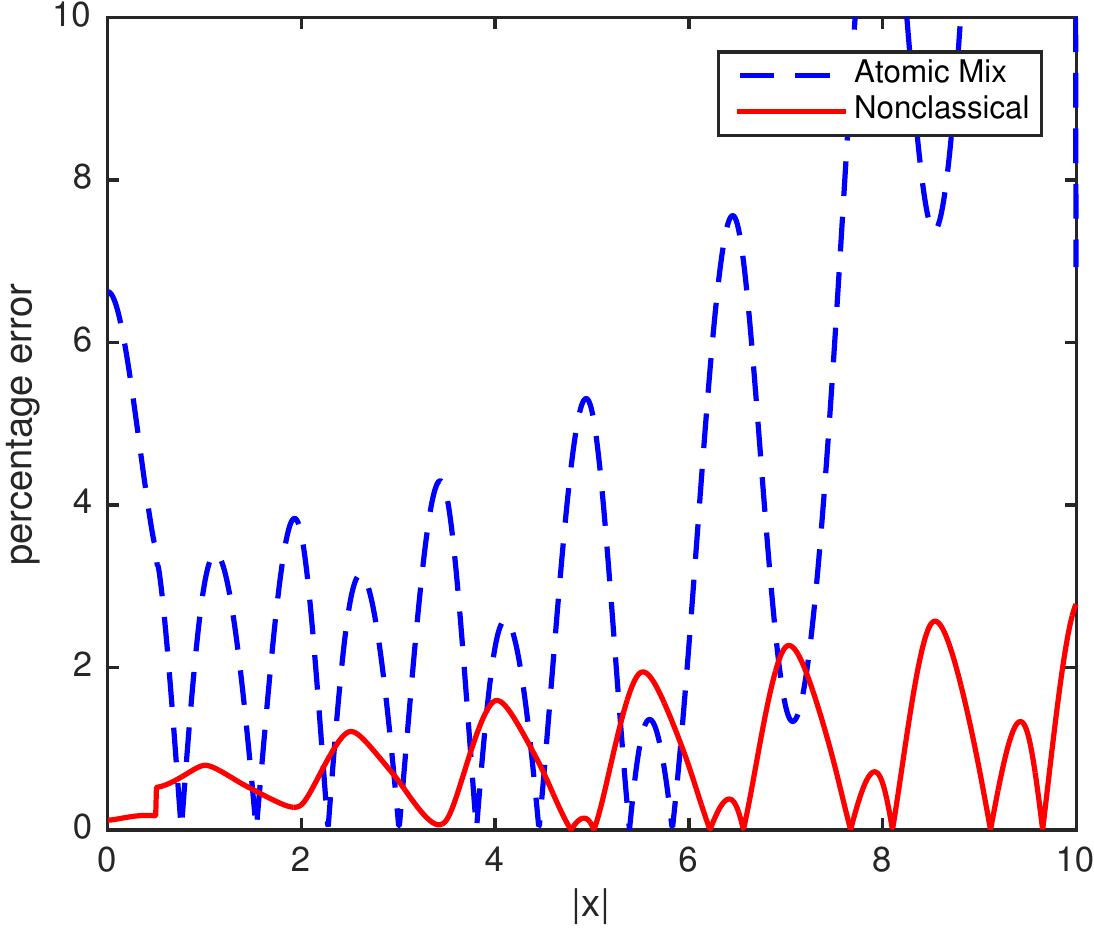}
        \caption{$c_1 = 0.4$}
        \label{figerrA40}
    \end{subfigure}
    \hfill
    \begin{subfigure}{0.495\textwidth}
        \centering
        \includegraphics[width=\textwidth]{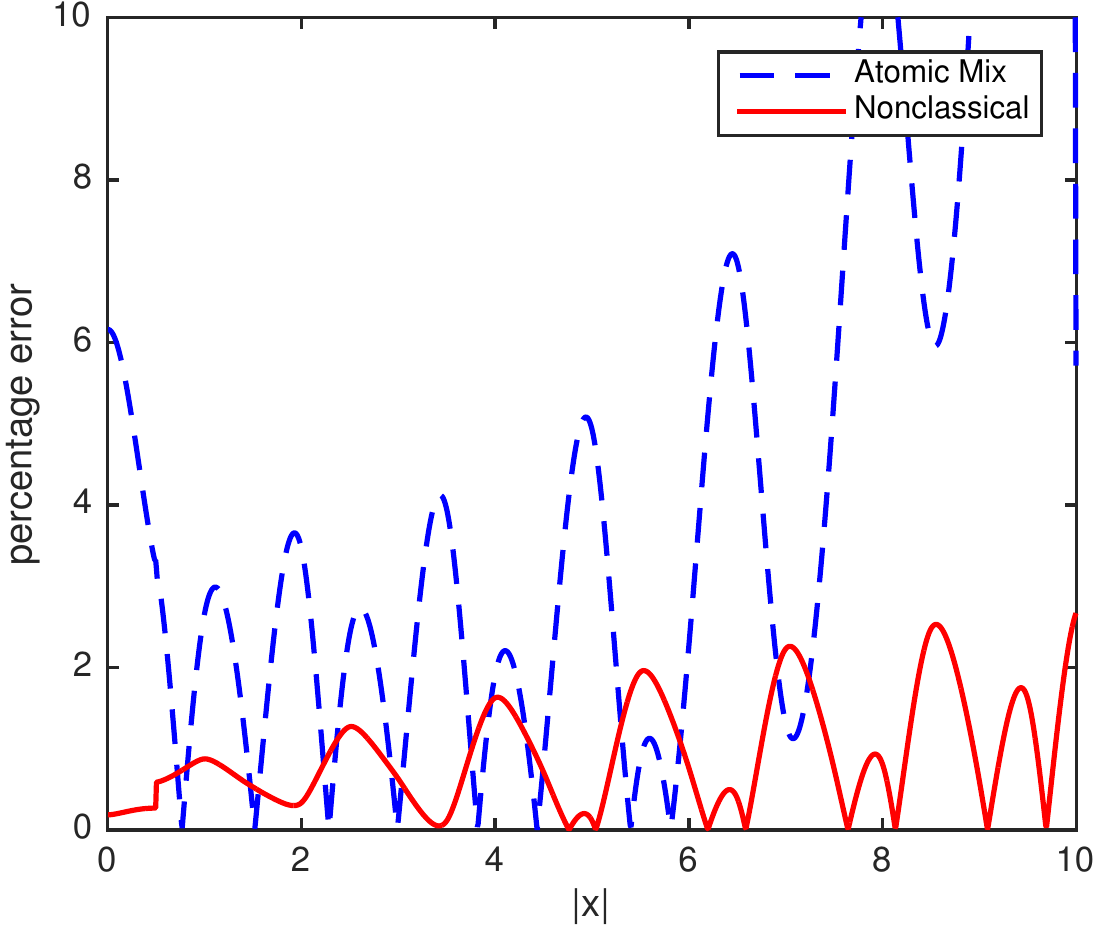}
        \caption{$c_1 = 0.5$}
        \label{figerrA50}
    \end{subfigure}
    \caption{Atomic mix and nonclassical percentage errors with respect to the benchmark solutions for problem set $\seta_1$}
    \label{figerrA1}
\end{figure}

\pagebreak
\begin{figure}[p]
    \centering
    \begin{subfigure}{0.495\textwidth}
        \centering
        \includegraphics[width=\textwidth]{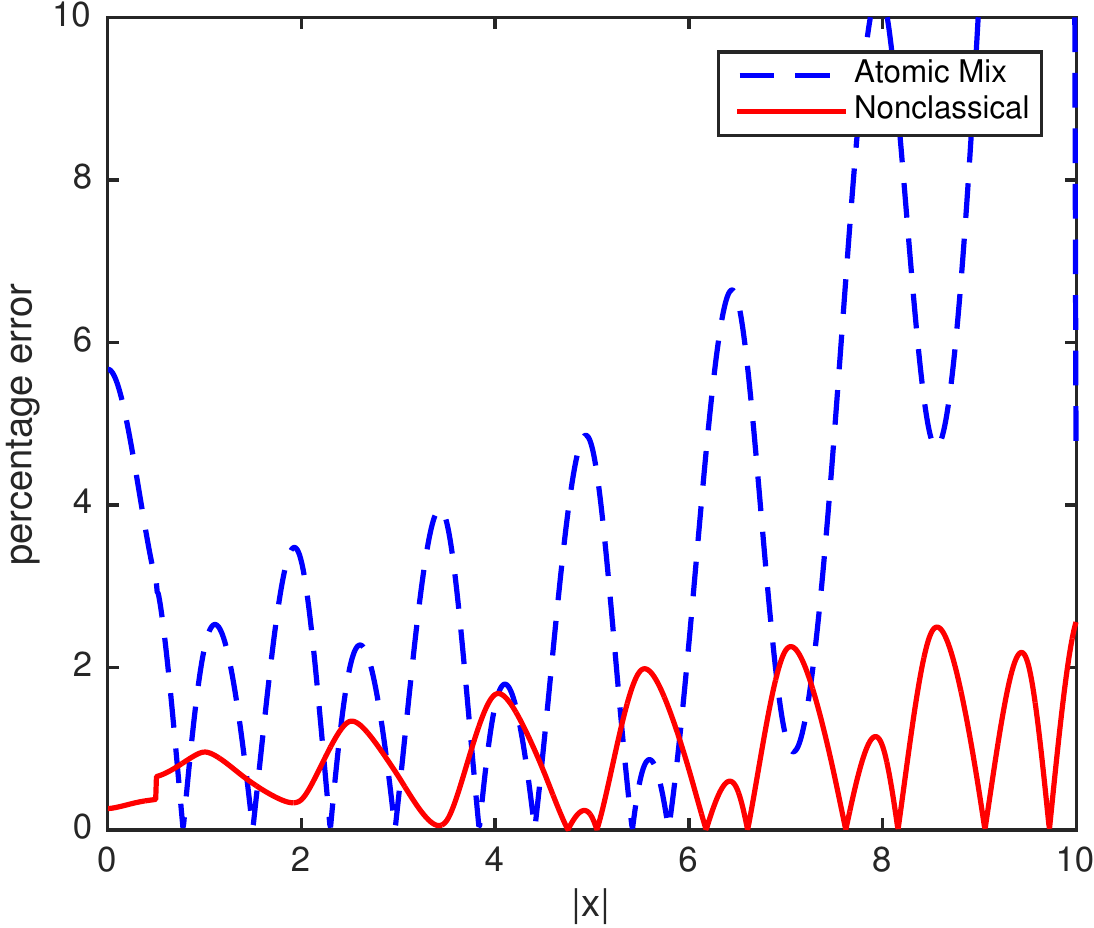}
        \caption{$c_1 = 0.6$}
        \label{figerrA60}
    \end{subfigure}
    \hfill
    \begin{subfigure}{0.495\textwidth}
        \centering
        \includegraphics[width=\textwidth]{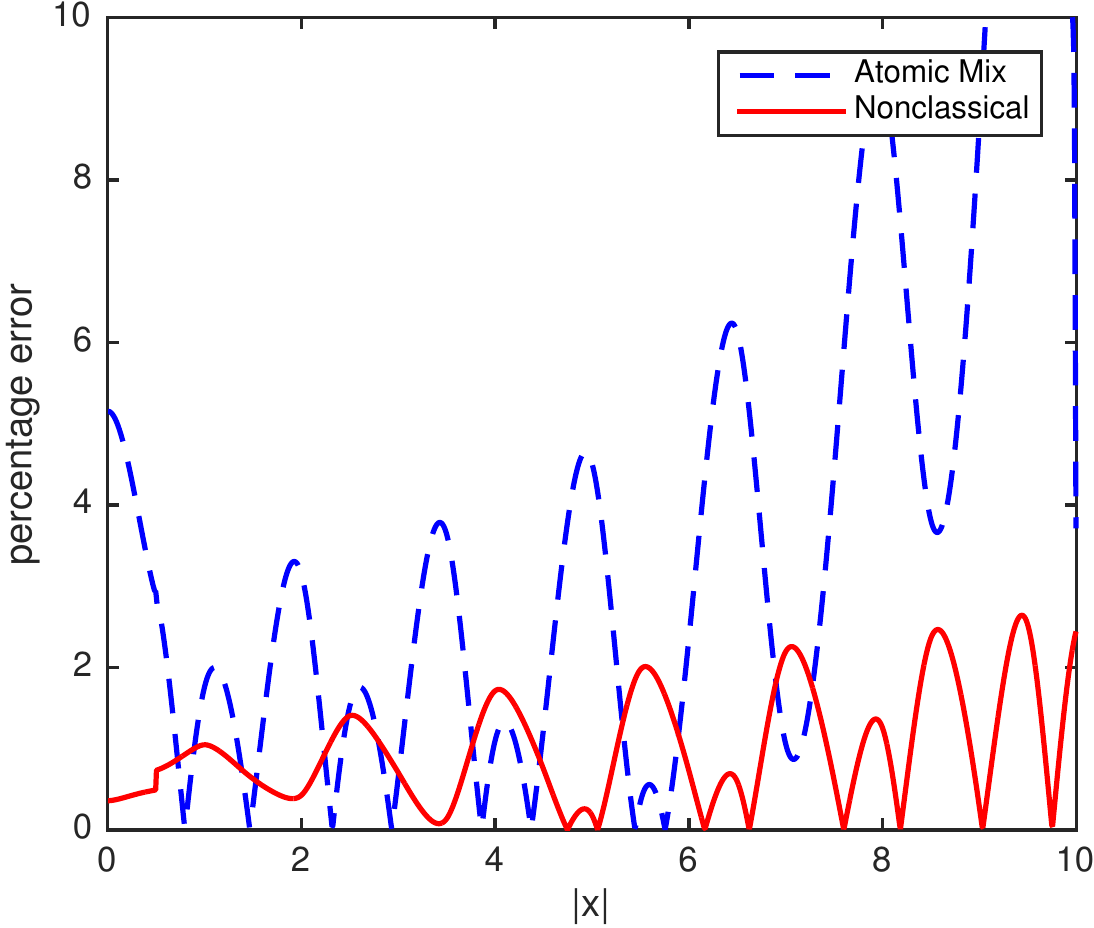}
        \caption{$c_1 = 0.7$}
        \label{figerrA70}
    \end{subfigure}
    \\
    \centering
    \begin{subfigure}{0.495\textwidth}
        \centering
        \includegraphics[width=\textwidth]{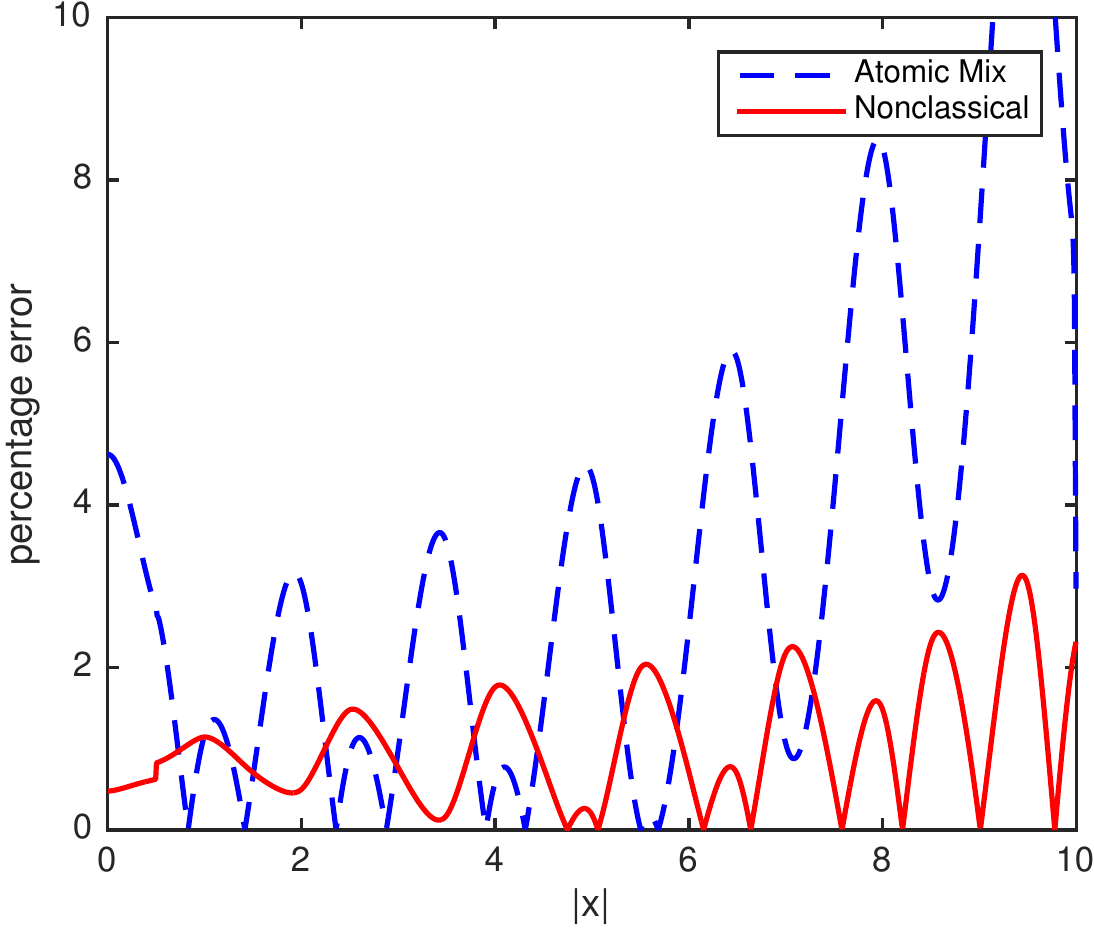}
        \caption{$c_1 = 0.8$}
        \label{figerrA80}
    \end{subfigure}
    \hfill
    \begin{subfigure}{0.495\textwidth}
        \centering
        \includegraphics[width=\textwidth]{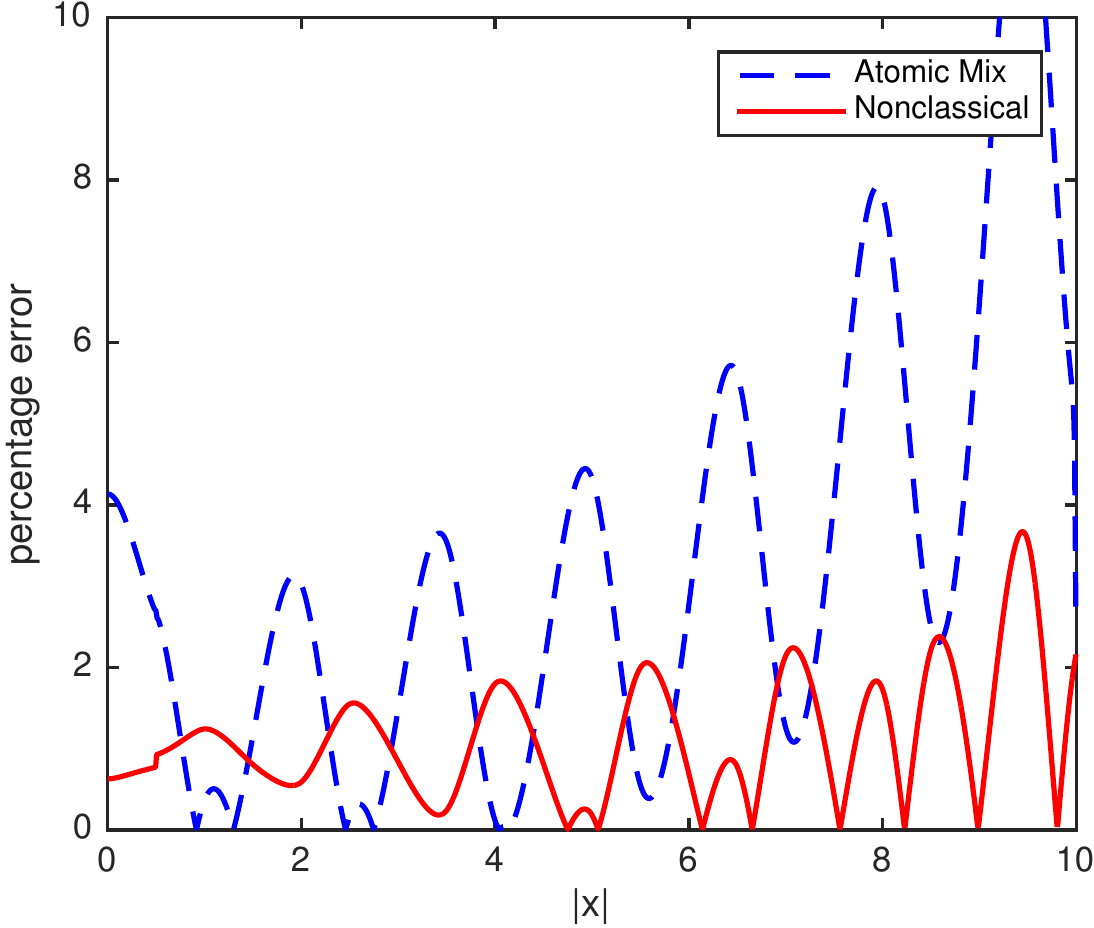}
        \caption{$c_1 = 0.9$}
        \label{figerrA90}
    \end{subfigure}
    \\
    \centering
    \begin{subfigure}{0.495\textwidth}
        \centering
        \includegraphics[width=\textwidth]{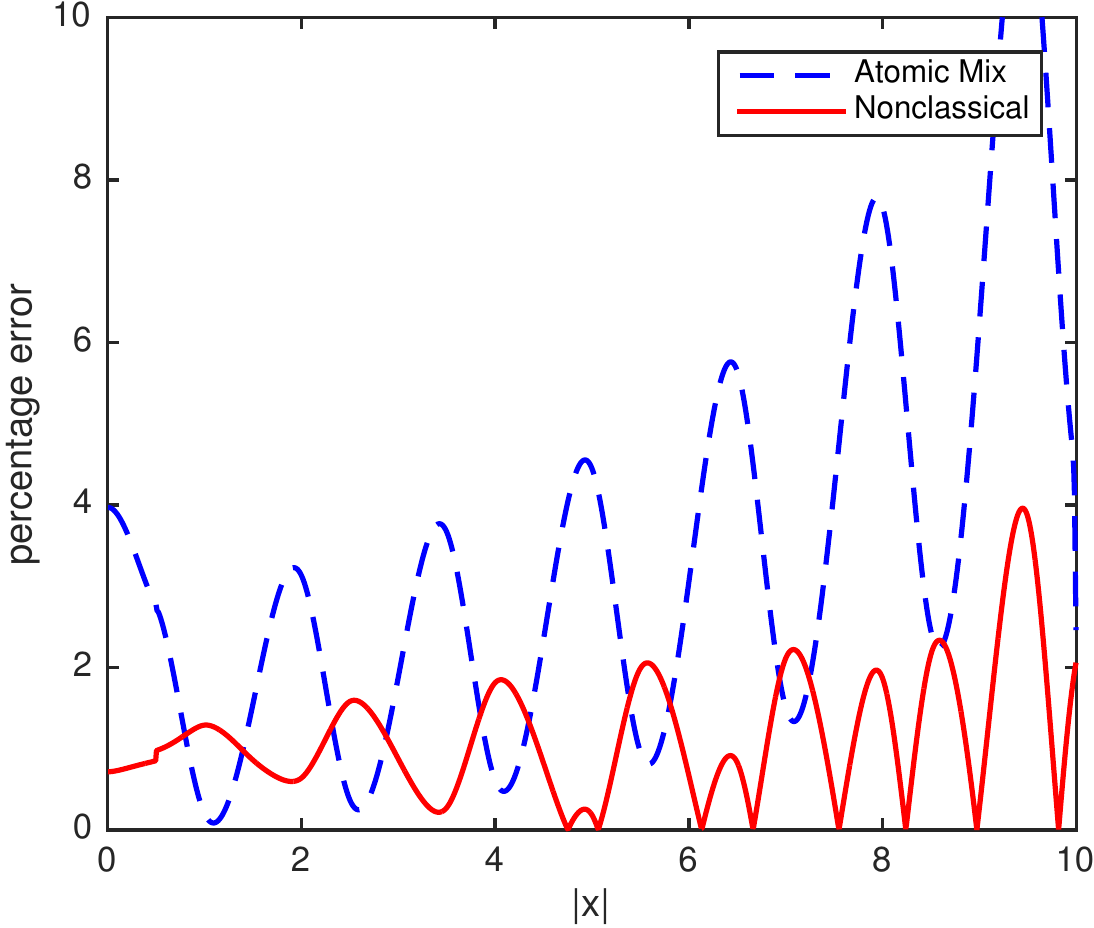}
        \caption{$c_1 = 0.95$}
        \label{figerrA95}
    \end{subfigure}
    \hfill
    \begin{subfigure}{0.495\textwidth}
        \centering
        \includegraphics[width=\textwidth]{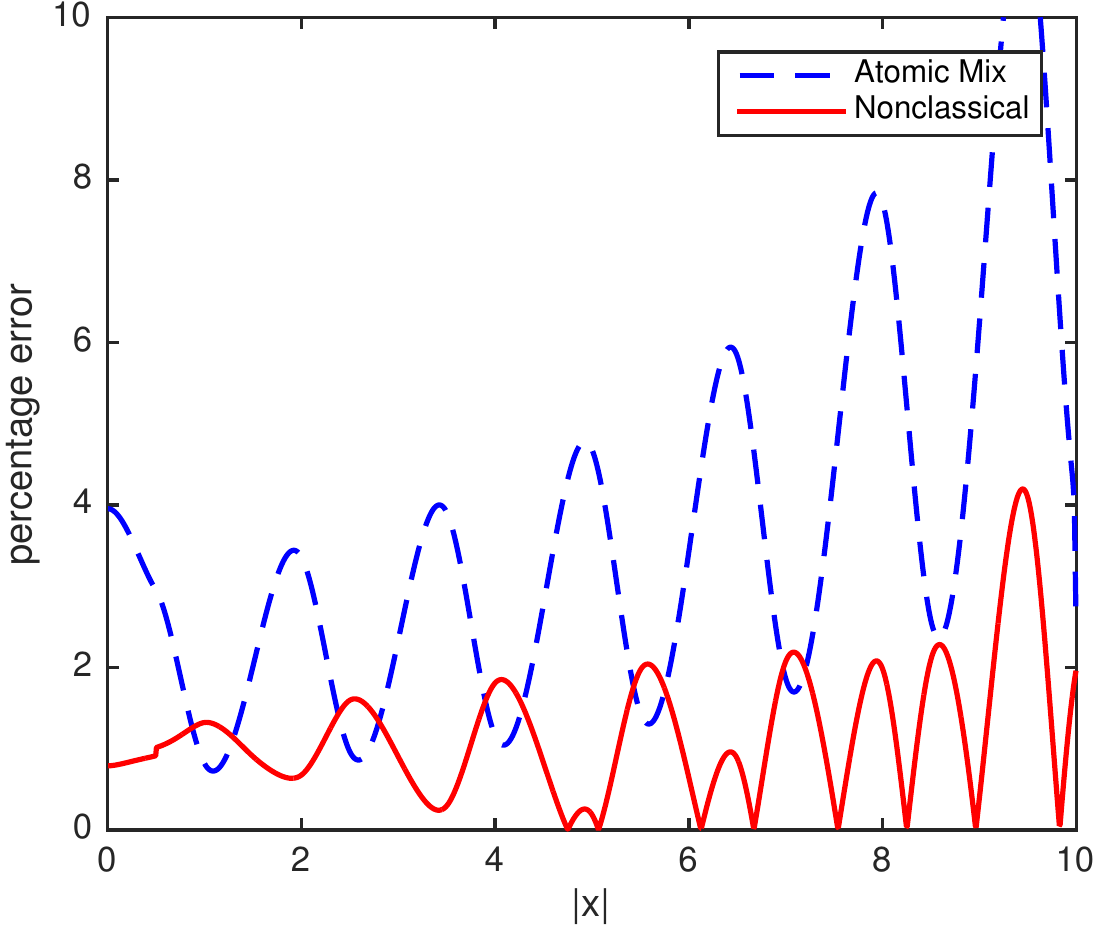}
        \caption{$c_1 = 0.99$}
        \label{figerrA99}
    \end{subfigure}
    \caption{Atomic mix and nonclassical percentage errors with respect to the benchmark solutions for problem set $\seta_1$}
    \label{figerrA2}
\end{figure}

\pagebreak
\begin{figure}[p]
    \centering
    \begin{subfigure}{0.495\textwidth}
        \centering
        \includegraphics[width=\textwidth]{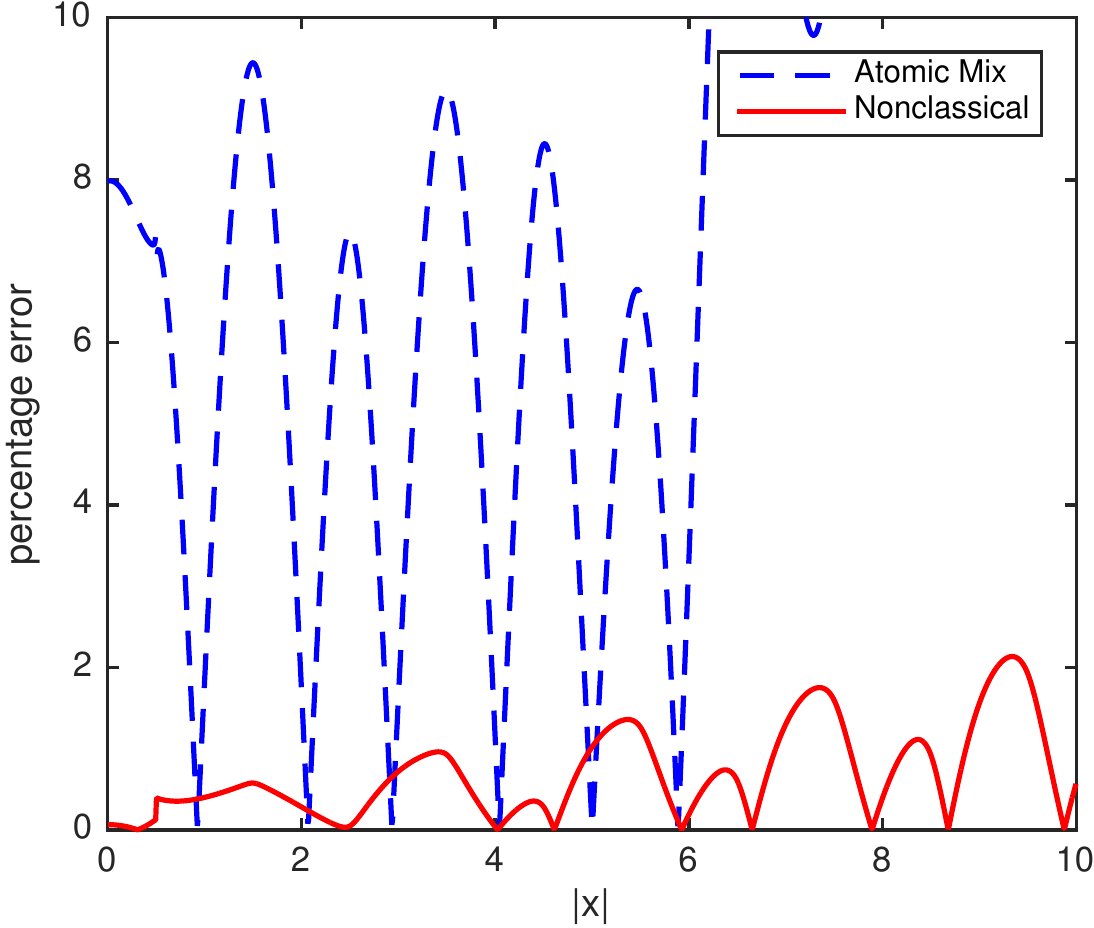}
        \caption{$c_1 = 0.0$}
        \label{figerrB00}
    \end{subfigure}
    \hfill
    \begin{subfigure}{0.495\textwidth}
        \centering
        \includegraphics[width=\textwidth]{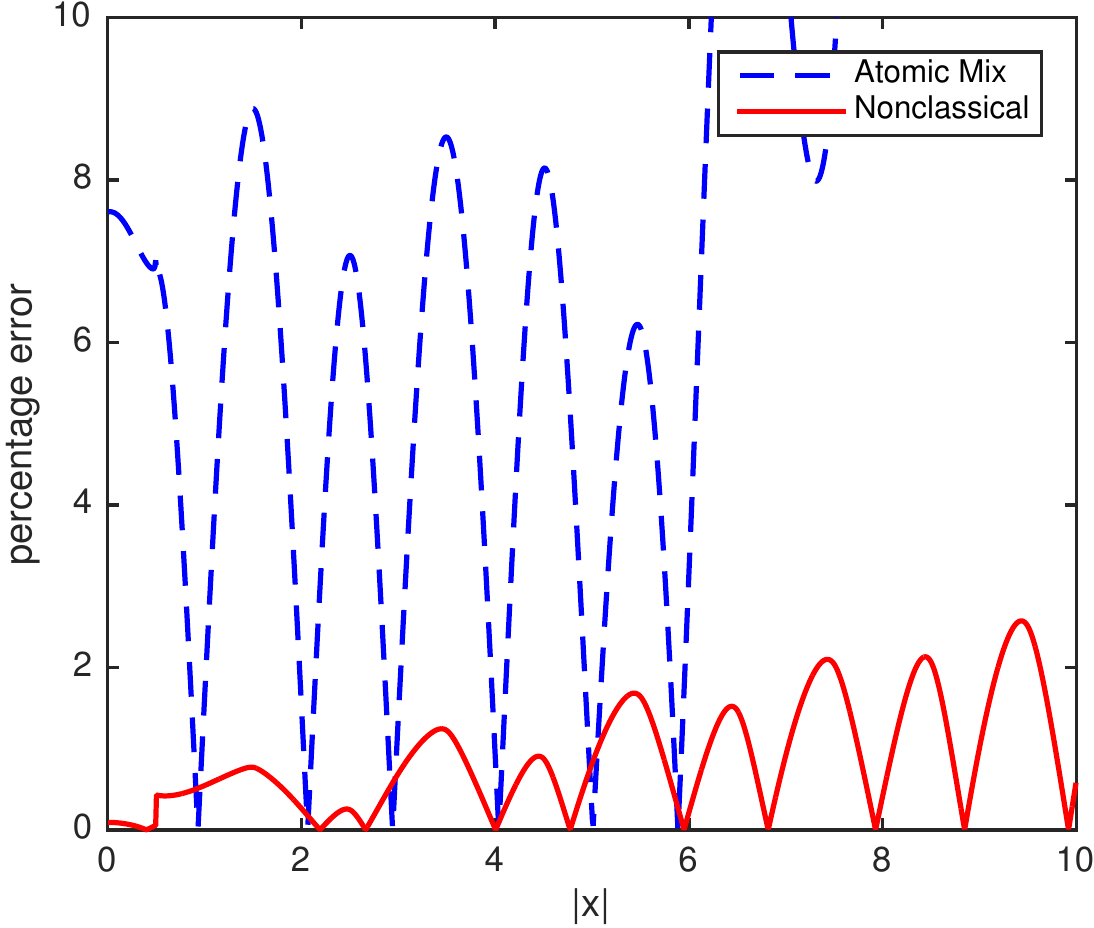}
        \caption{$c_1 = 0.1$}
        \label{figerrB10}
    \end{subfigure}
    \\
    \centering
    \begin{subfigure}{0.495\textwidth}
        \centering
        \includegraphics[width=\textwidth]{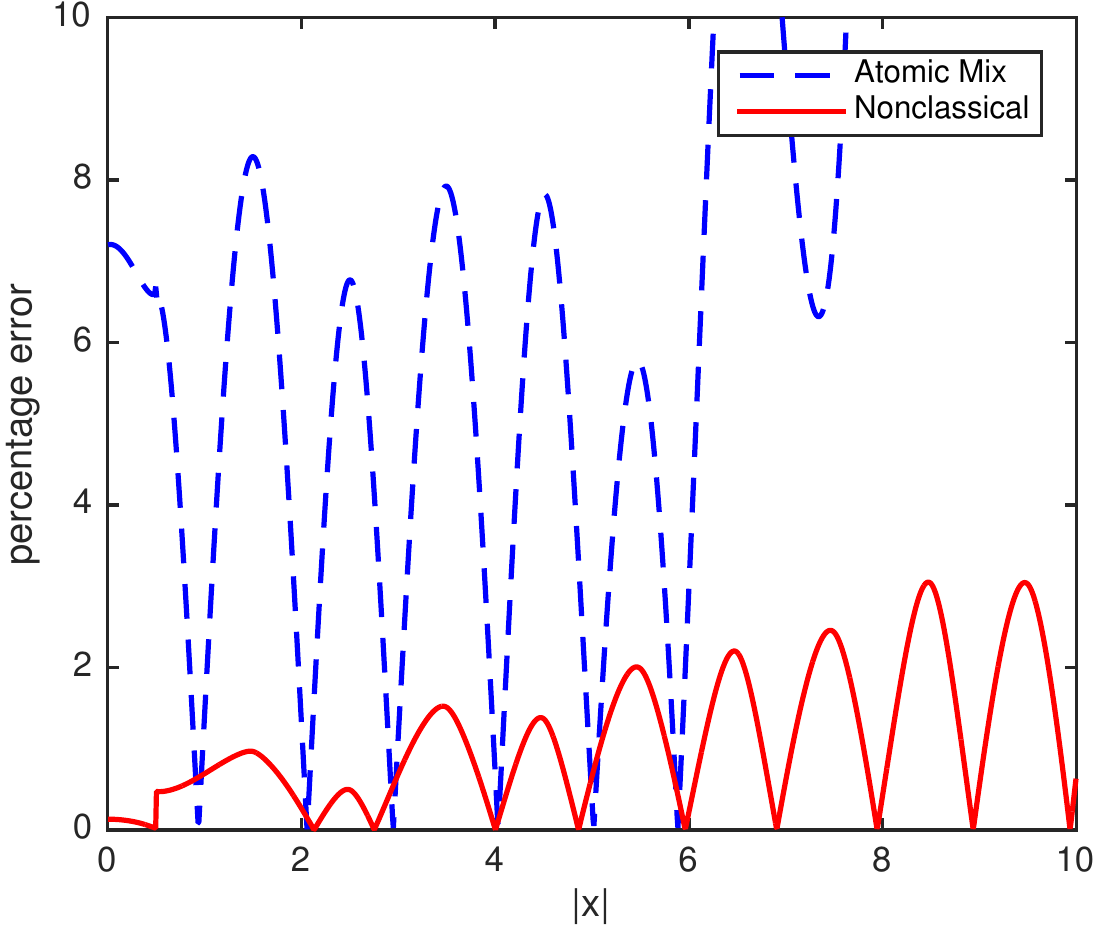}
        \caption{$c_1 = 0.2$}
        \label{figerrB20}
    \end{subfigure}
    \hfill
    \begin{subfigure}{0.495\textwidth}
        \centering
        \includegraphics[width=\textwidth]{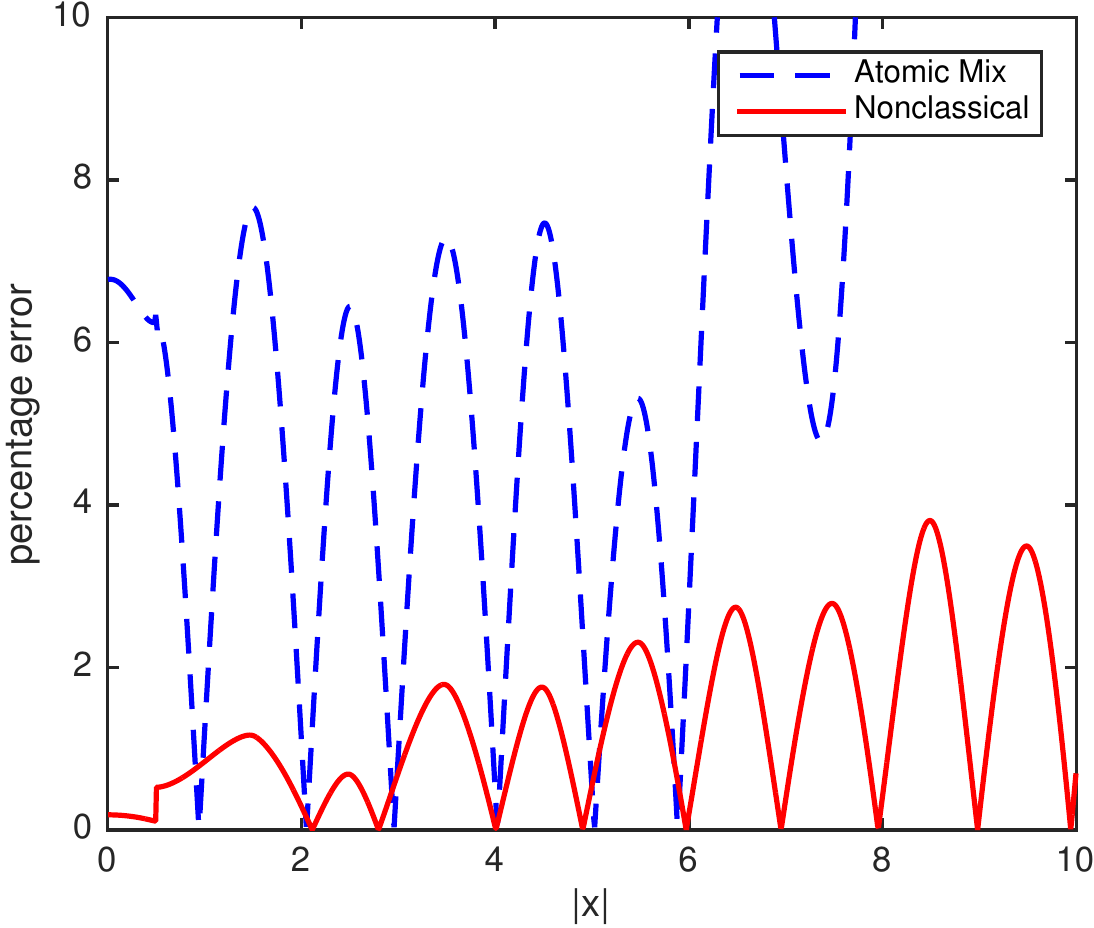}
        \caption{$c_1 = 0.3$}
        \label{figerrB30}
    \end{subfigure}
    \\
    \centering
    \begin{subfigure}{0.495\textwidth}
        \centering
        \includegraphics[width=\textwidth]{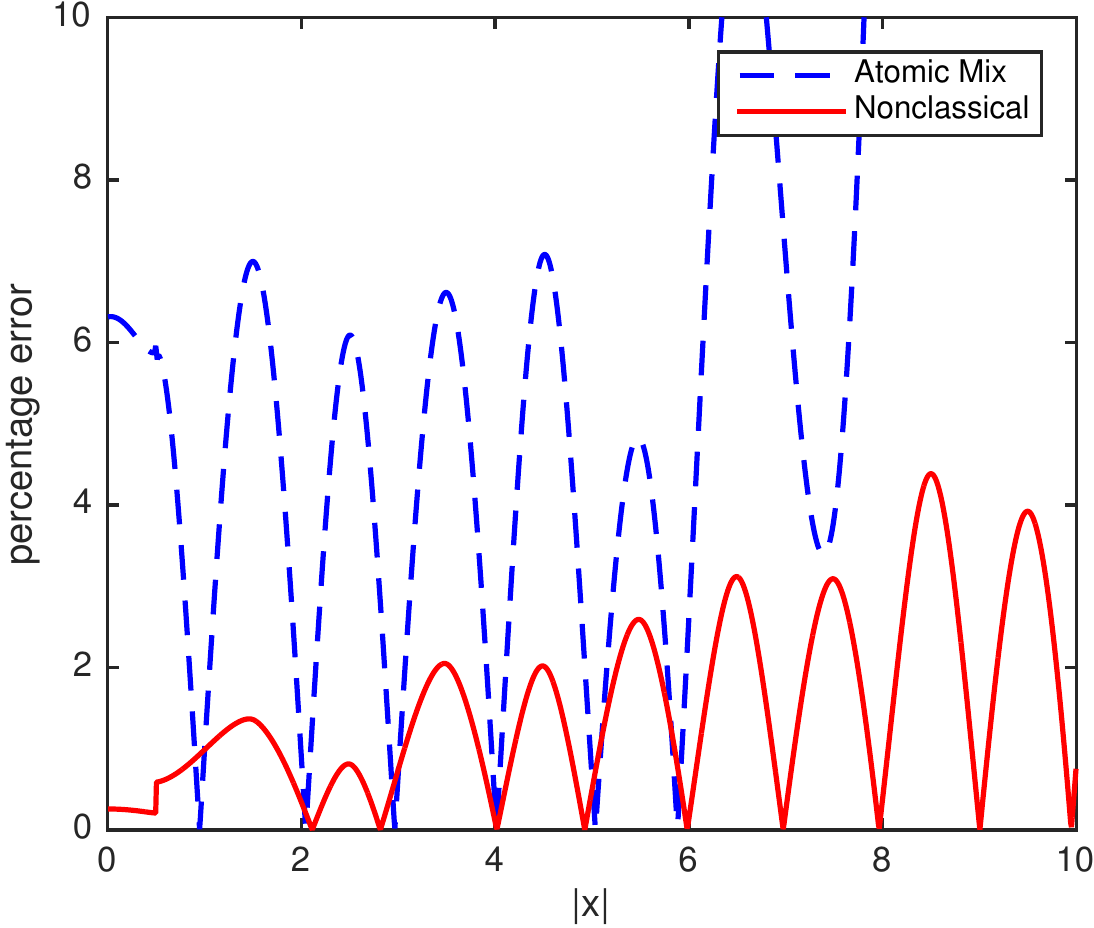}
        \caption{$c_1 = 0.4$}
        \label{figerrB40}
    \end{subfigure}
    \hfill
    \begin{subfigure}{0.495\textwidth}
        \centering
        \includegraphics[width=\textwidth]{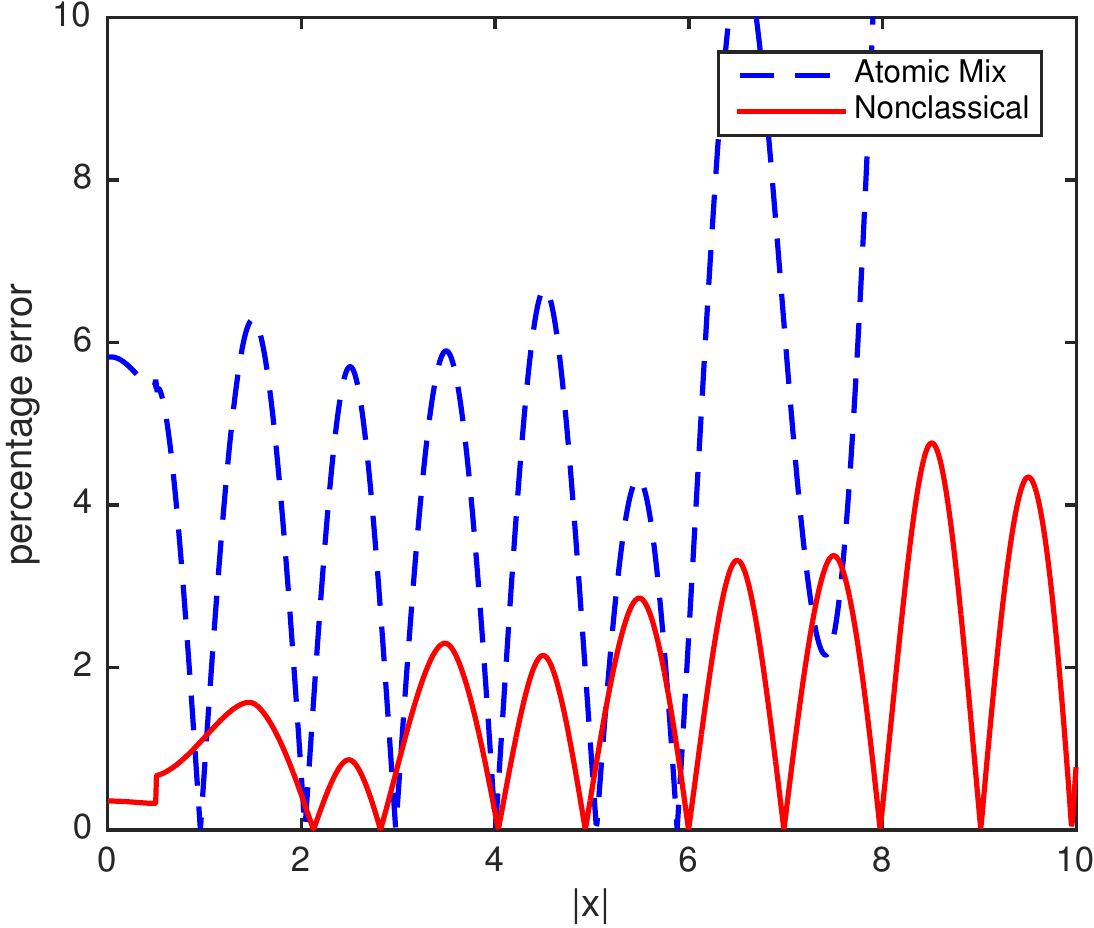}
        \caption{$c_1 = 0.5$}
        \label{figerrB50}
    \end{subfigure}
    \caption{Atomic mix and nonclassical percentage errors with respect to the benchmark solutions for problem set $\seta_2$}
    \label{figerrB1}
\end{figure}

\pagebreak
\begin{figure}[p]
    \centering
    \begin{subfigure}{0.495\textwidth}
        \centering
        \includegraphics[width=\textwidth]{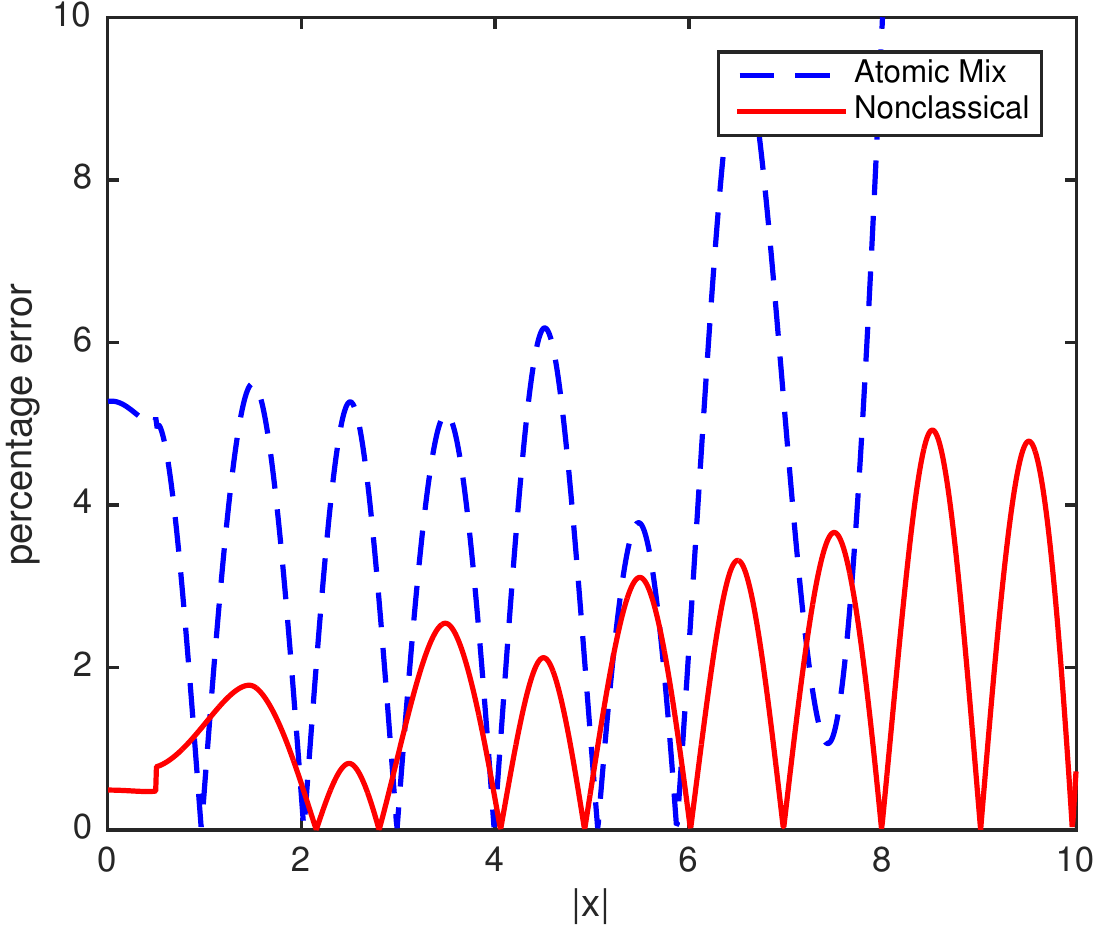}
        \caption{$c_1 = 0.6$}
        \label{figerrB60}
    \end{subfigure}
    \hfill
    \begin{subfigure}{0.495\textwidth}
        \centering
        \includegraphics[width=\textwidth]{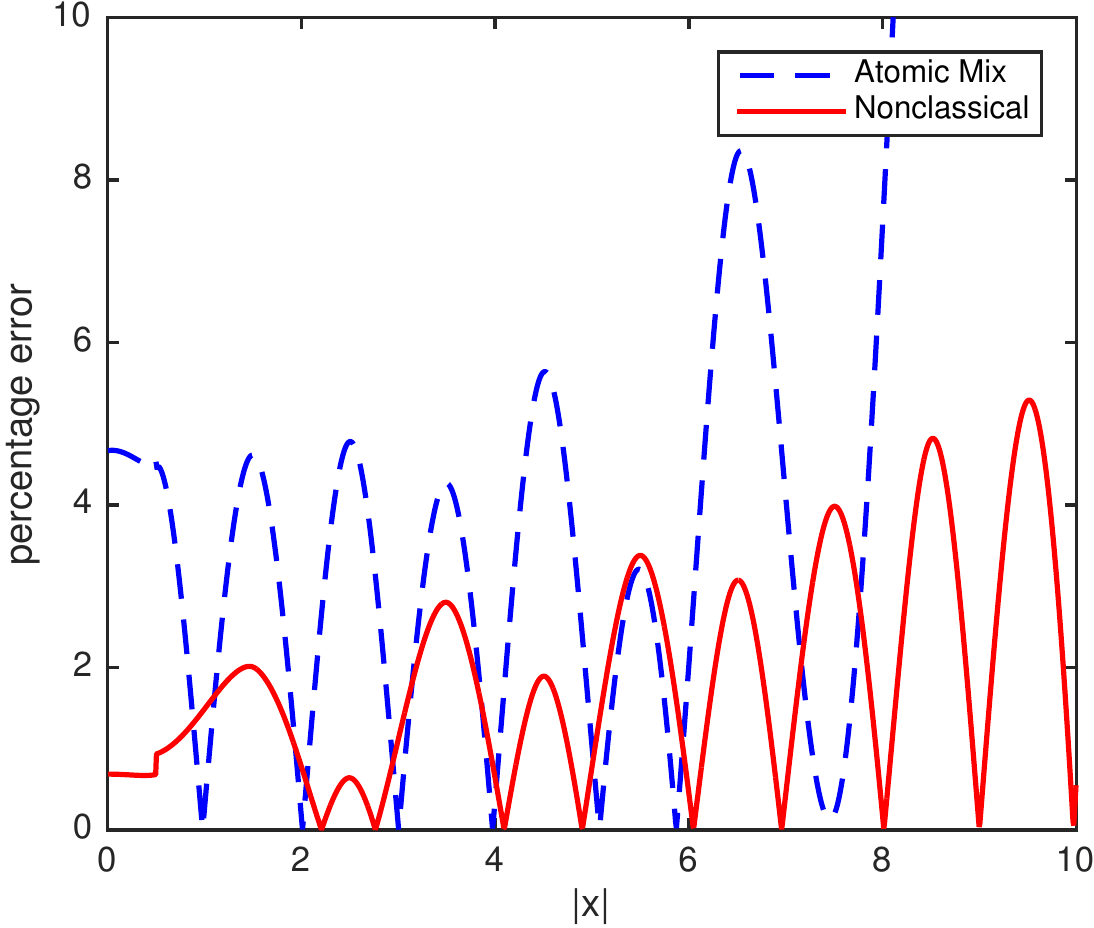}
        \caption{$c_1 = 0.7$}
        \label{figerrB70}
    \end{subfigure}
    \\
    \centering
    \begin{subfigure}{0.495\textwidth}
        \centering
        \includegraphics[width=\textwidth]{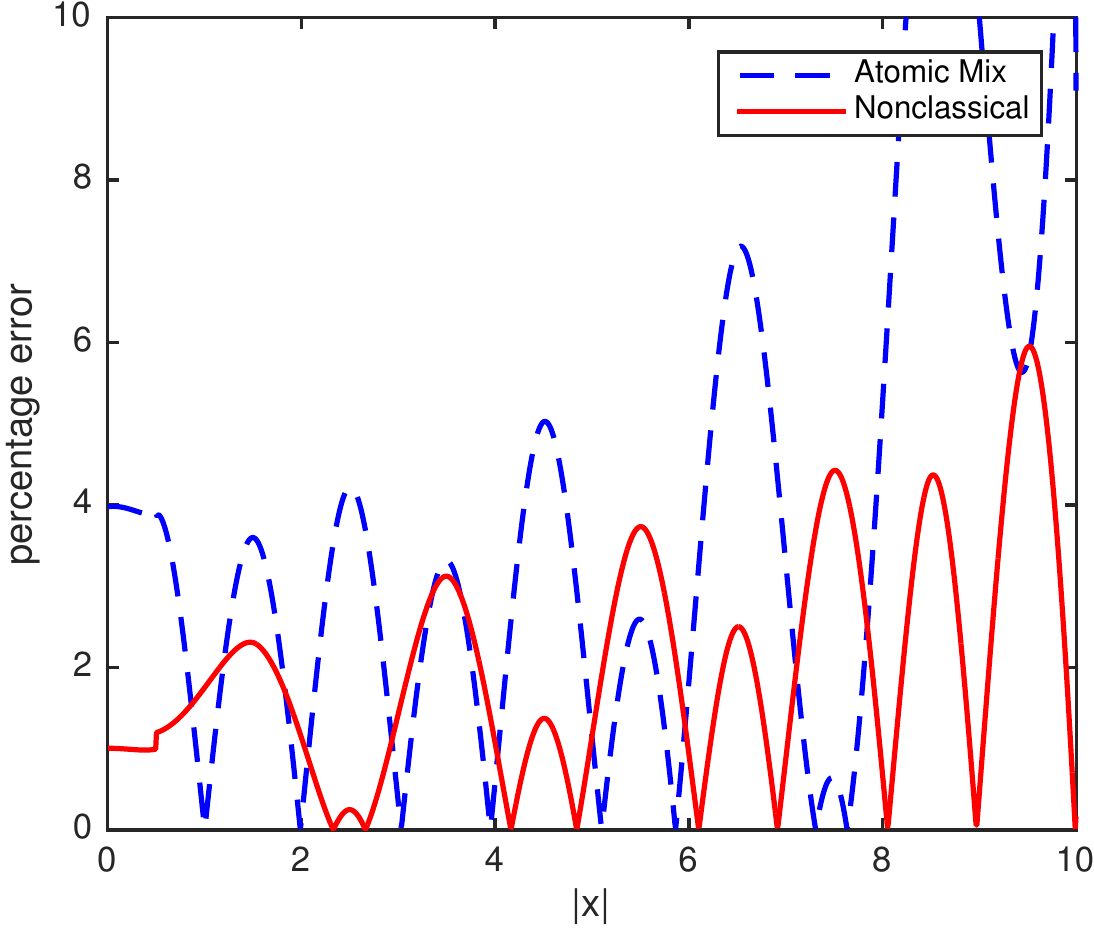}
        \caption{$c_1 = 0.8$}
        \label{figerrB80}
    \end{subfigure}
    \hfill
    \begin{subfigure}{0.495\textwidth}
        \centering
        \includegraphics[width=\textwidth]{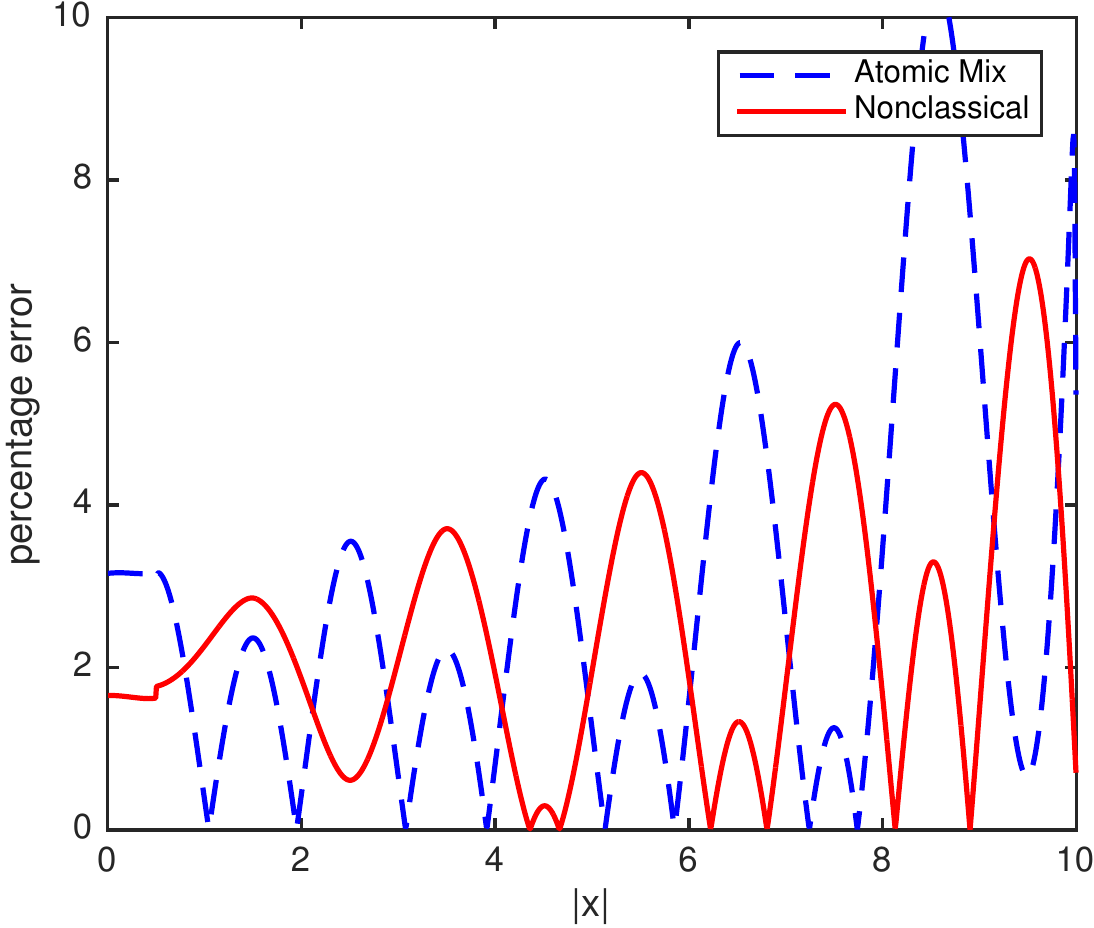}
        \caption{$c_1 = 0.9$}
        \label{figerrB90}
    \end{subfigure}
    \\
    \centering
    \begin{subfigure}{0.495\textwidth}
        \centering
        \includegraphics[width=\textwidth]{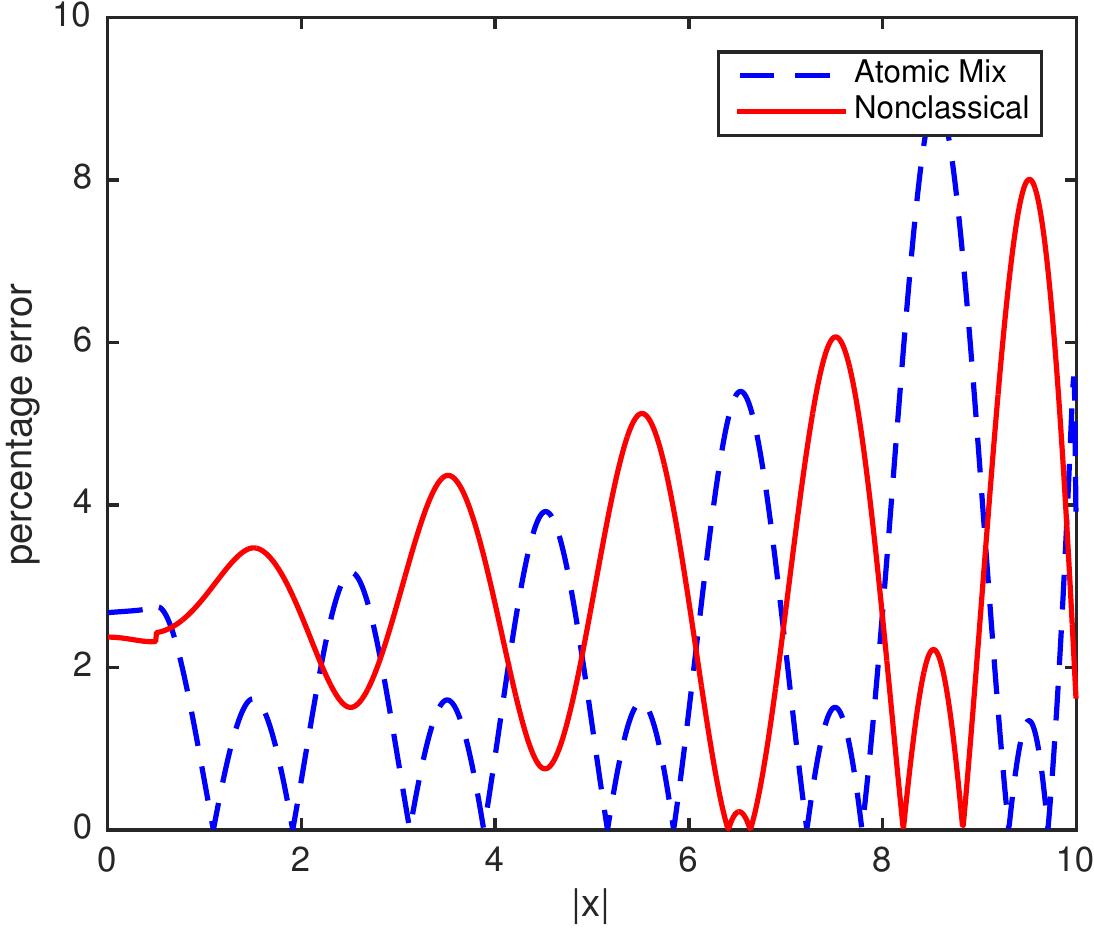}
        \caption{$c_1 = 0.95$}
        \label{figerrB95}
    \end{subfigure}
    \hfill
    \begin{subfigure}{0.495\textwidth}
        \centering
        \includegraphics[width=\textwidth]{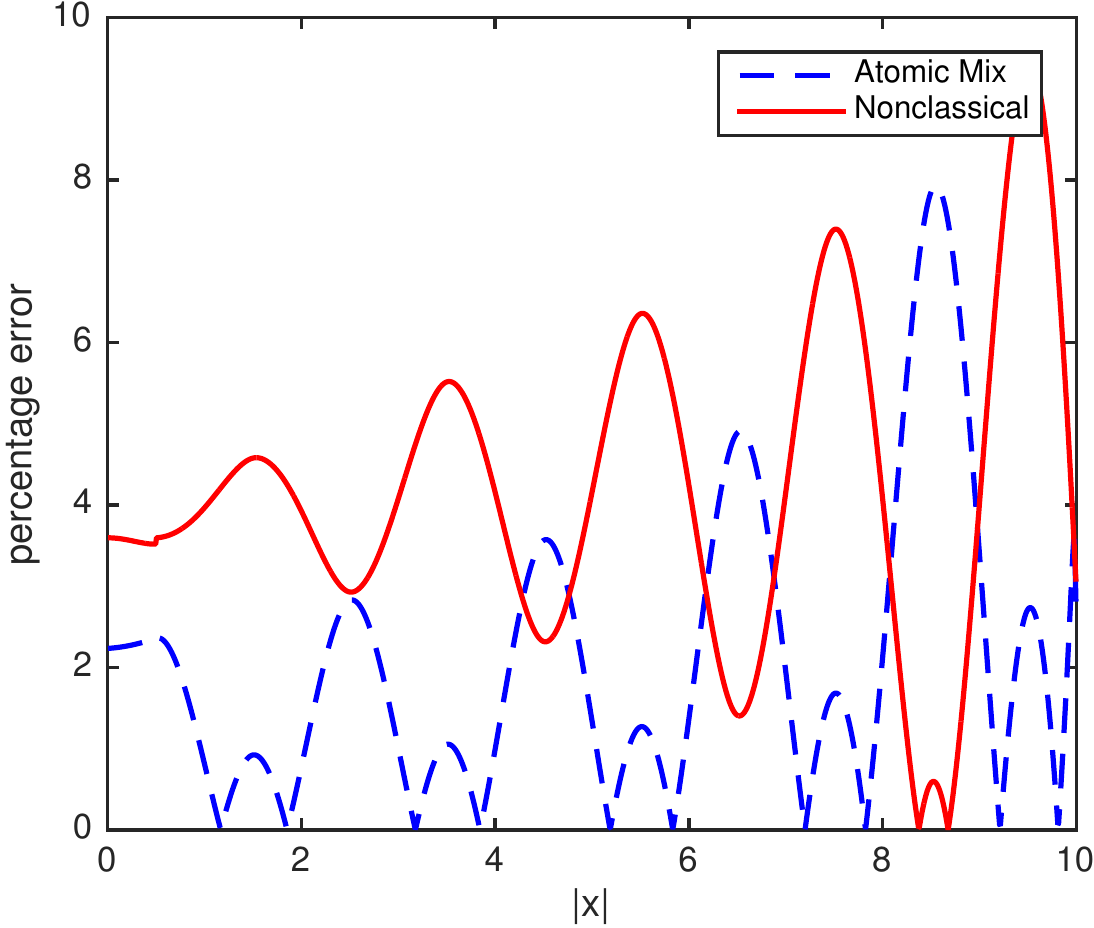}
        \caption{$c_1 = 0.99$}
        \label{figerrB99}
    \end{subfigure}
    \caption{Atomic mix and nonclassical percentage errors with respect to the benchmark solutions for problem set $\seta_2$}
    \label{figerrB2}
\end{figure}

\pagebreak
\begin{figure}[p]
    \centering
    \begin{subfigure}{0.495\textwidth}
        \centering
        \includegraphics[width=\textwidth]{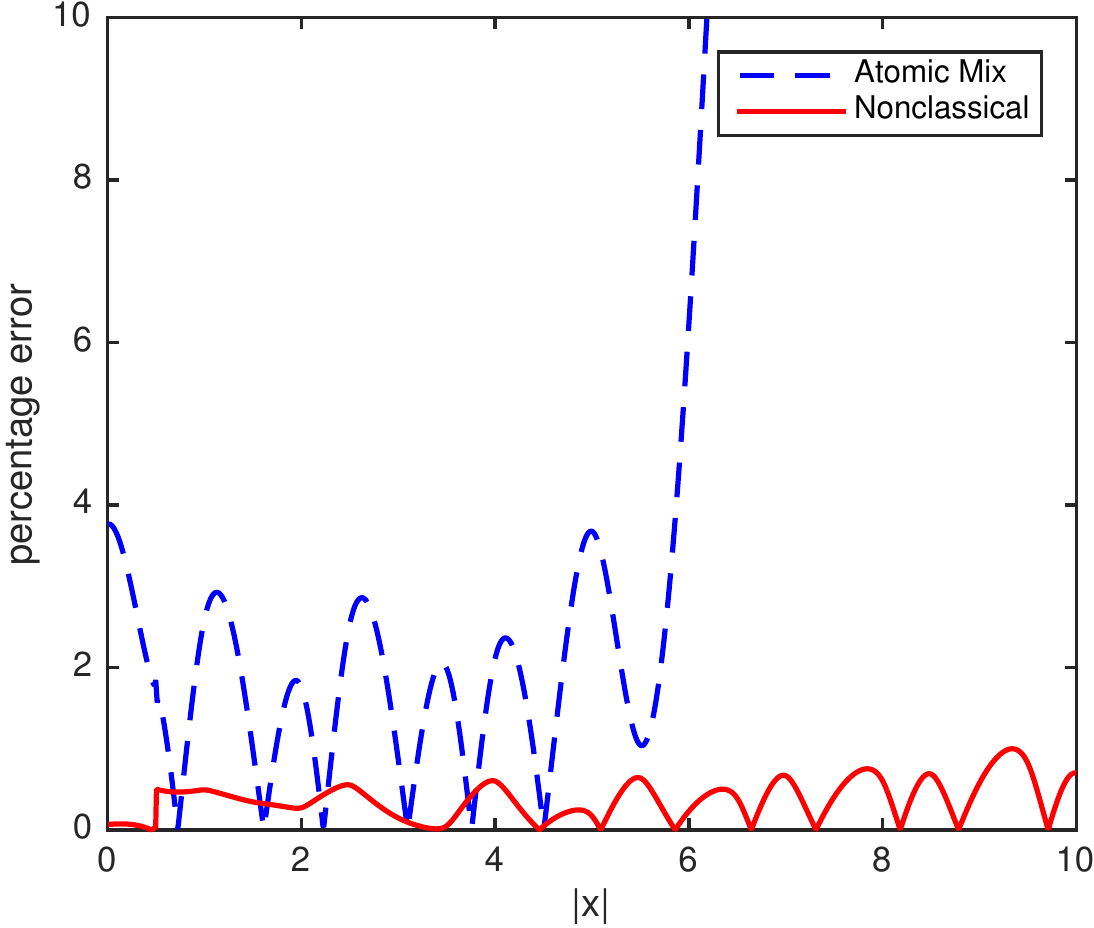}
        \caption{$c_1 = 0.0$}
        \label{figerrC00}
    \end{subfigure}
    \hfill
    \begin{subfigure}{0.495\textwidth}
        \centering
        \includegraphics[width=\textwidth]{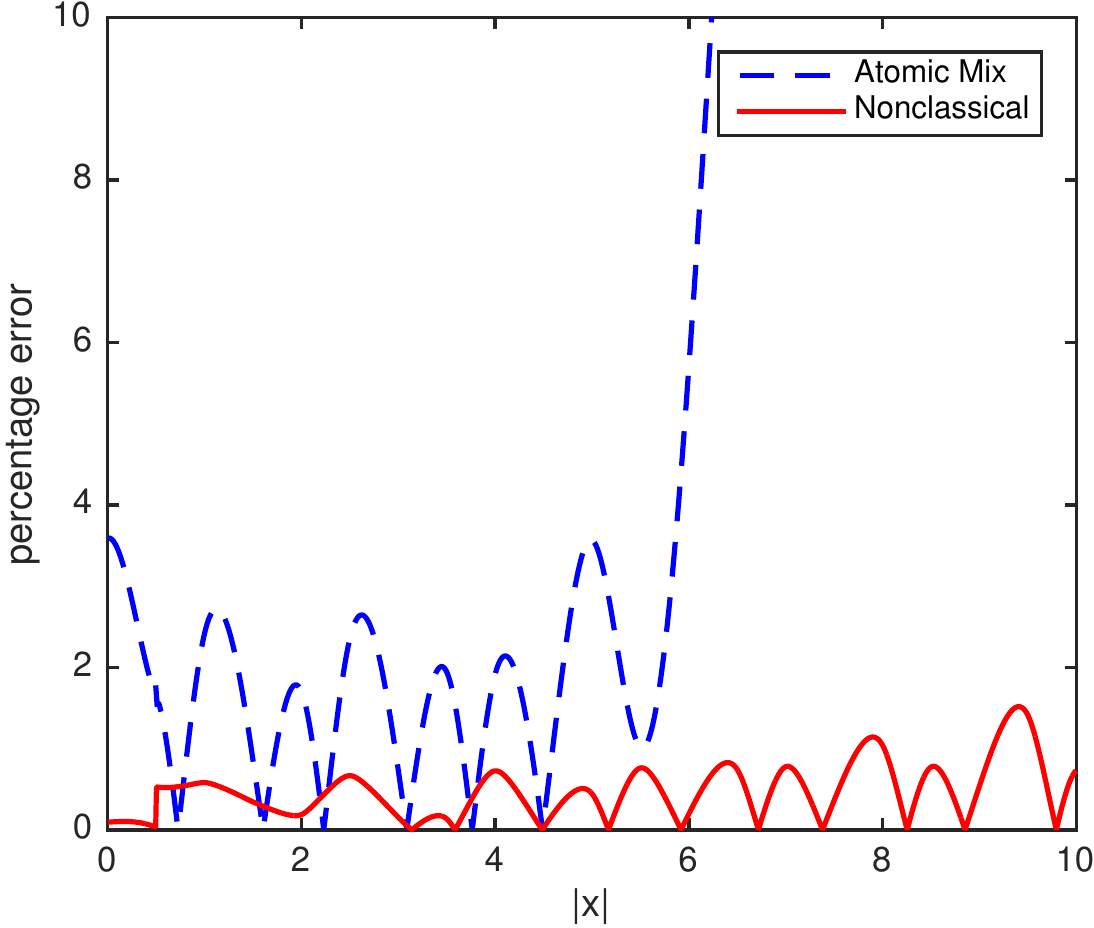}
        \caption{$c_1 = 0.1$}
        \label{figerrC10}
    \end{subfigure}
    \\
    \centering
    \begin{subfigure}{0.495\textwidth}
        \centering
        \includegraphics[width=\textwidth]{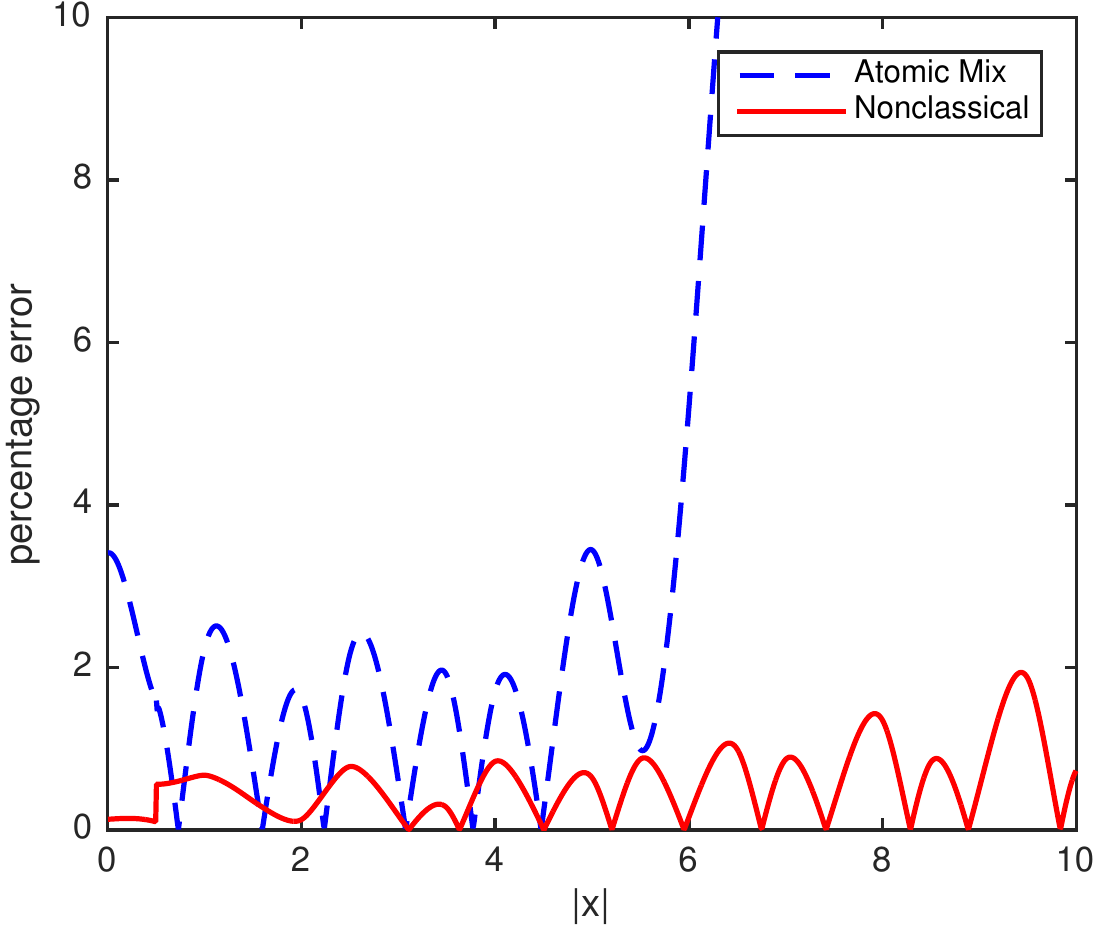}
        \caption{$c_1 = 0.2$}
        \label{figerrC20}
    \end{subfigure}
    \hfill
    \begin{subfigure}{0.495\textwidth}
        \centering
        \includegraphics[width=\textwidth]{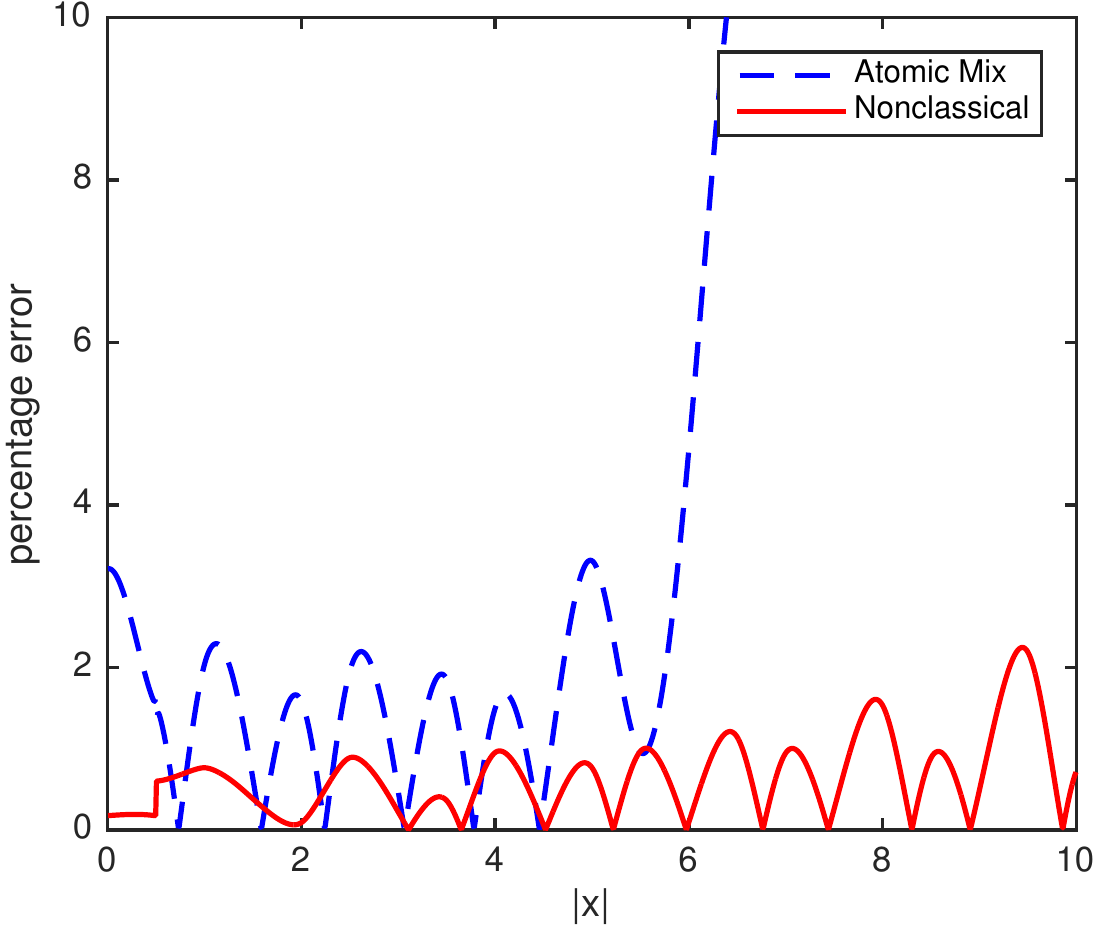}
        \caption{$c_1 = 0.3$}
        \label{figerrC30}
    \end{subfigure}
    \\
    \centering
    \begin{subfigure}{0.495\textwidth}
        \centering
        \includegraphics[width=\textwidth]{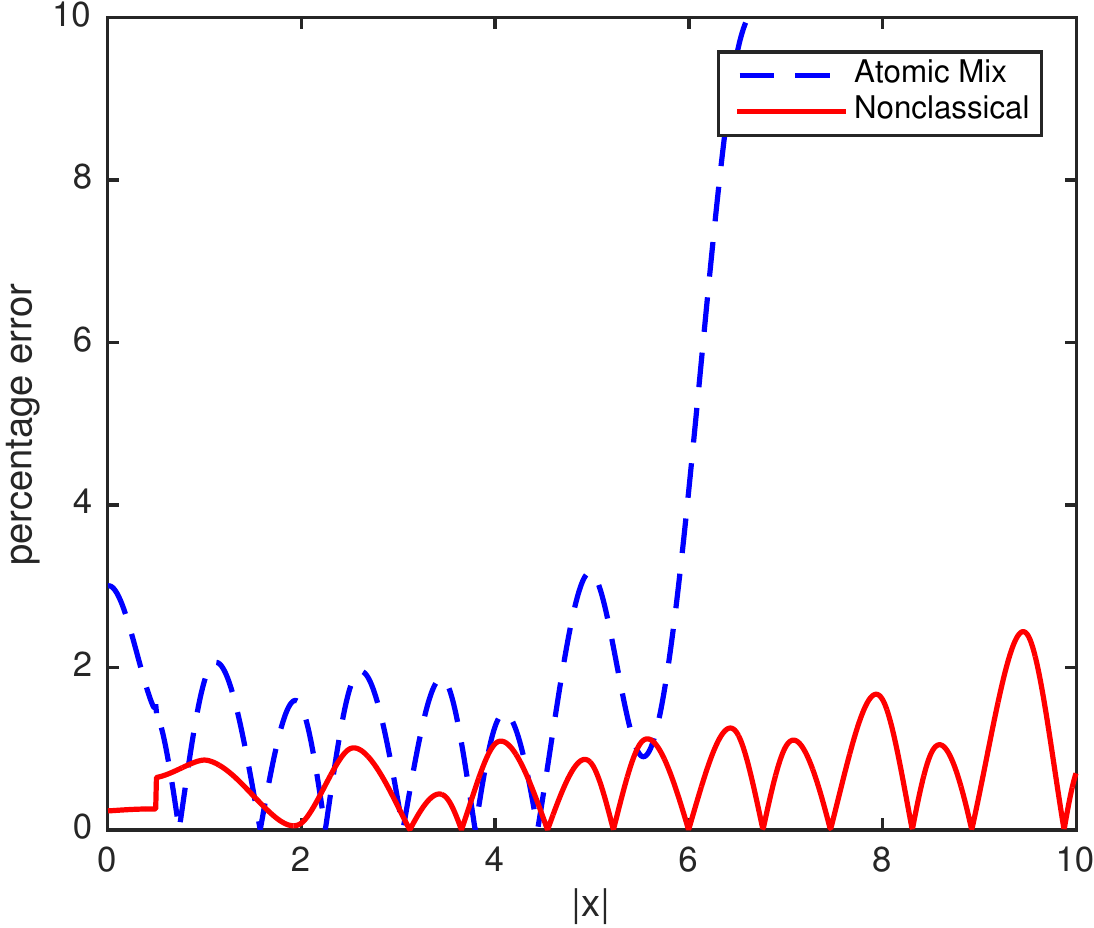}
        \caption{$c_1 = 0.4$}
        \label{figerrC40}
    \end{subfigure}
    \hfill
    \begin{subfigure}{0.495\textwidth}
        \centering
        \includegraphics[width=\textwidth]{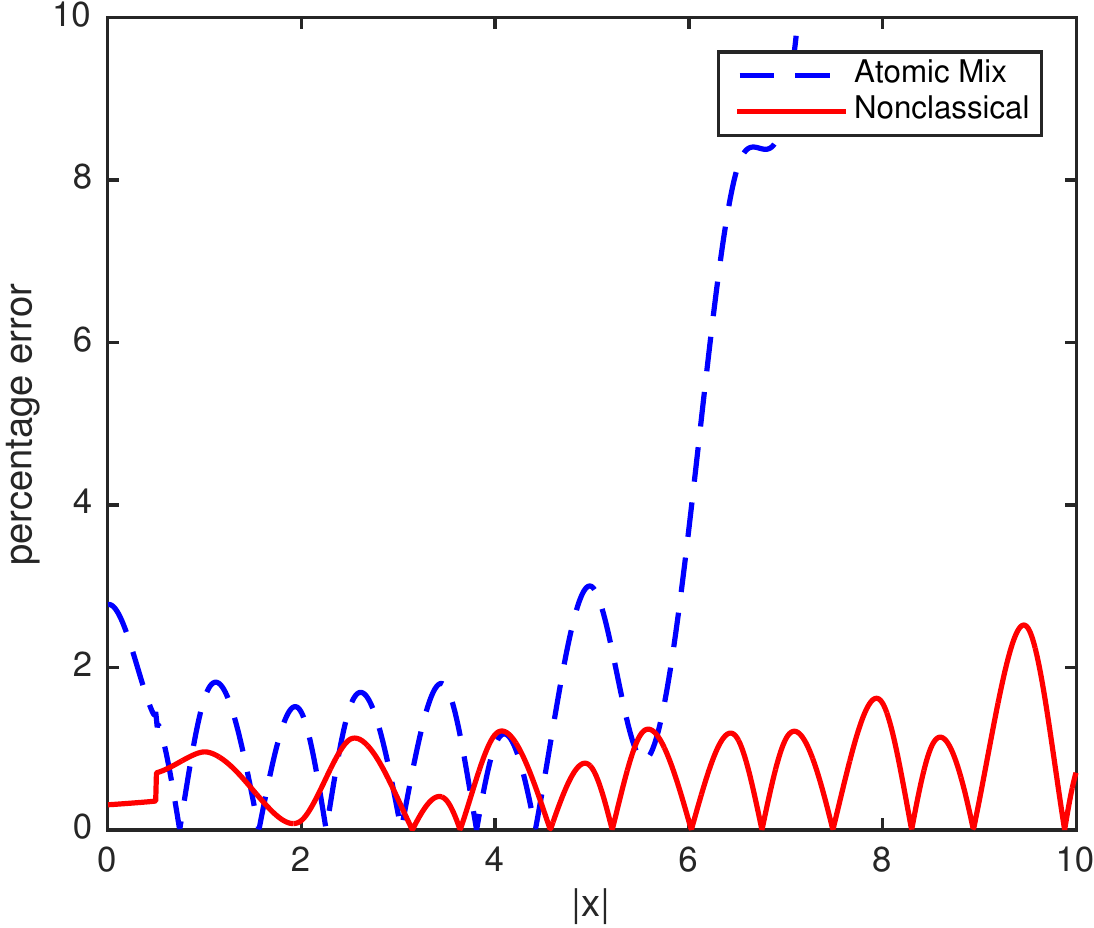}
        \caption{$c_1 = 0.5$}
        \label{figerrC50}
    \end{subfigure}
    \caption{Atomic mix and nonclassical percentage errors with respect to the benchmark solutions for problem set $\seta_3$}
    \label{figerrC1}
\end{figure}

\pagebreak
\begin{figure}[p]
    \centering
    \begin{subfigure}{0.495\textwidth}
        \centering
        \includegraphics[width=\textwidth]{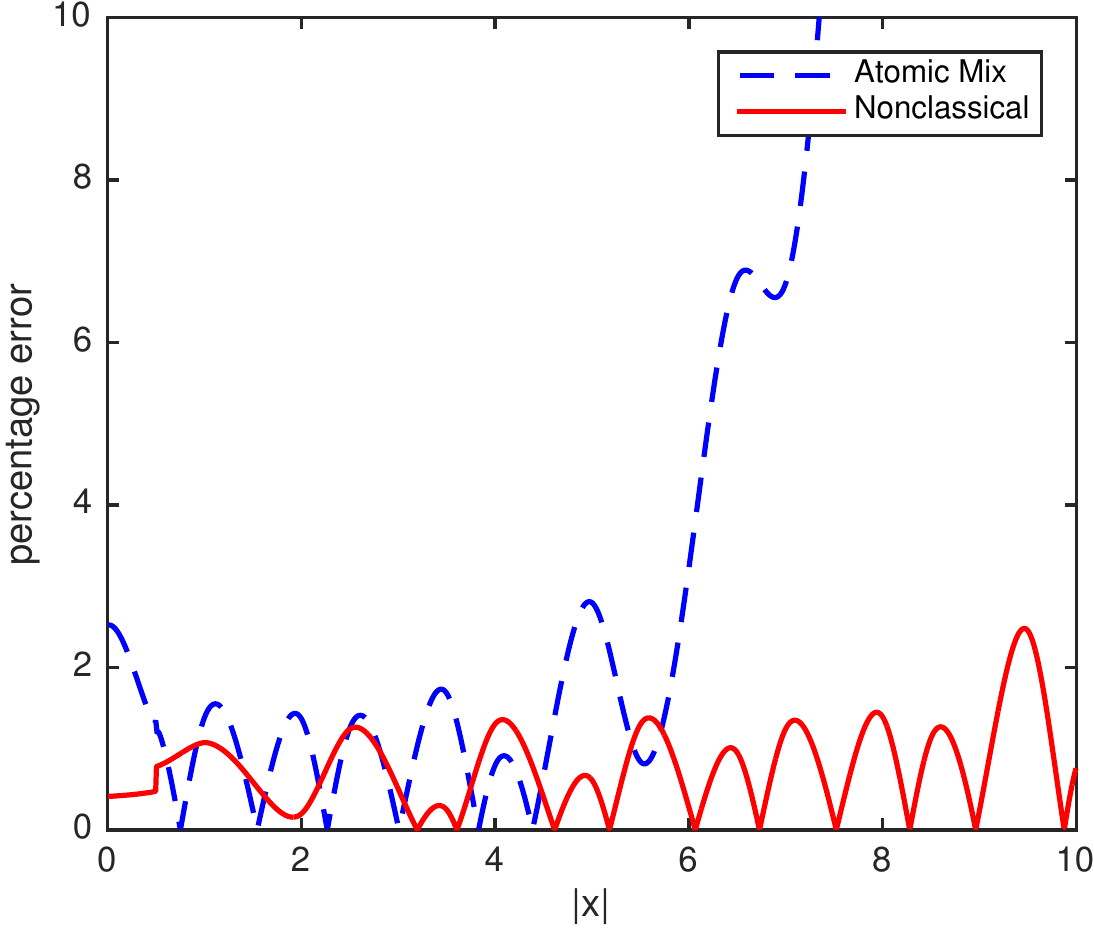}
        \caption{$c_1 = 0.6$}
        \label{figerrC60}
    \end{subfigure}
    \hfill
    \begin{subfigure}{0.495\textwidth}
        \centering
        \includegraphics[width=\textwidth]{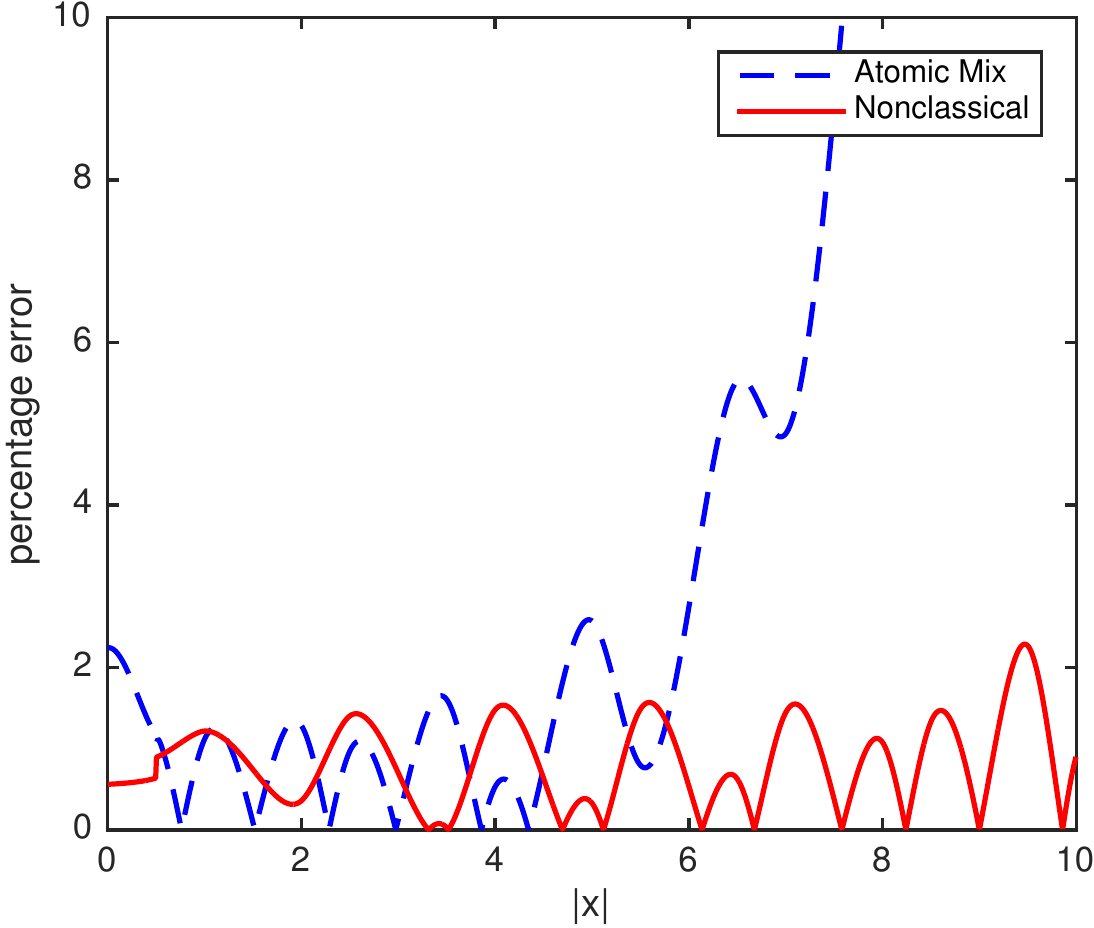}
        \caption{$c_1 = 0.7$}
        \label{figerrC70}
    \end{subfigure}
    \\
    \centering
    \begin{subfigure}{0.495\textwidth}
        \centering
        \includegraphics[width=\textwidth]{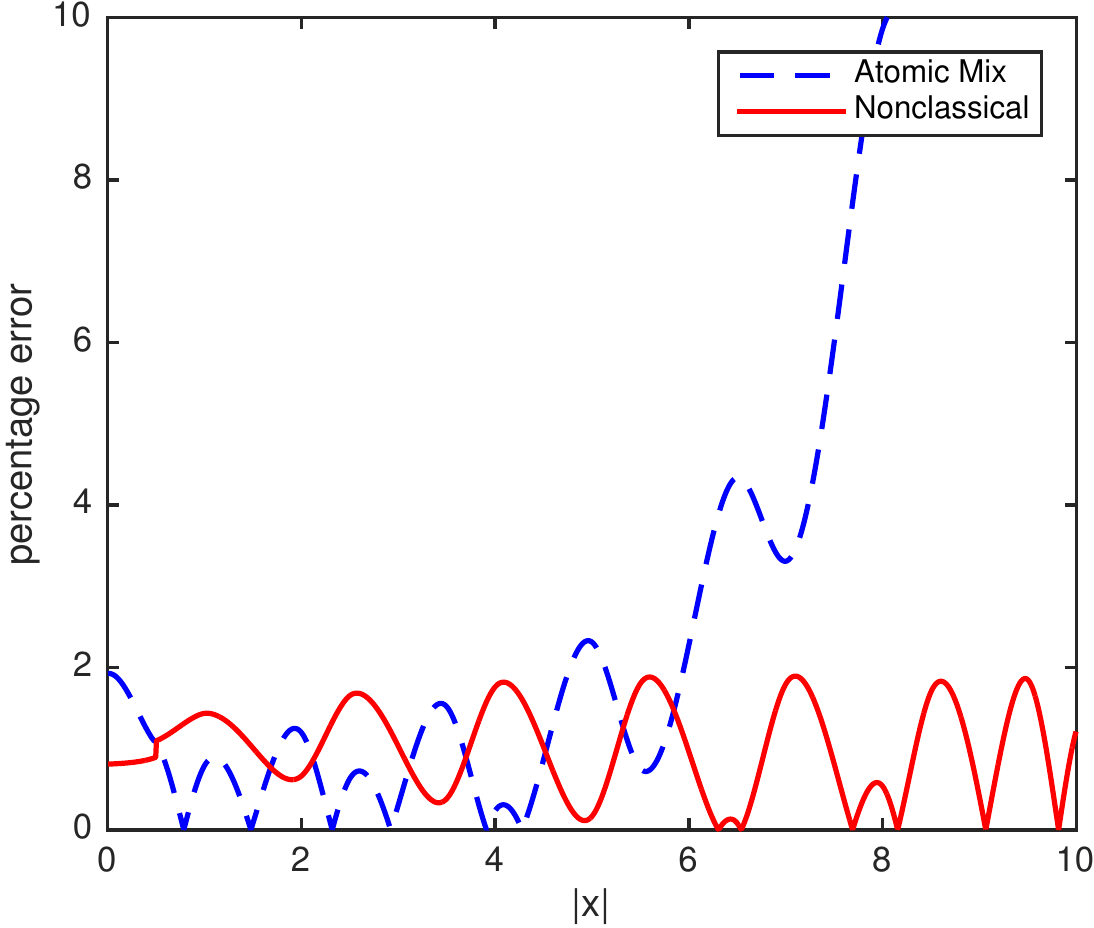}
        \caption{$c_1 = 0.8$}
        \label{figerrC80}
    \end{subfigure}
    \hfill
    \begin{subfigure}{0.495\textwidth}
        \centering
        \includegraphics[width=\textwidth]{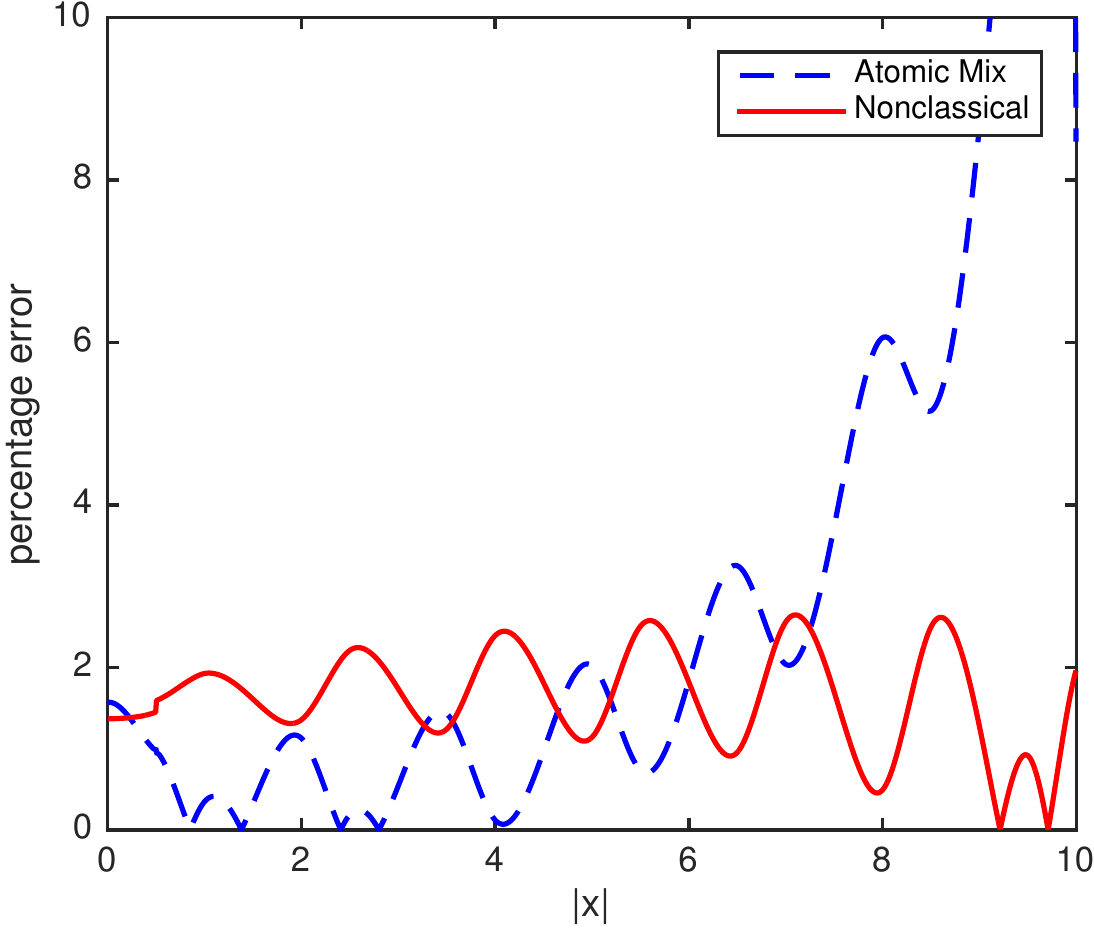}
        \caption{$c_1 = 0.9$}
        \label{figerrC90}
    \end{subfigure}
    \\
    \centering
    \begin{subfigure}{0.495\textwidth}
        \centering
        \includegraphics[width=\textwidth]{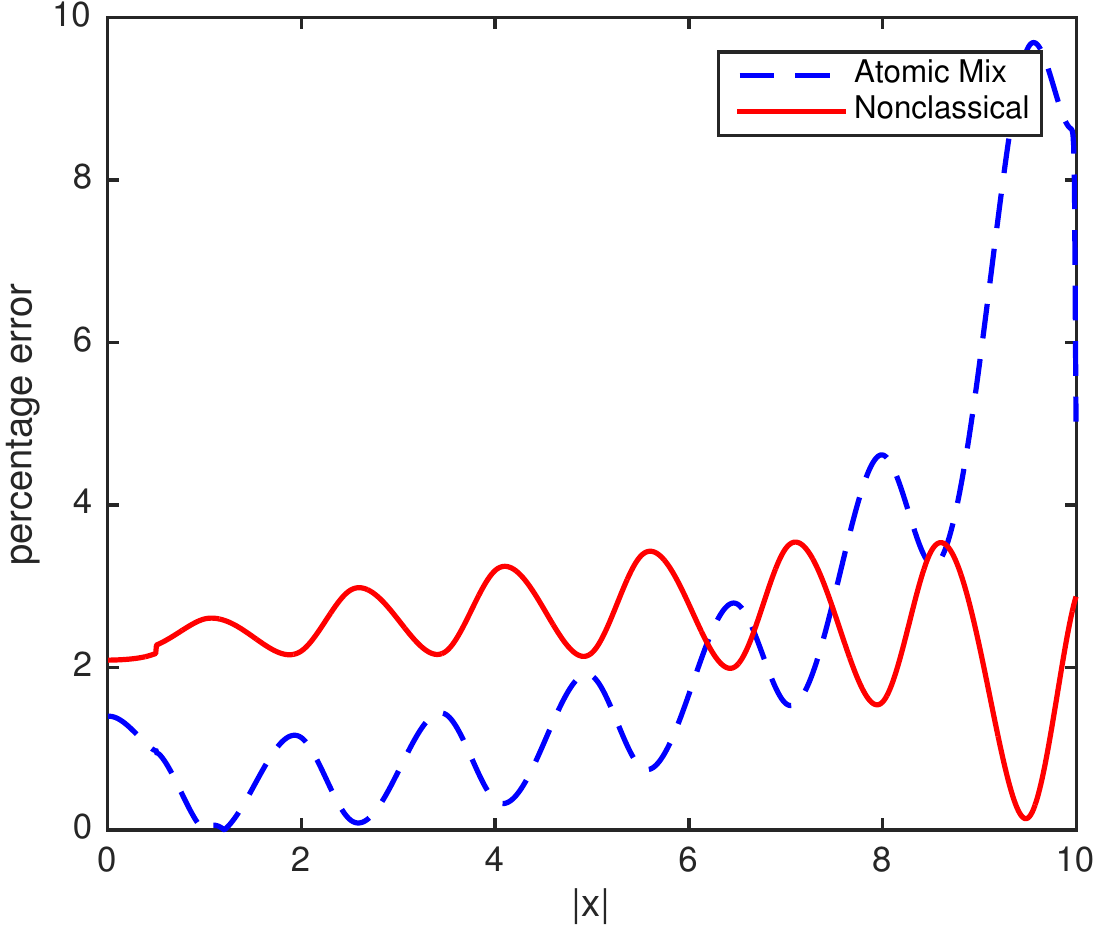}
        \caption{$c_1 = 0.95$}
        \label{figerrC95}
    \end{subfigure}
    \hfill
    \begin{subfigure}{0.495\textwidth}
        \centering
        \includegraphics[width=\textwidth]{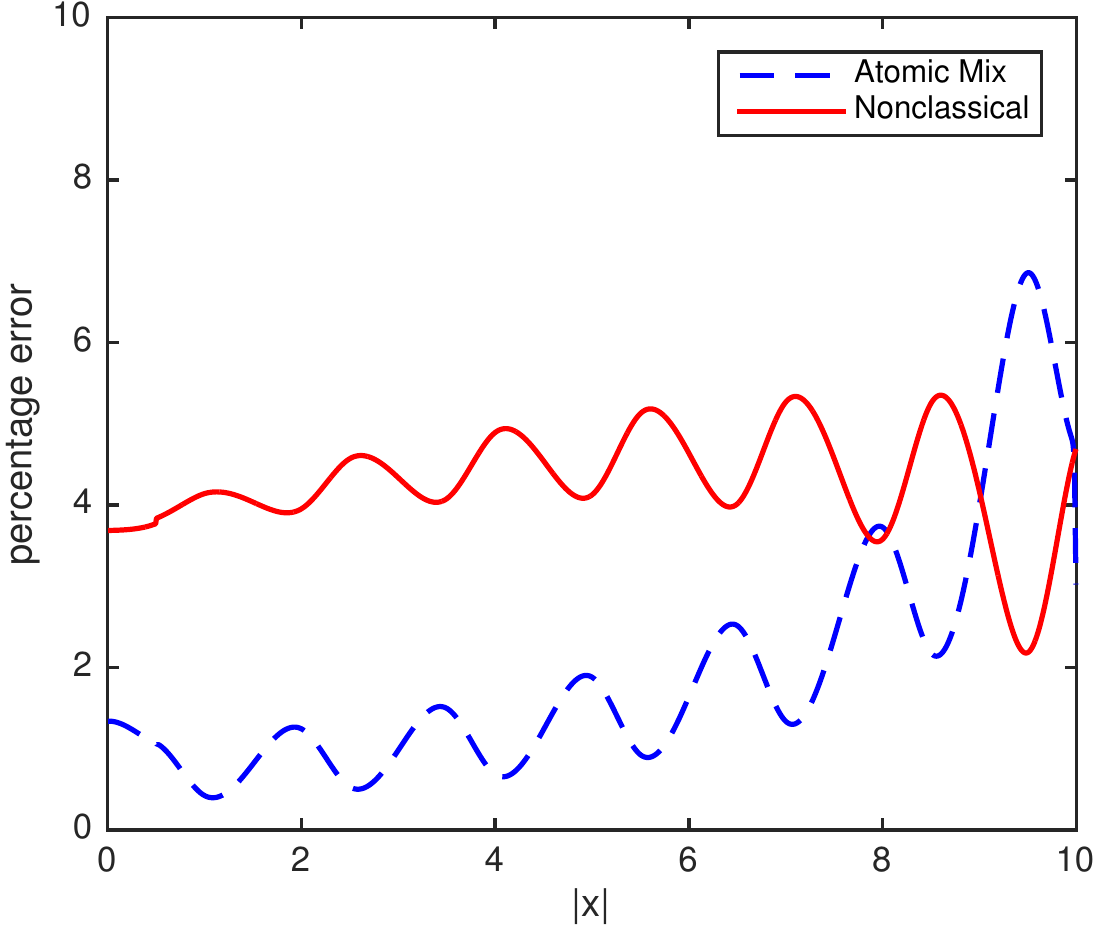}
        \caption{$c_1 = 0.99$}
        \label{figerrC99}
    \end{subfigure}
    \caption{Atomic mix and nonclassical percentage errors with respect to the benchmark solutions for problem set $\seta_3$}
    \label{figerrC2}
\end{figure}

\pagebreak
\setcounter{subfigure}{0}
\begin{figure}[p]
    \centering
    \begin{subsubcaption}
    \begin{subfigure}{0.495\textwidth}
        \centering
        \includegraphics[width=\textwidth]{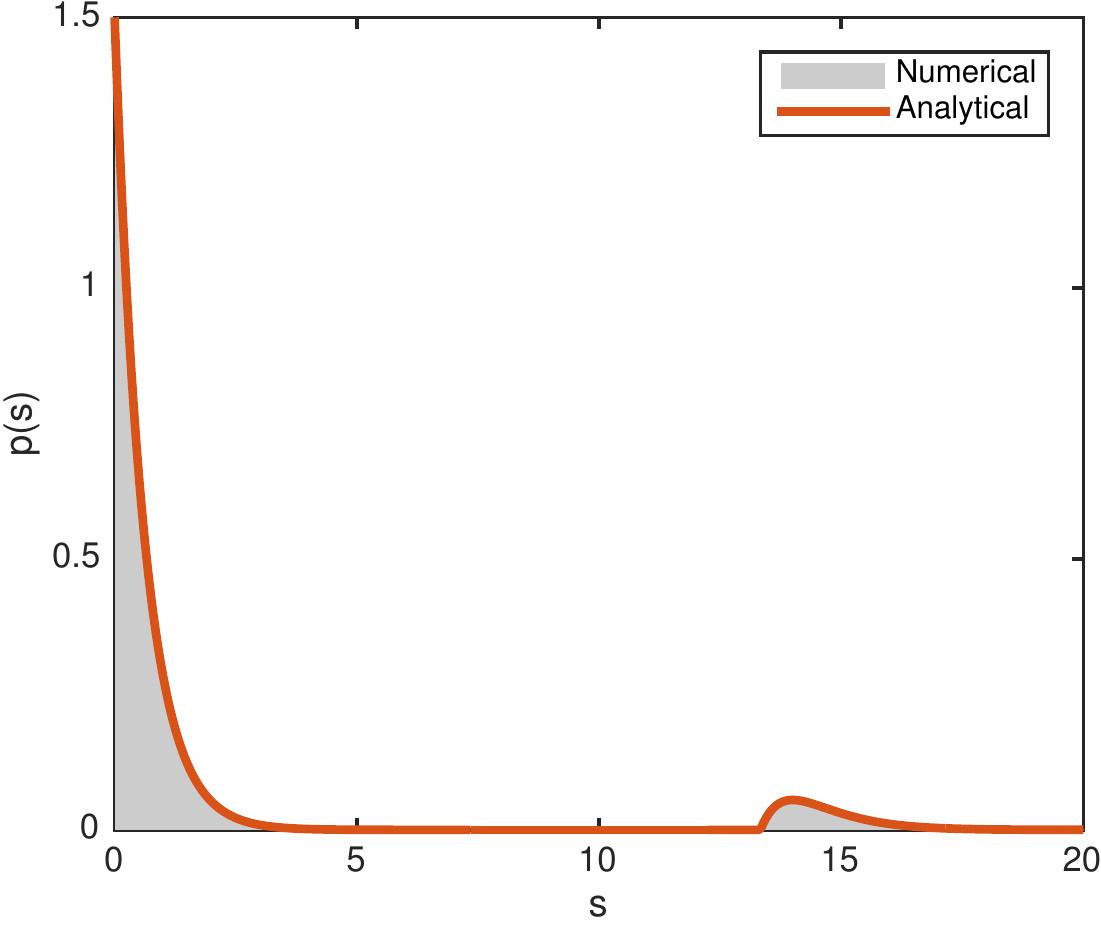}
        \caption{Set $\setb_1$: $\ell_1=20/3$, $\ell_2=40/3$}
        \label{fig14a}
    \end{subfigure}
    \hfill
    \begin{subfigure}{0.495\textwidth}
        \centering
        \includegraphics[width=\textwidth]{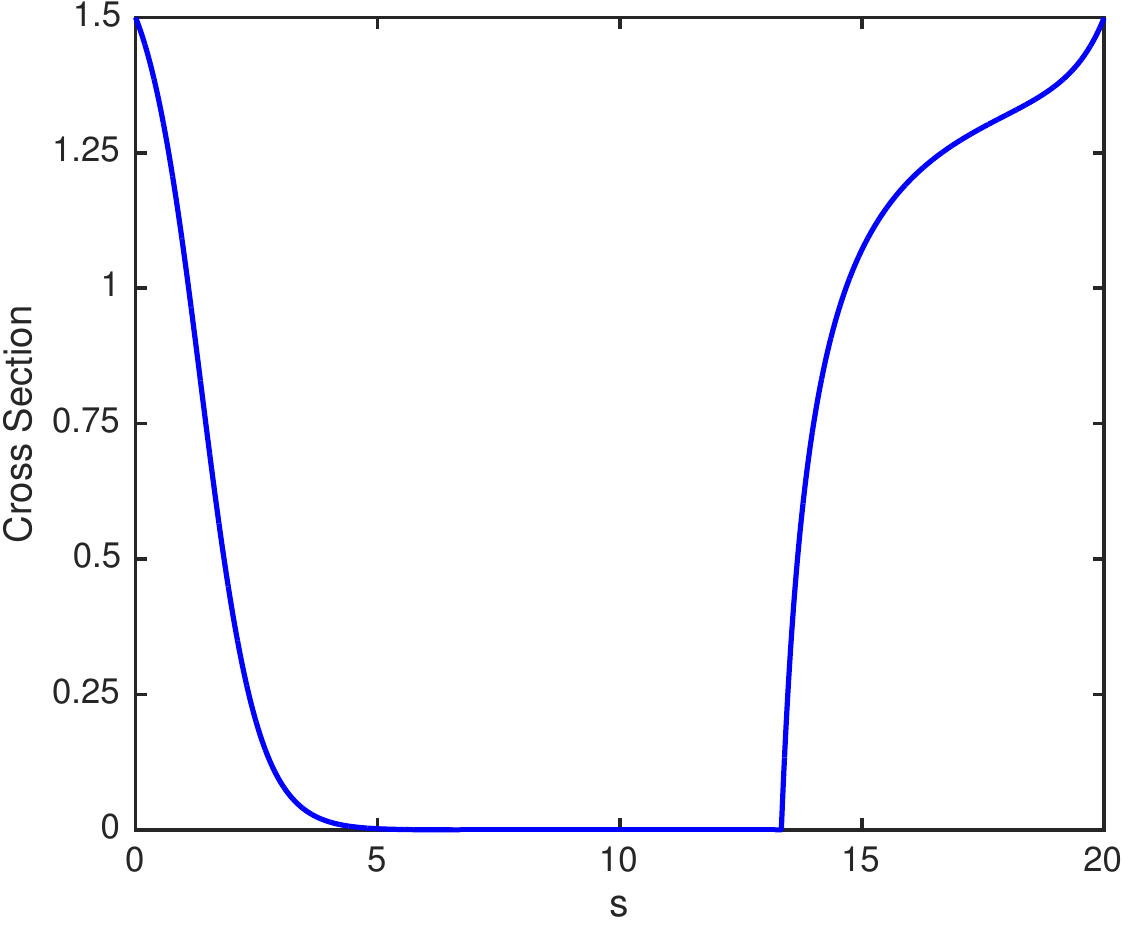}
        \caption{Set $\setb_1$: $\Sigma_{t1} = 1.5$}
        \label{fig14b}
    \end{subfigure}
    \end{subsubcaption}
    \\
    \begin{subsubcaption}
    \centering
    \begin{subfigure}{0.495\textwidth}
        \centering
        \includegraphics[width=\textwidth]{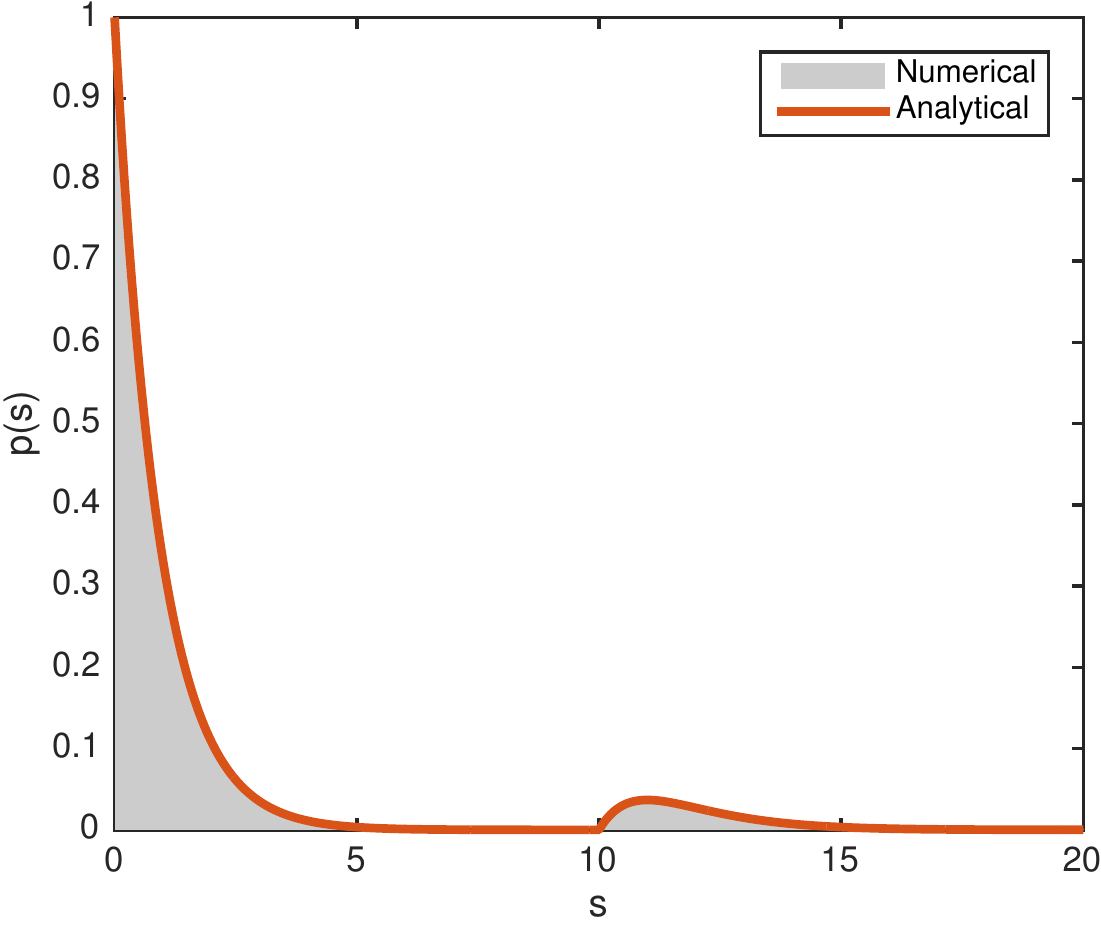}
      \caption{Set $\setb_2$: $\ell_1=10$, $\ell_2=10$}
        \label{fig14c}
    \end{subfigure}
    \hfill
    \begin{subfigure}{0.495\textwidth}
        \centering
        \includegraphics[width=\textwidth]{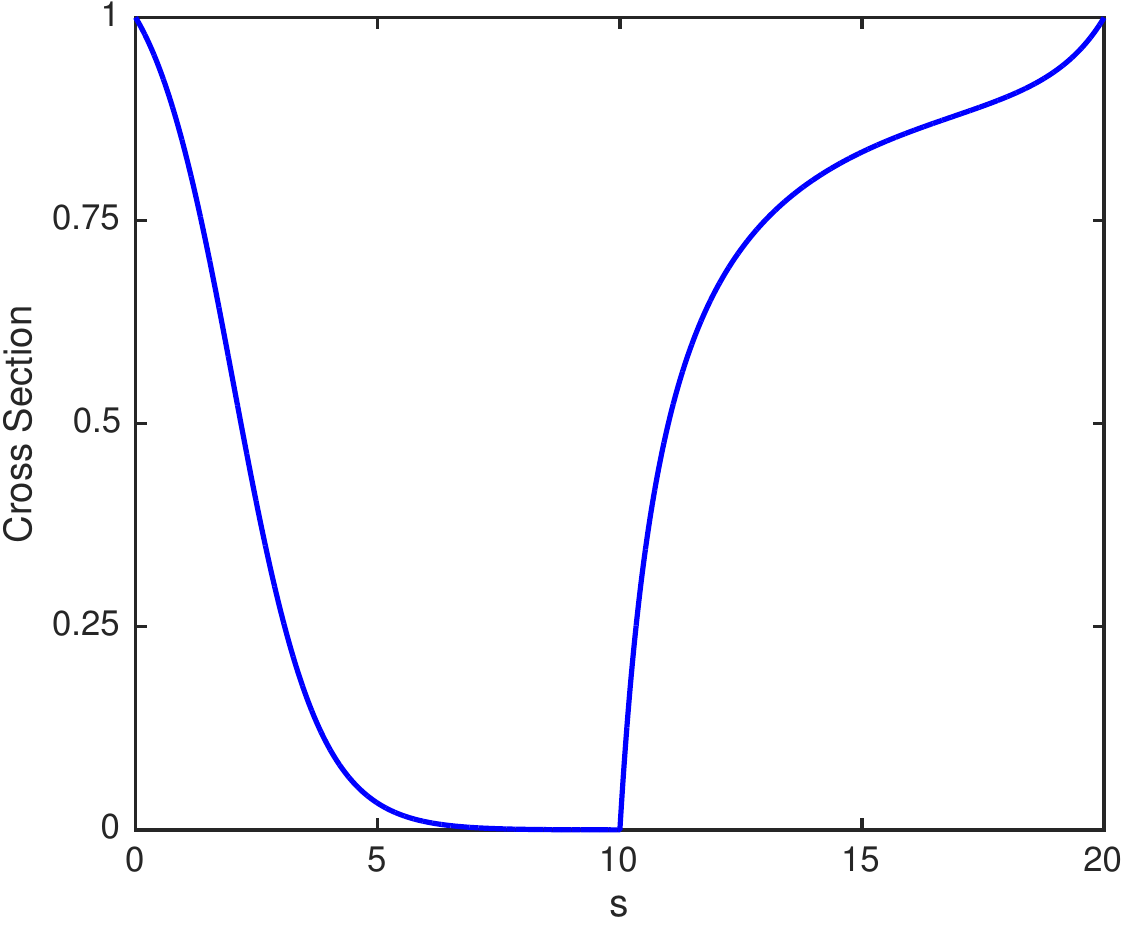}
         \caption{Set $\setb_2$: $\Sigma_{t1} = 1.0$}
       \label{fig14d}
    \end{subfigure}
    \end{subsubcaption}
    \\
    \begin{subsubcaption}
    \centering
    \begin{subfigure}{0.495\textwidth}
        \centering
        \includegraphics[width=\textwidth]{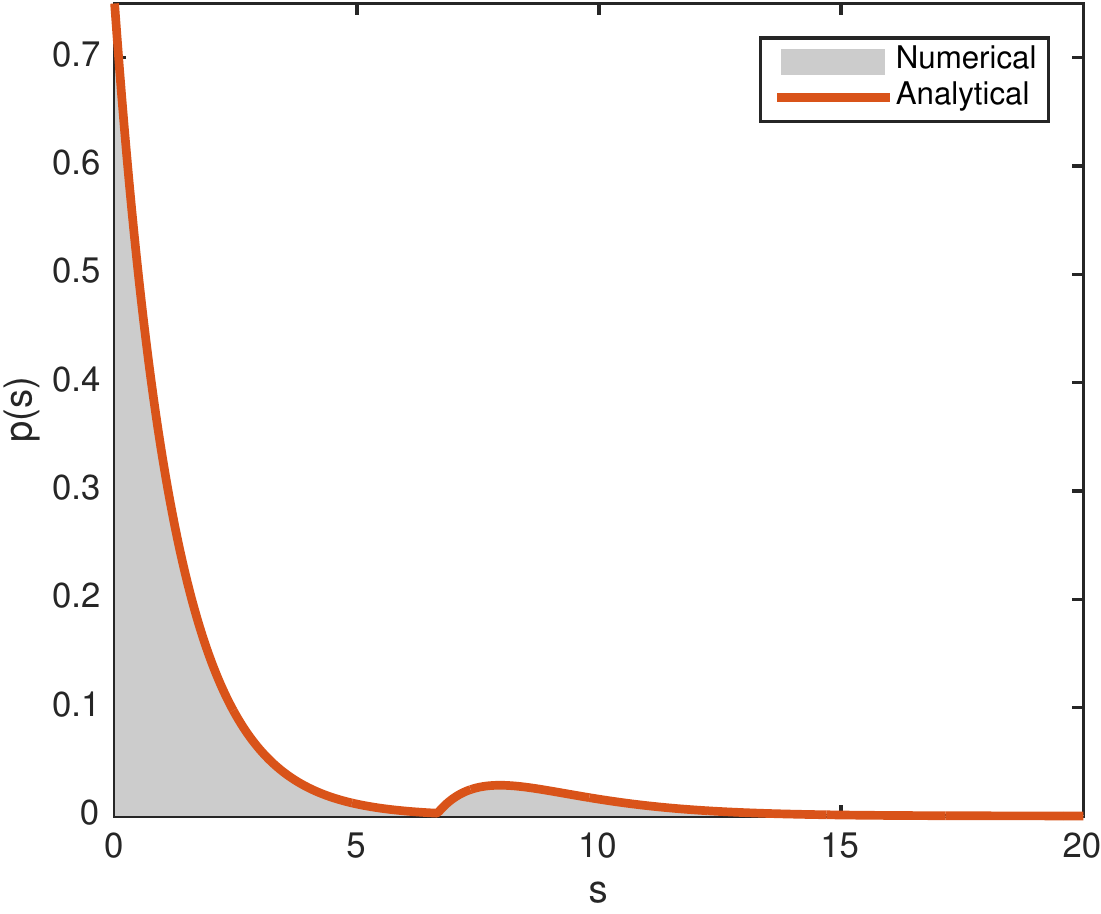}
       \caption{Set $\setb_3$: $\ell_1=40/3$, $\ell_2=20/3$}
         \label{fig14e}
    \end{subfigure}
    \hfill
    \begin{subfigure}{0.495\textwidth}
        \centering
        \includegraphics[width=\textwidth]{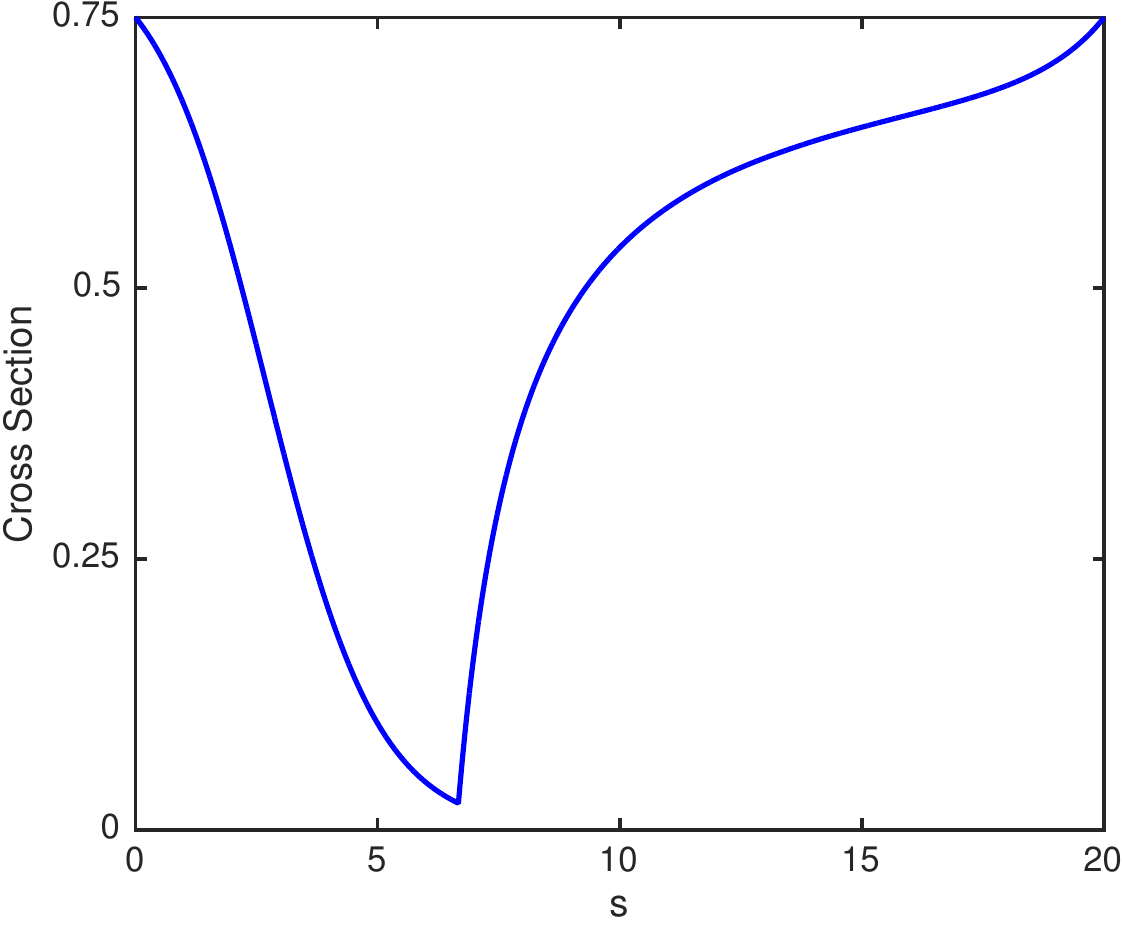}
      \caption{Set $\setb_3$: $\Sigma_{t1} = 0.75$}
          \label{fig14f}
    \end{subfigure}
    \end{subsubcaption}
    \caption{Path-length distribution functions and corresponding nonclassical cross sections for problem set $\setb$ }
    \label{fig14}
\end{figure}

\pagebreak
\setcounter{subfigure}{0}
\begin{figure}[p]
    \centering
    \begin{subsubcaption}
    \begin{subfigure}{0.495\textwidth}
        \centering
        \includegraphics[width=\textwidth]{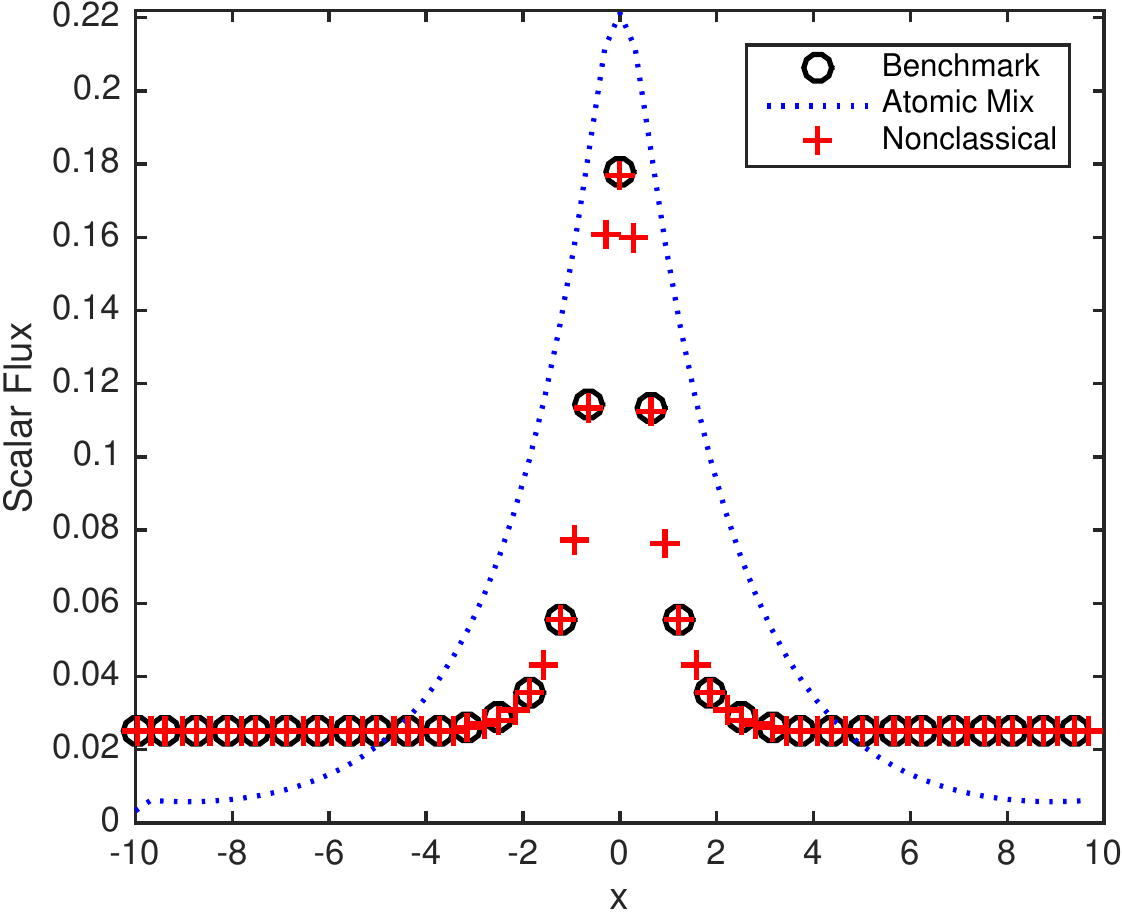}
        \caption{Problem set $\setb_1$ with $c_1=00$}
        \label{fig15a}
    \end{subfigure}
    \hfill
    \begin{subfigure}{0.495\textwidth}
        \centering
        \includegraphics[width=\textwidth]{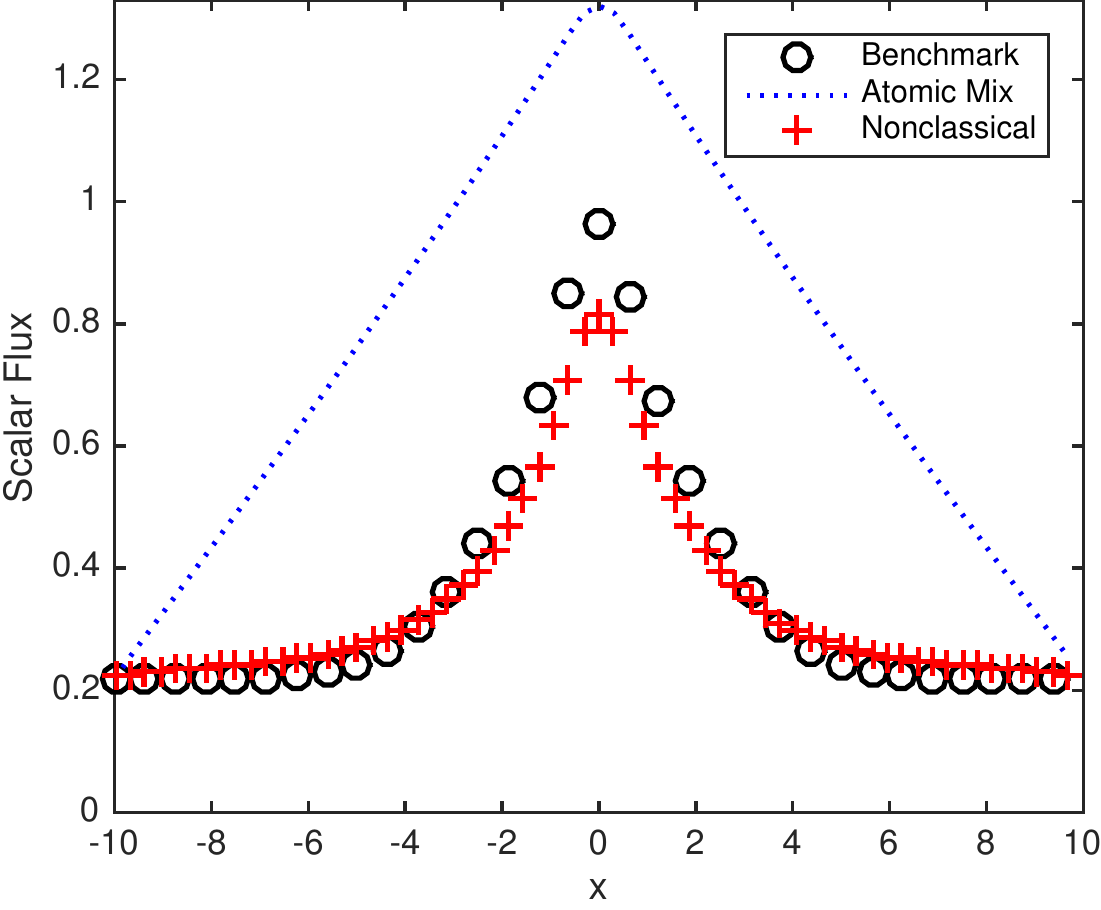}
        \caption{Problem set $\setb_1$ with $c_1=0.99$}
        \label{fig15b}
    \end{subfigure}
        \end{subsubcaption}
    \\
    \begin{subsubcaption}
    \centering
    \begin{subfigure}{0.495\textwidth}
        \includegraphics[width=\textwidth]{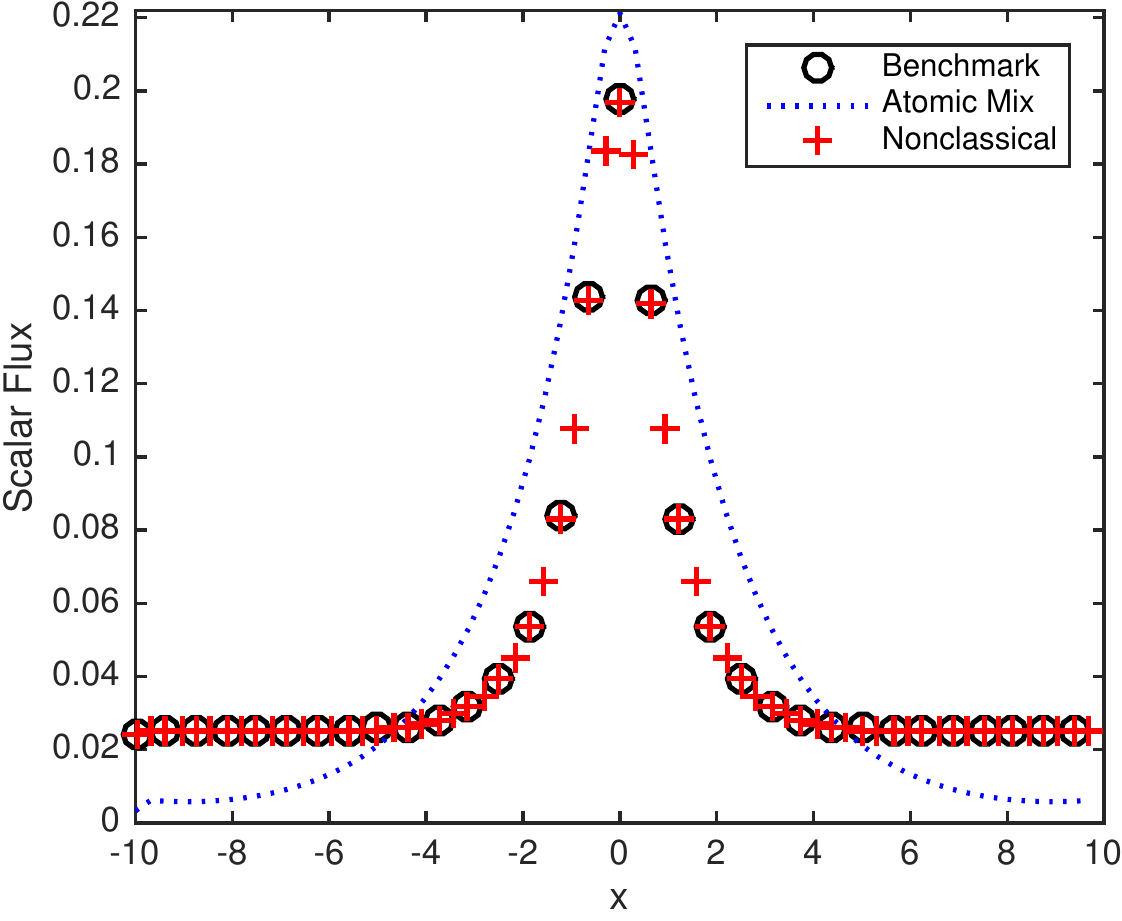}
        \caption{Problem set $\setb_2$ with $c_1=00$}
        \label{fig15c}
    \end{subfigure}
    \hfill
    \begin{subfigure}{0.495\textwidth}
        \centering
        \includegraphics[width=\textwidth]{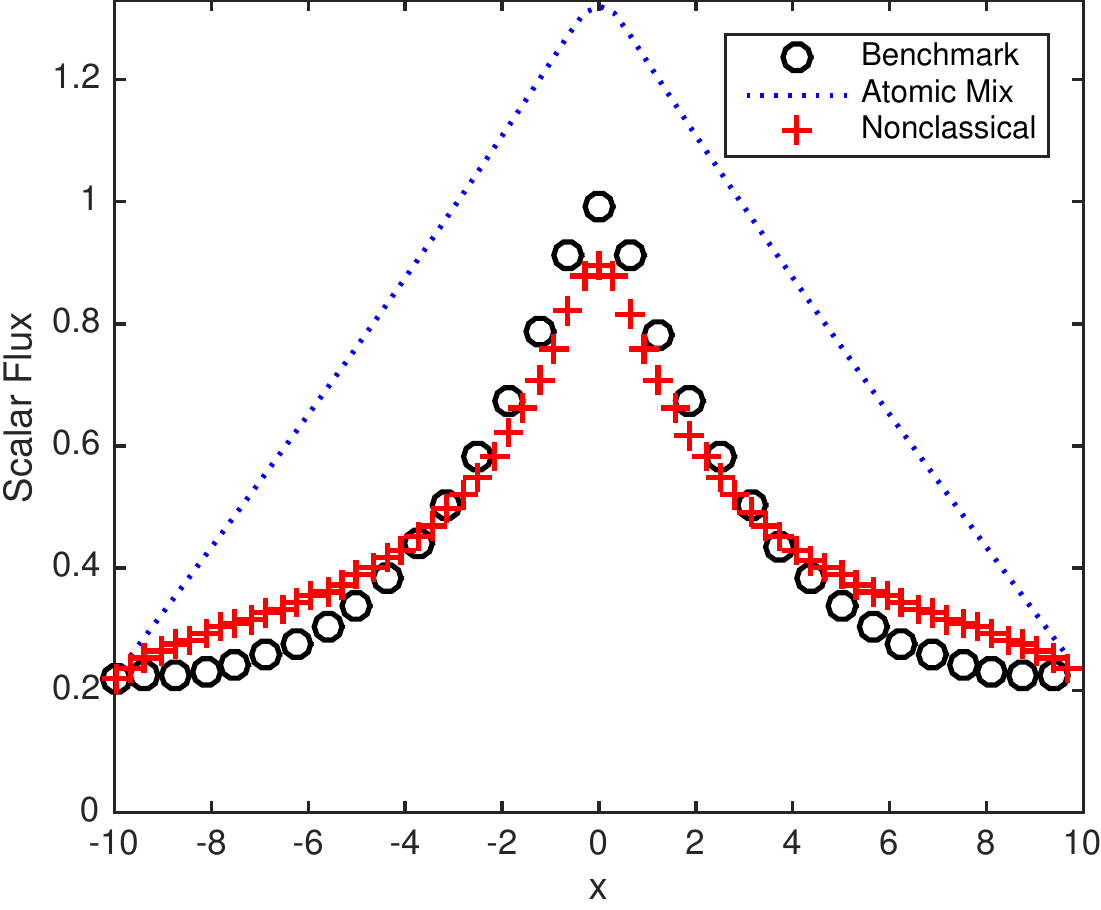}
        \caption{Problem set $\setb_2$ with $c_1=0.99$}
        \label{fig15d}
    \end{subfigure}
        \end{subsubcaption}
    \\
    \begin{subsubcaption}
    \centering
    \begin{subfigure}{0.495\textwidth}
        \includegraphics[width=\textwidth]{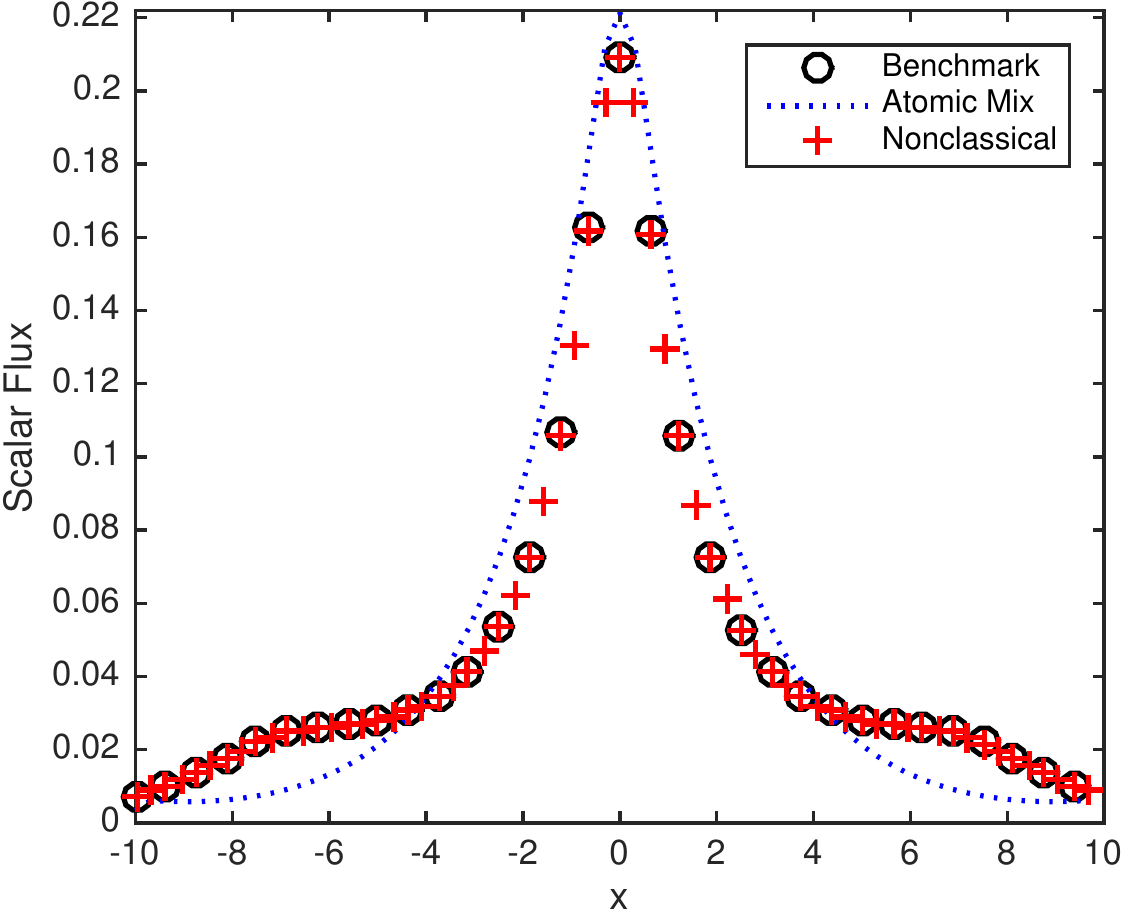}
        \caption{Problem set $\setb_3$ with $c_1=00$}
        \label{fig15e}
    \end{subfigure}
    \hfill
    \begin{subfigure}{0.495\textwidth}
        \centering
        \includegraphics[width=\textwidth]{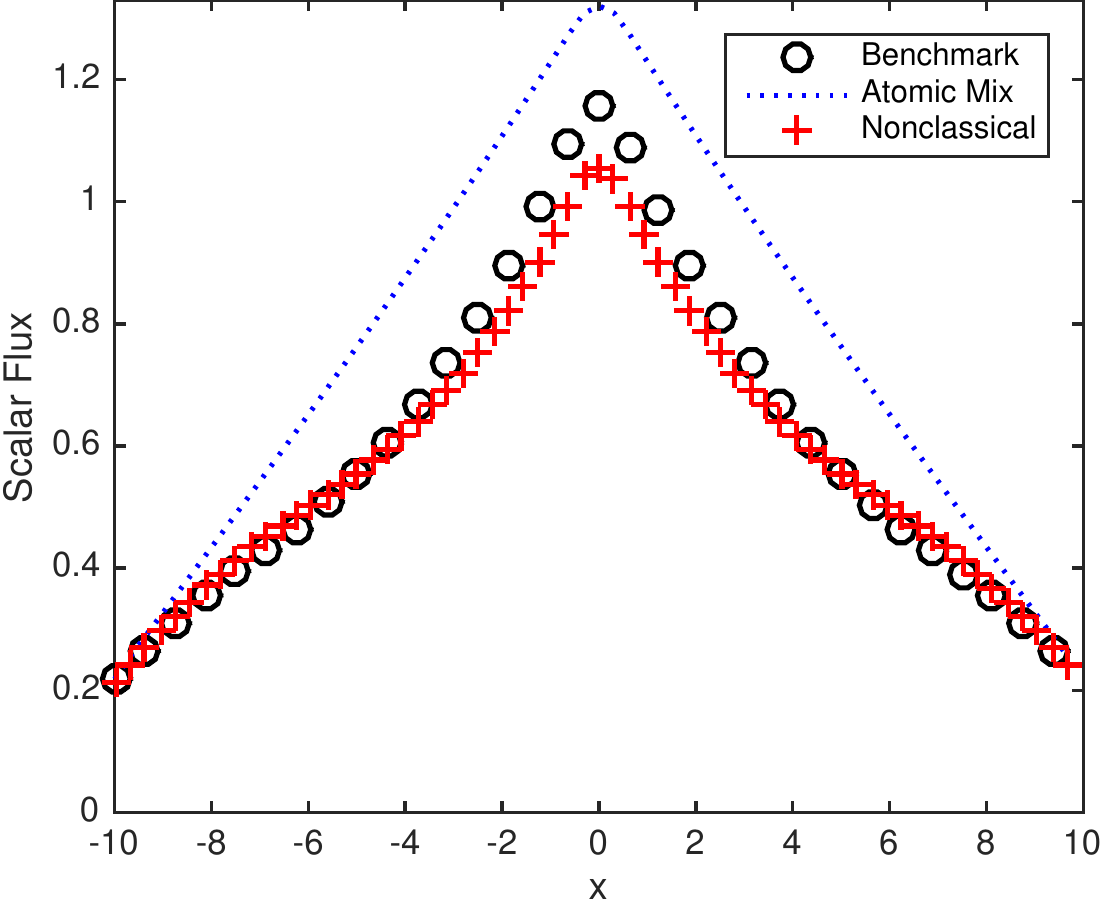}
        \caption{Problem set $\setb_3$ with $c_1=0.99$}
        \label{fig15f}
    \end{subfigure}
    \end{subsubcaption}
    \\
        \caption{Ensemble-averaged scalar fluxes for problem set $\setb$}
    \label{fig15}
\end{figure}

\pagebreak
\begin{figure}[p]
    \centering
    \begin{subfigure}{0.495\textwidth}
        \centering
        \includegraphics[width=\textwidth]{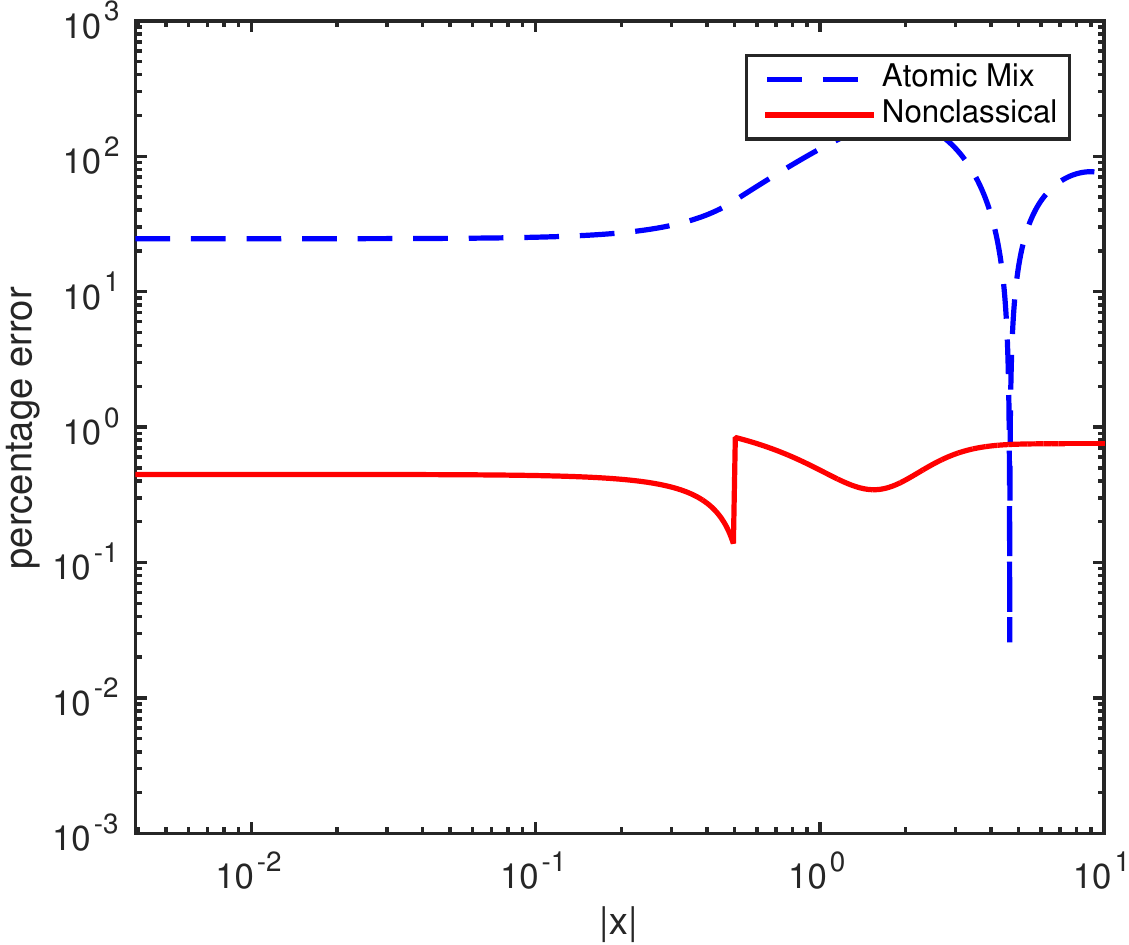}
        \caption{$c_1 = 0.0$}
        \label{figerrD00}
    \end{subfigure}
    \hfill
    \begin{subfigure}{0.495\textwidth}
        \centering
        \includegraphics[width=\textwidth]{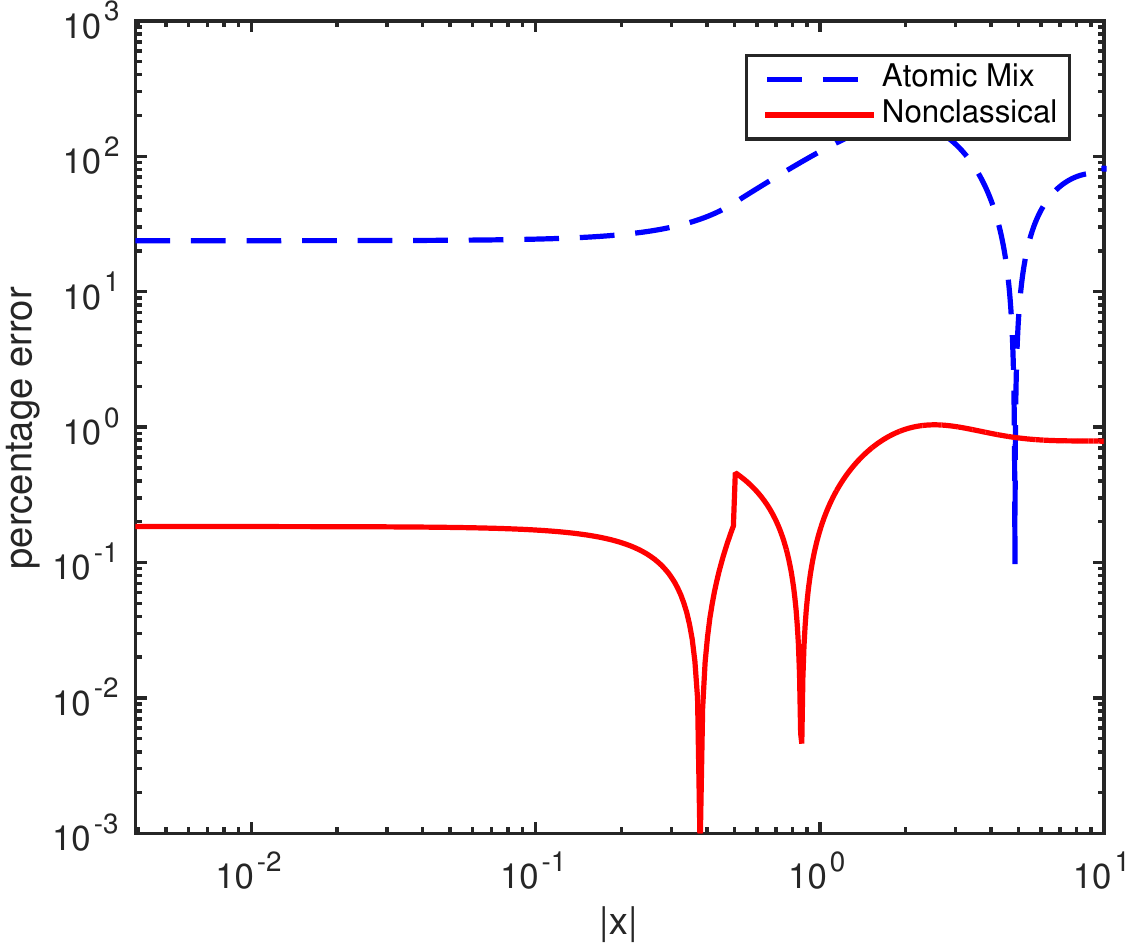}
        \caption{$c_1 = 0.1$}
        \label{figerrD10}
    \end{subfigure}
    \\
    \centering
    \begin{subfigure}{0.495\textwidth}
        \centering
        \includegraphics[width=\textwidth]{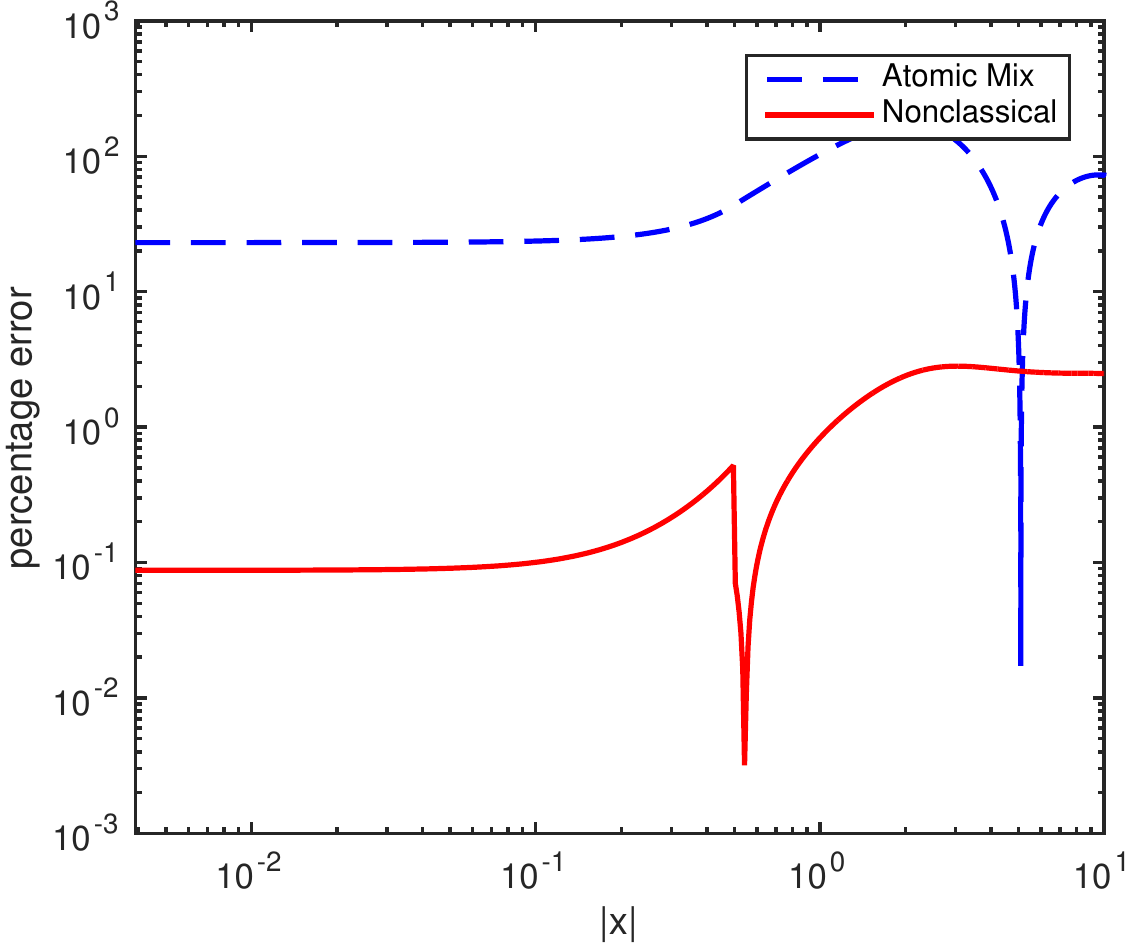}
        \caption{$c_1 = 0.2$}
        \label{figerrD20}
    \end{subfigure}
    \hfill
    \begin{subfigure}{0.495\textwidth}
        \centering
        \includegraphics[width=\textwidth]{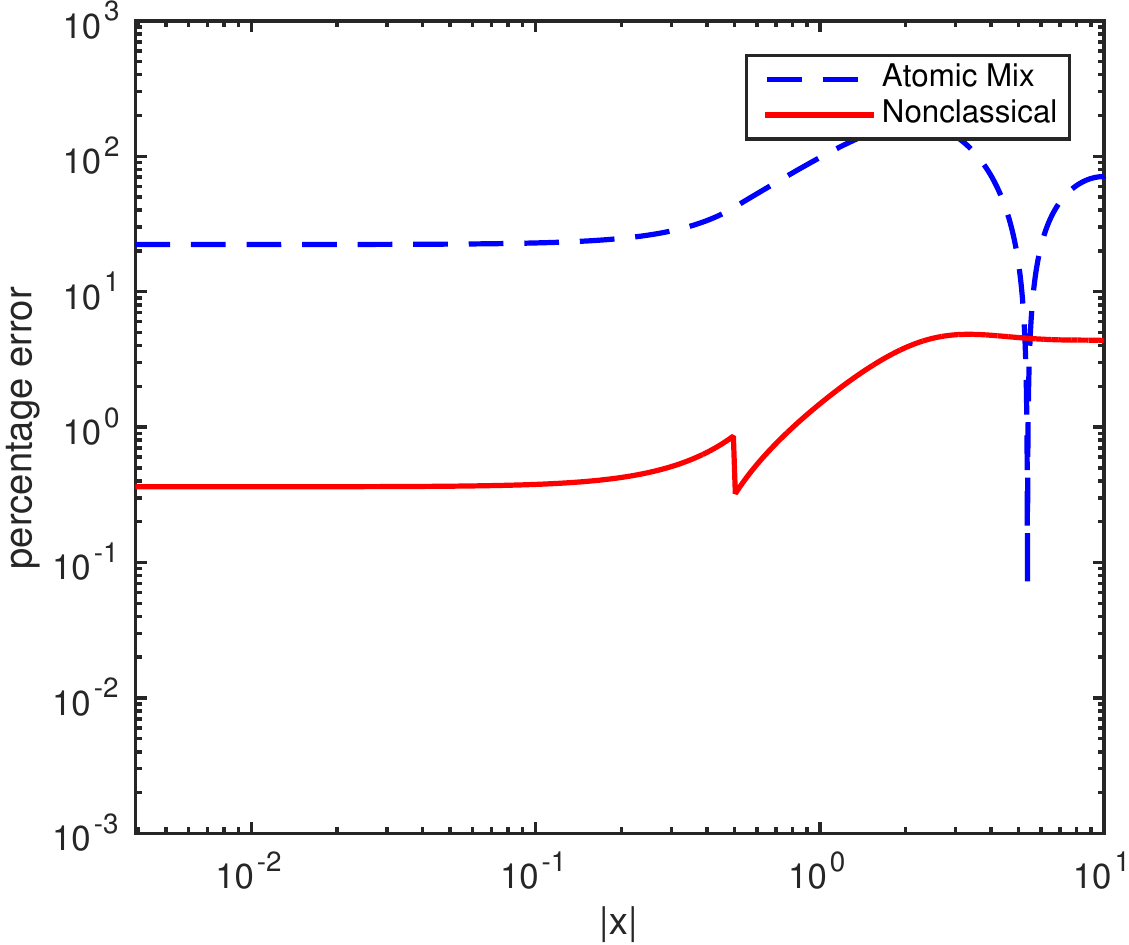}
        \caption{$c_1 = 0.3$}
        \label{figerrD30}
    \end{subfigure}
    \\
    \centering
    \begin{subfigure}{0.495\textwidth}
        \centering
        \includegraphics[width=\textwidth]{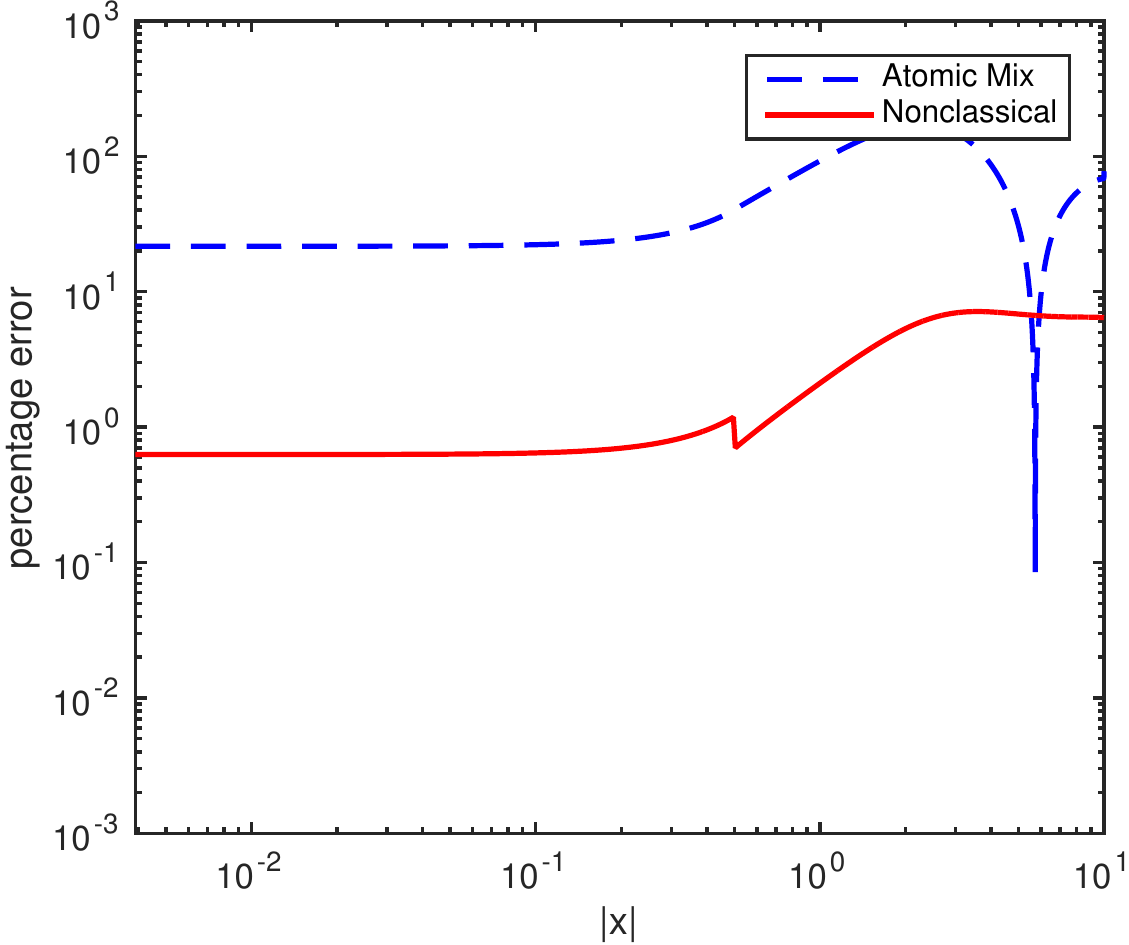}
        \caption{$c_1 = 0.4$}
        \label{figerrD40}
    \end{subfigure}
    \hfill
    \begin{subfigure}{0.495\textwidth}
        \centering
        \includegraphics[width=\textwidth]{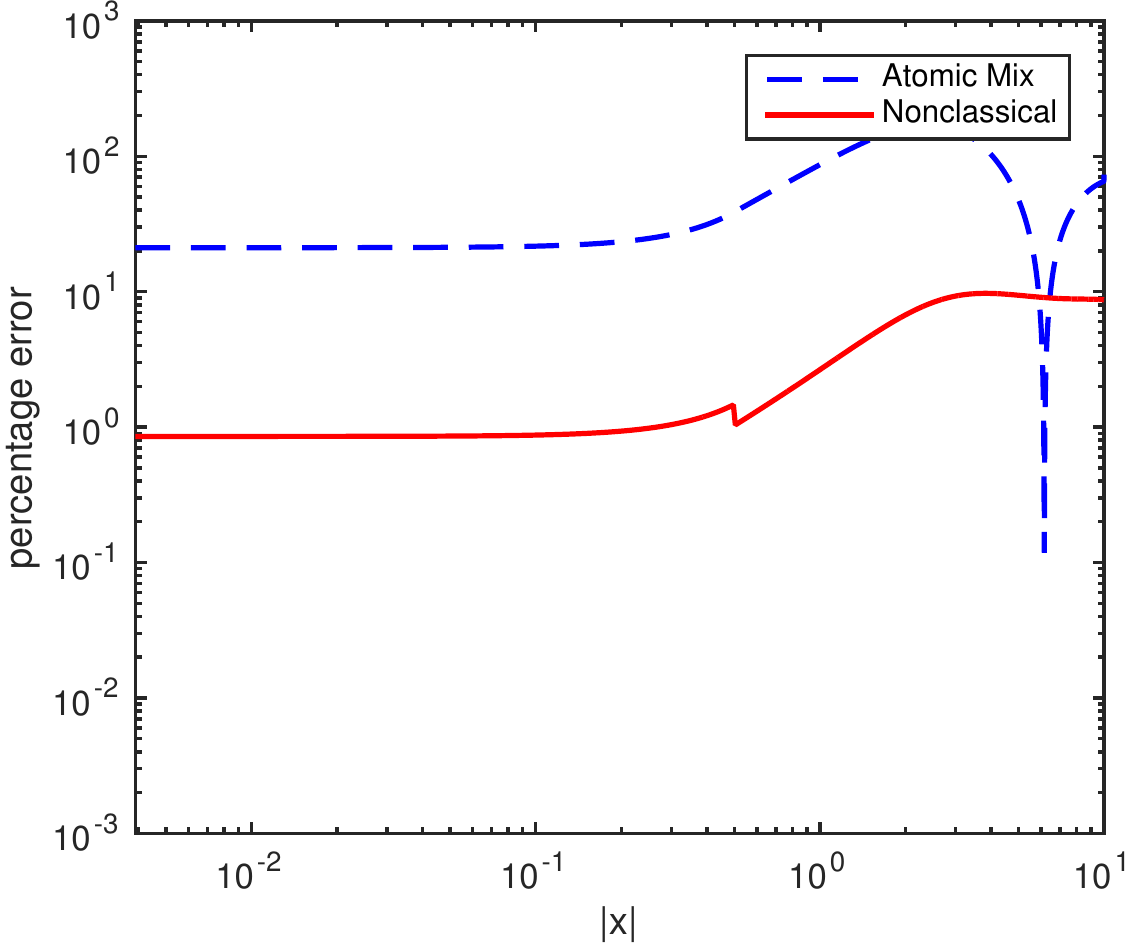}
        \caption{$c_1 = 0.5$}
        \label{figerrD50}
    \end{subfigure}
    \caption{Atomic mix and nonclassical percentage errors with respect to the benchmark solutions for problem set $\setb_1$ (log scale)}
    \label{figerrD1}
\end{figure}

\pagebreak
\begin{figure}[p]
    \centering
    \begin{subfigure}{0.495\textwidth}
        \centering
        \includegraphics[width=\textwidth]{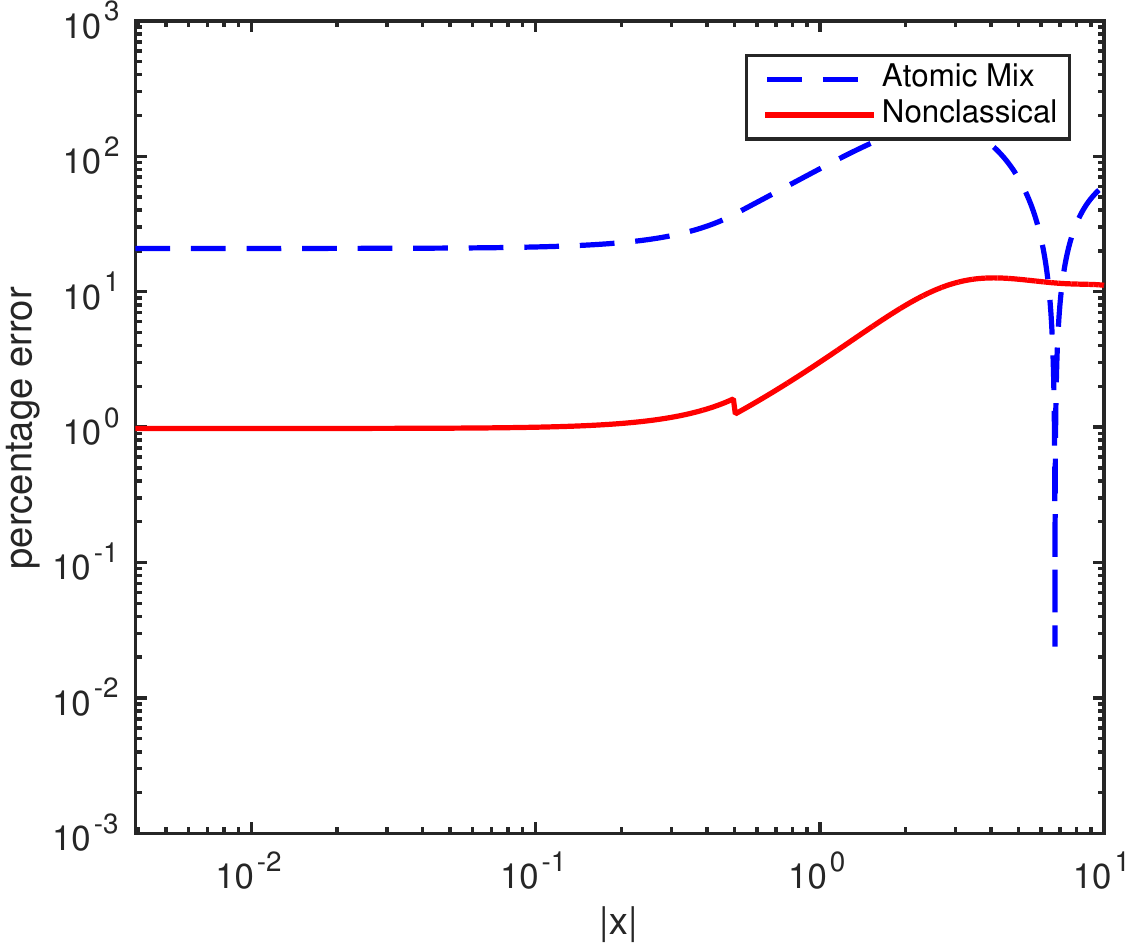}
        \caption{$c_1 = 0.6$}
        \label{figerrD60}
    \end{subfigure}
    \hfill
    \begin{subfigure}{0.495\textwidth}
        \centering
        \includegraphics[width=\textwidth]{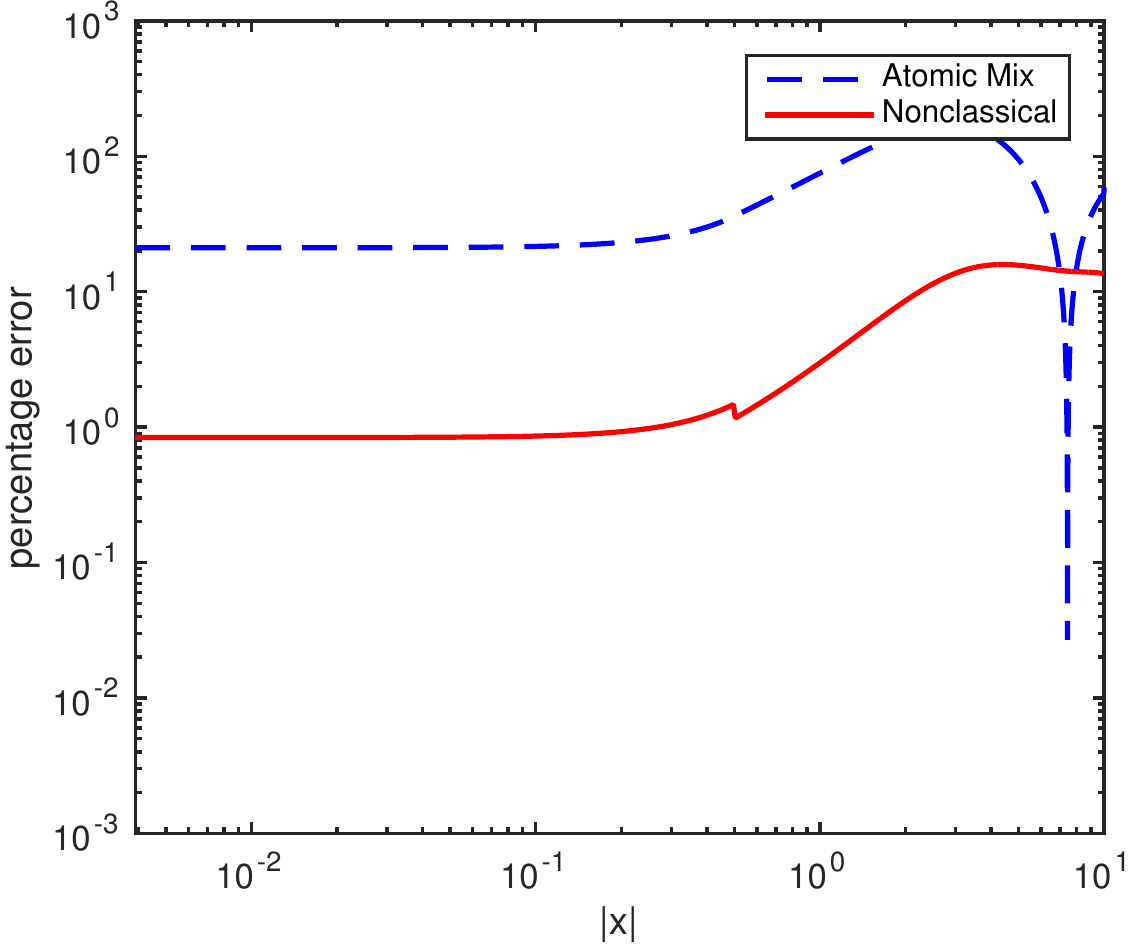}
        \caption{$c_1 = 0.7$}
        \label{figerrD70}
    \end{subfigure}
    \\
    \centering
    \begin{subfigure}{0.495\textwidth}
        \centering
        \includegraphics[width=\textwidth]{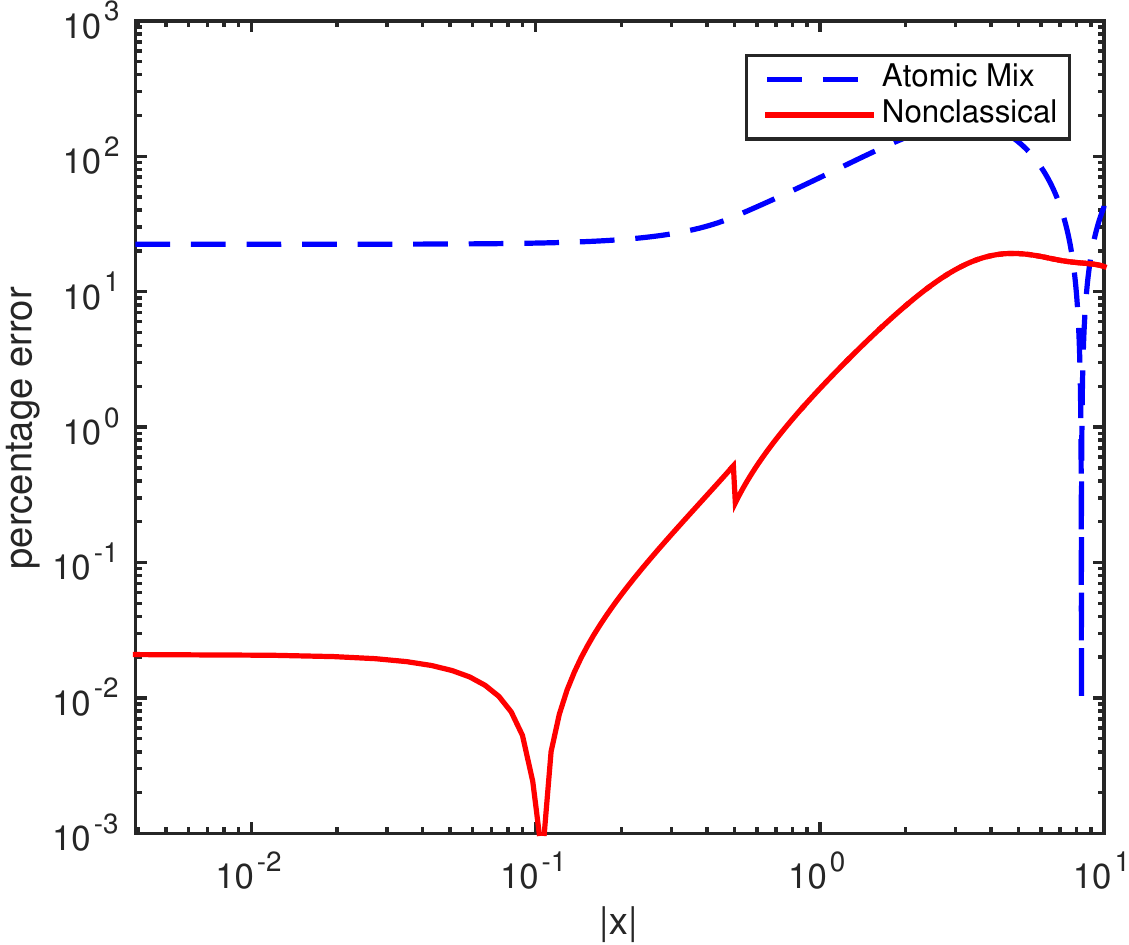}
        \caption{$c_1 = 0.8$}
        \label{figerrD80}
    \end{subfigure}
    \hfill
    \begin{subfigure}{0.495\textwidth}
        \centering
        \includegraphics[width=\textwidth]{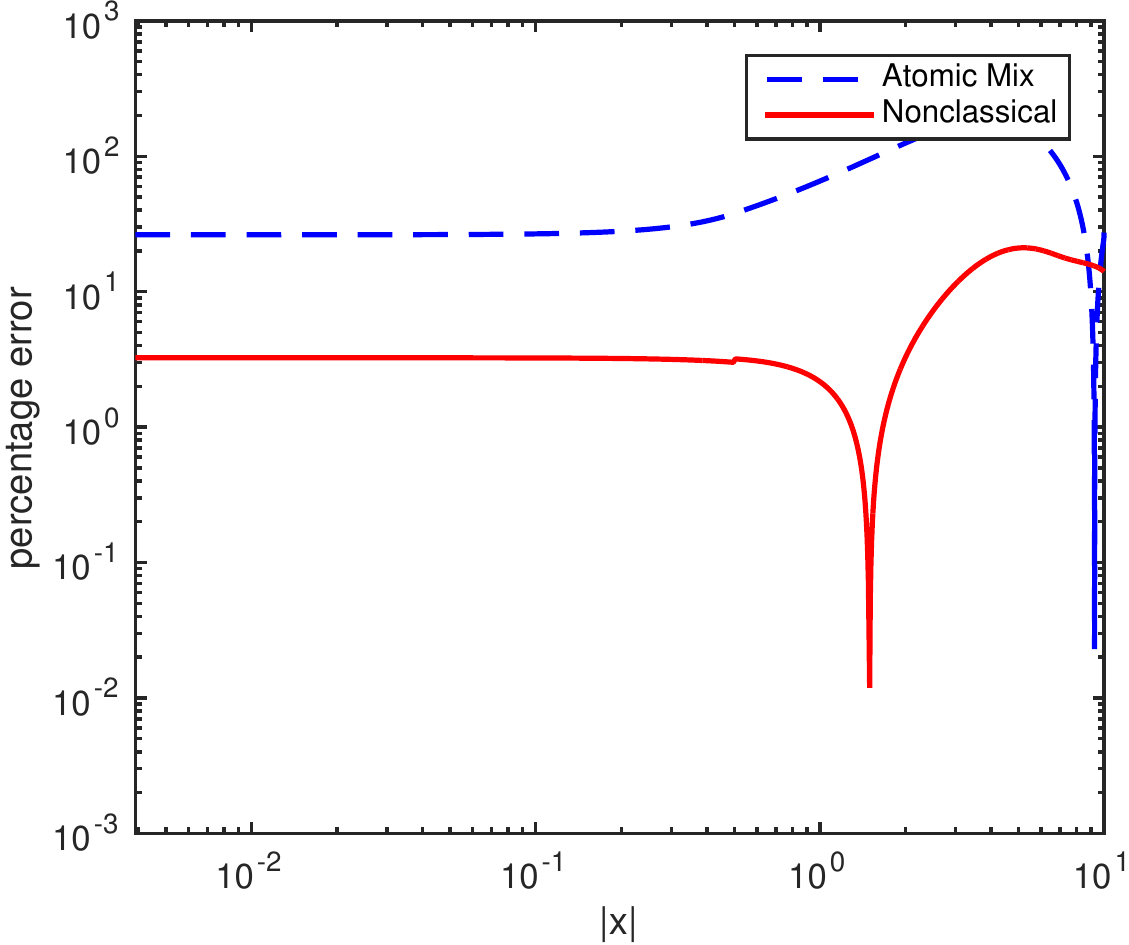}
        \caption{$c_1 = 0.9$}
        \label{figerrD90}
    \end{subfigure}
    \\
    \centering
    \begin{subfigure}{0.495\textwidth}
        \centering
        \includegraphics[width=\textwidth]{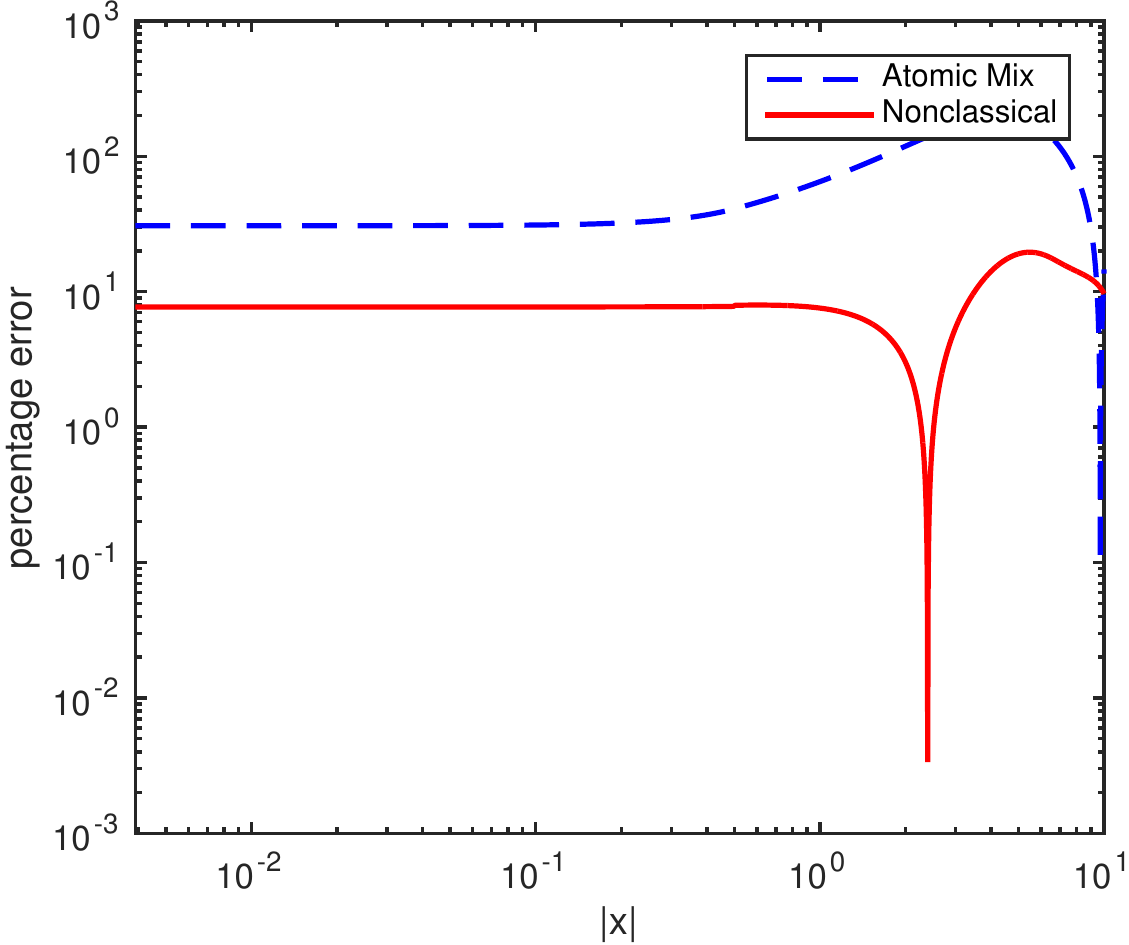}
        \caption{$c_1 = 0.95$}
        \label{figerrD95}
    \end{subfigure}
    \hfill
    \begin{subfigure}{0.495\textwidth}
        \centering
        \includegraphics[width=\textwidth]{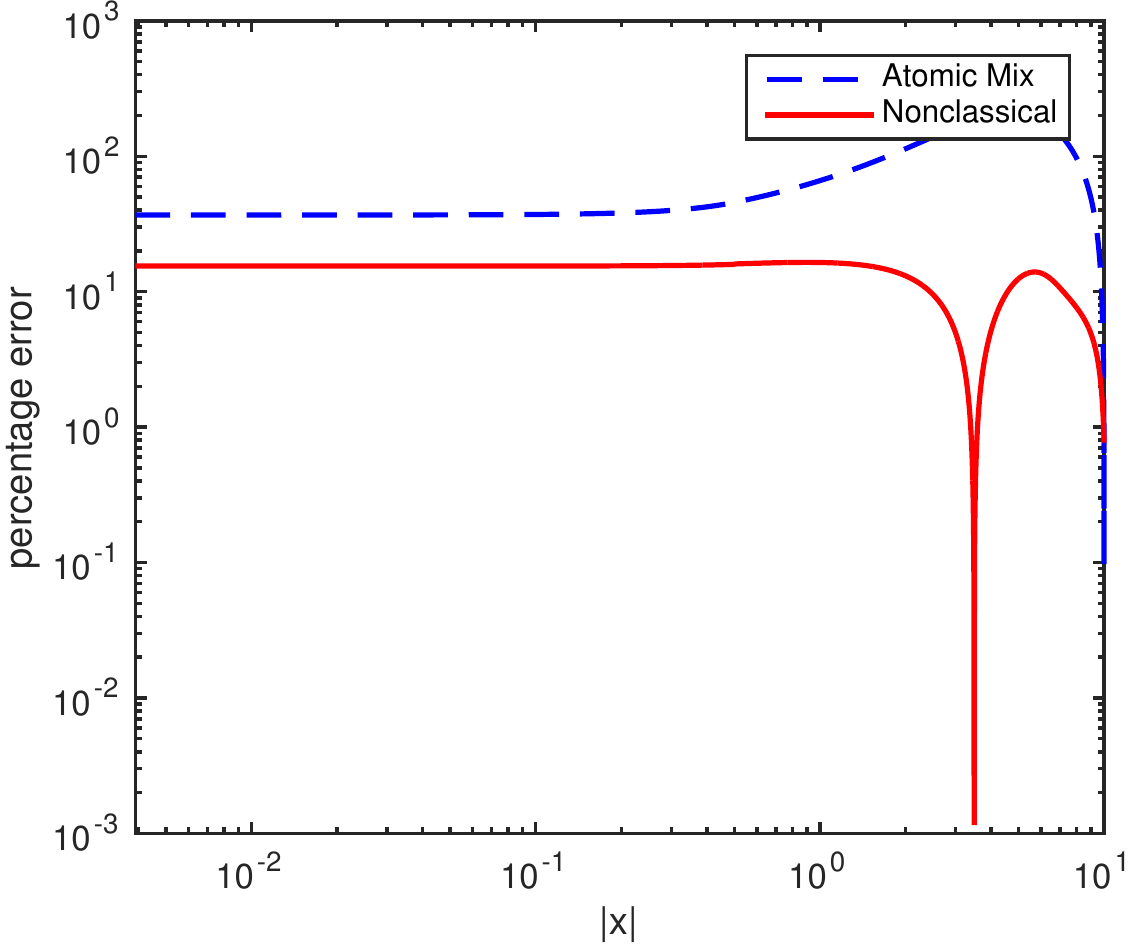}
        \caption{$c_1 = 0.99$}
        \label{figerrD99}
    \end{subfigure}
    \caption{Atomic mix and nonclassical percentage errors with respect to the benchmark solutions for problem set $\setb_1$ (log scale)}
    \label{figerrD2}
\end{figure}

\pagebreak
\begin{figure}[p]
    \centering
    \begin{subfigure}{0.495\textwidth}
        \centering
        \includegraphics[width=\textwidth]{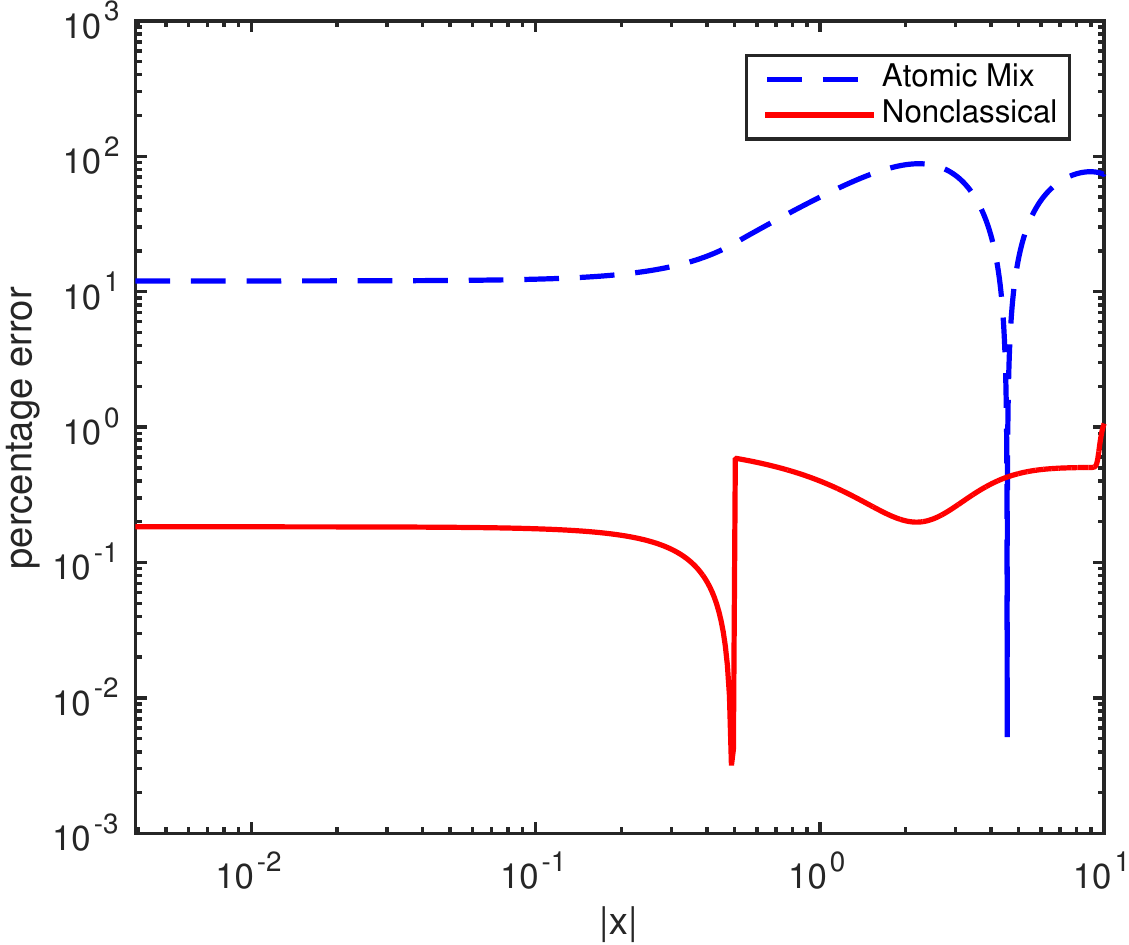}
        \caption{$c_1 = 0.0$}
        \label{figerrE00}
    \end{subfigure}
    \hfill
    \begin{subfigure}{0.495\textwidth}
        \centering
        \includegraphics[width=\textwidth]{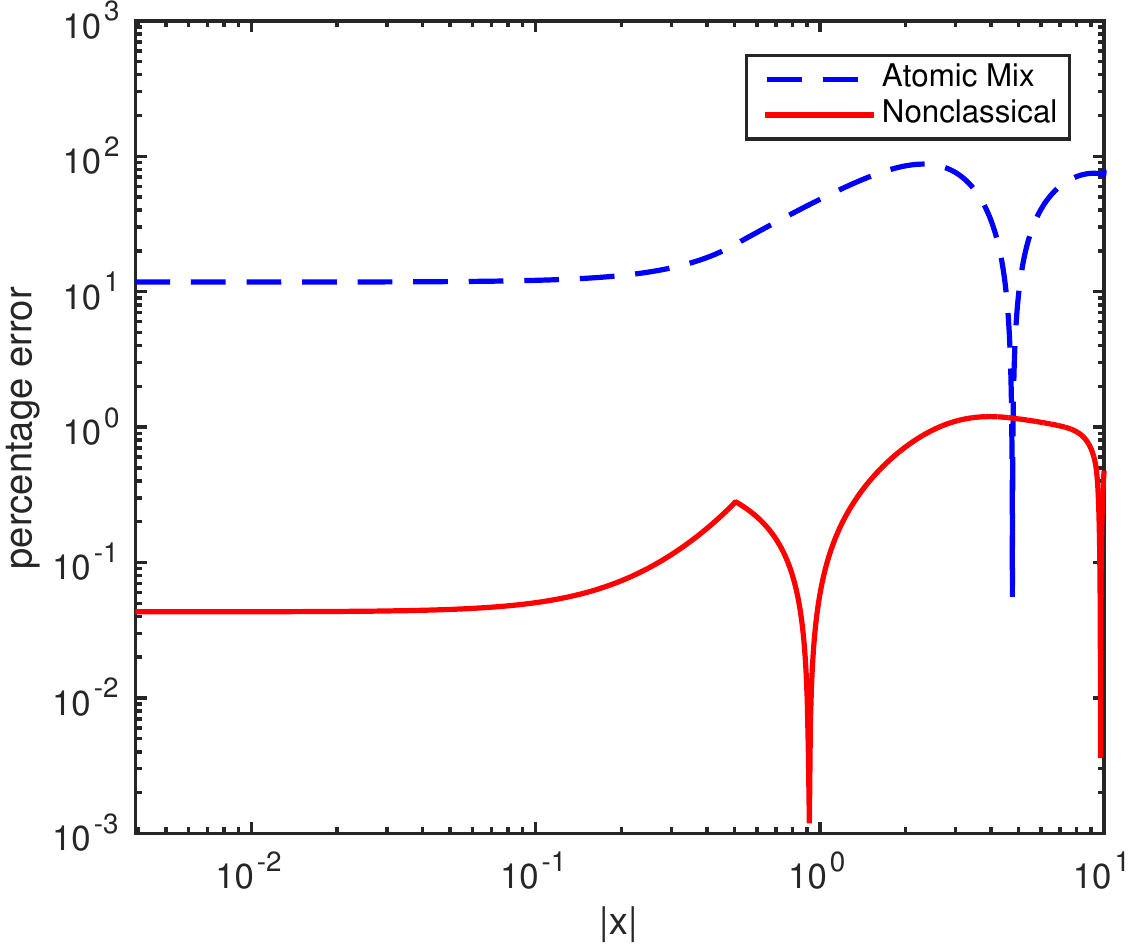}
        \caption{$c_1 = 0.1$}
        \label{figerrE10}
    \end{subfigure}
    \\
    \centering
    \begin{subfigure}{0.495\textwidth}
        \centering
        \includegraphics[width=\textwidth]{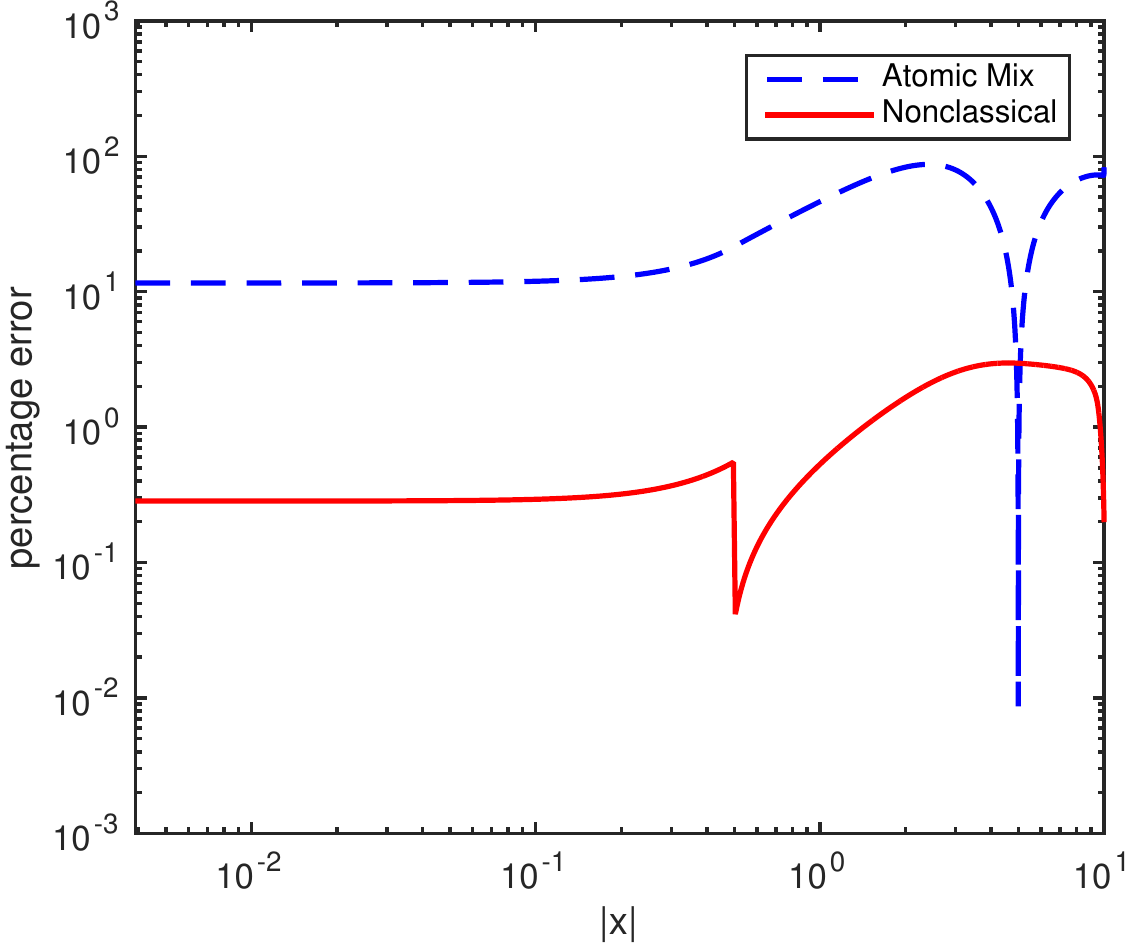}
        \caption{$c_1 = 0.2$}
        \label{figerrE20}
    \end{subfigure}
    \hfill
    \begin{subfigure}{0.495\textwidth}
        \centering
        \includegraphics[width=\textwidth]{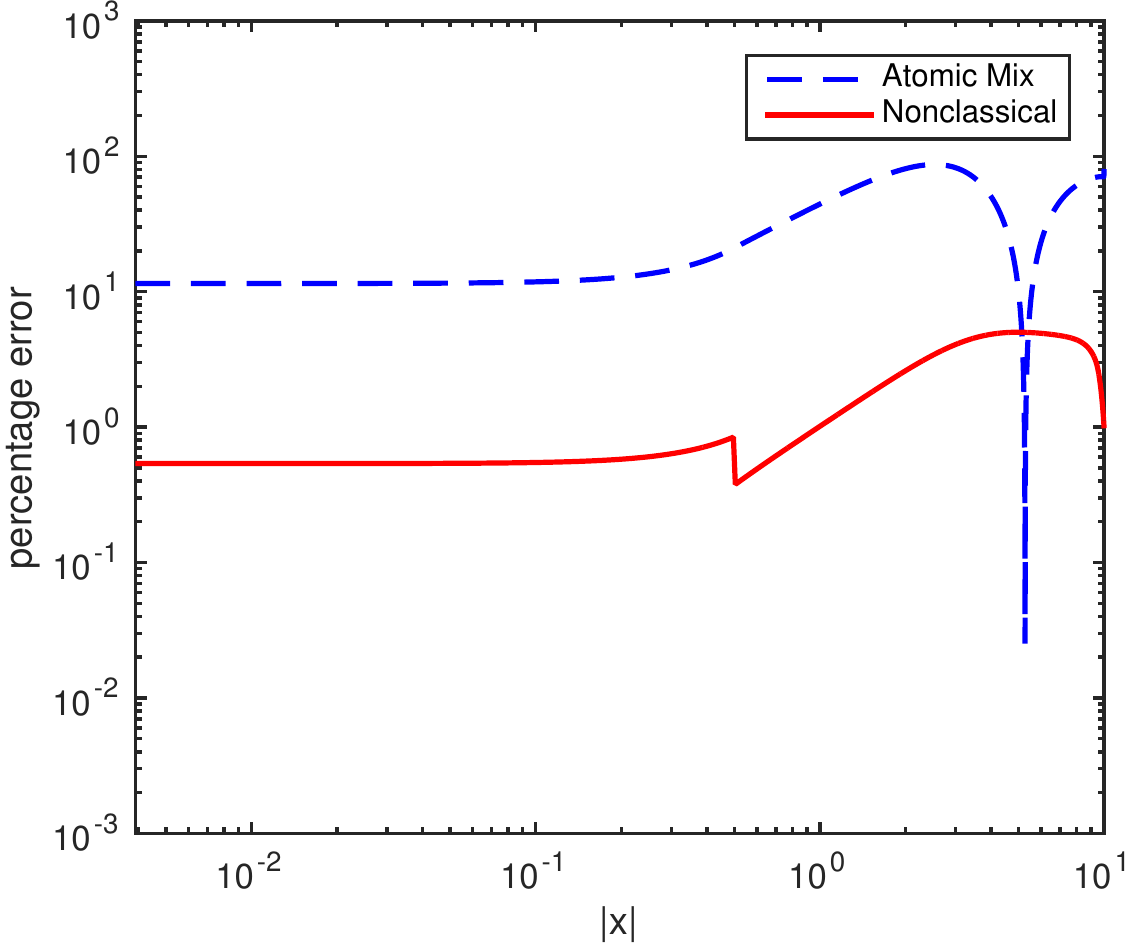}
        \caption{$c_1 = 0.3$}
        \label{figerrE30}
    \end{subfigure}
    \\
    \centering
    \begin{subfigure}{0.495\textwidth}
        \centering
        \includegraphics[width=\textwidth]{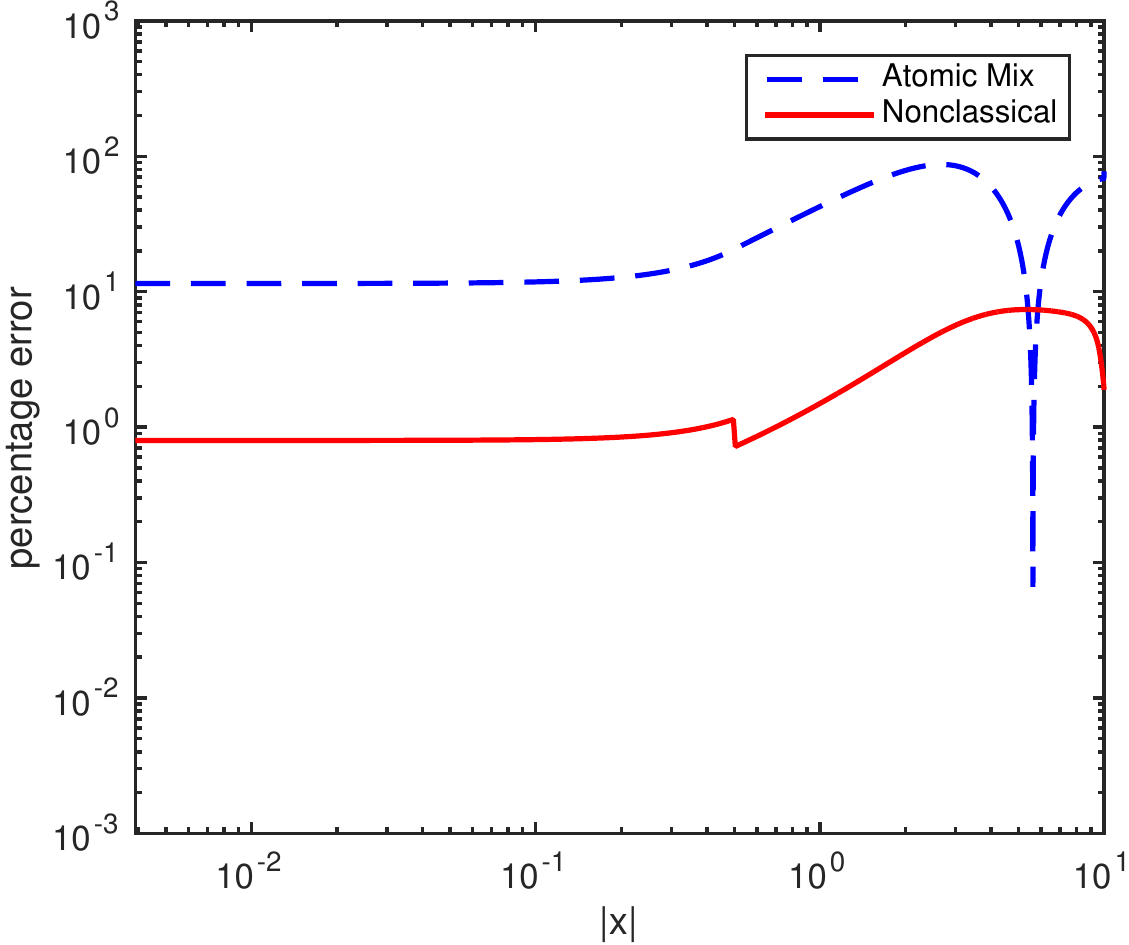}
        \caption{$c_1 = 0.4$}
        \label{figerrE40}
    \end{subfigure}
    \hfill
    \begin{subfigure}{0.495\textwidth}
        \centering
        \includegraphics[width=\textwidth]{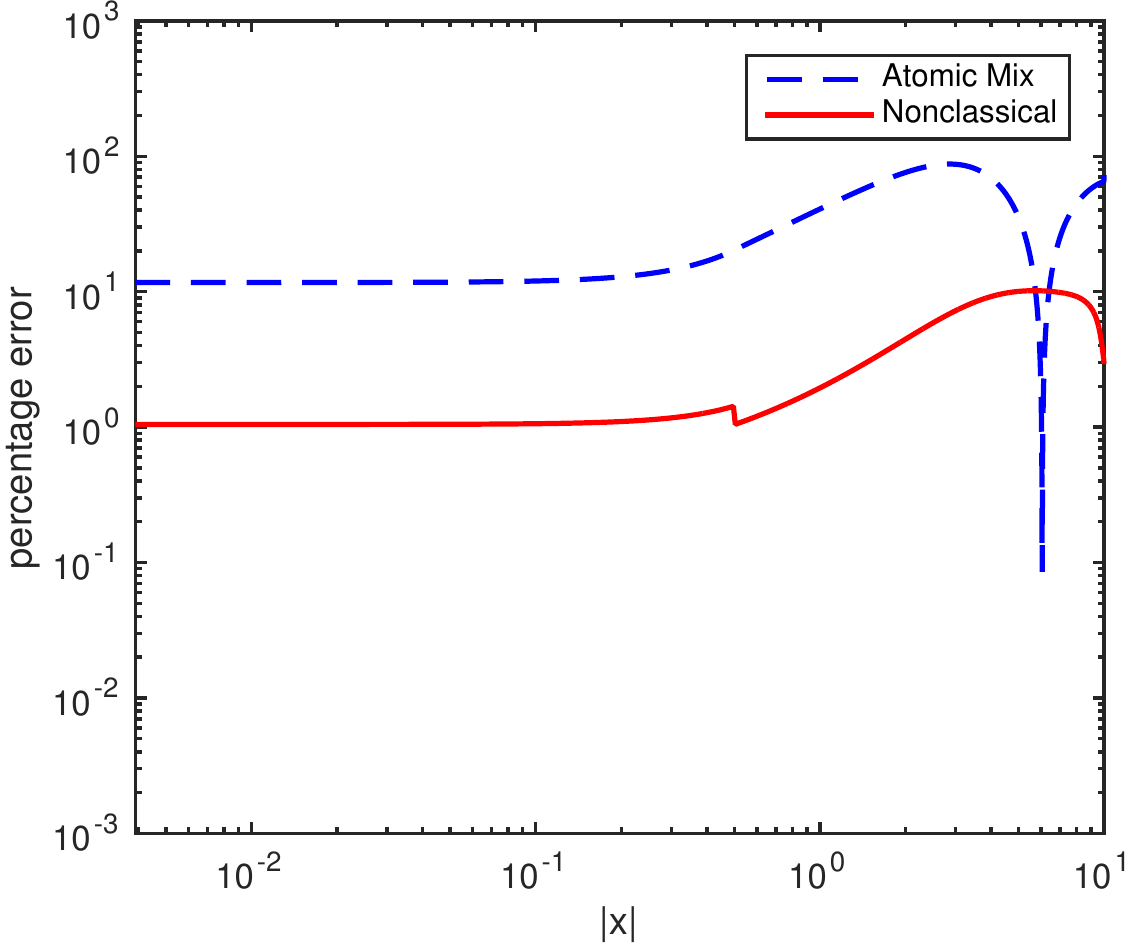}
        \caption{$c_1 = 0.5$}
        \label{figerrE50}
    \end{subfigure}
    \caption{Atomic mix and nonclassical percentage errors with respect to the benchmark solutions for problem set $\setb_2$ (log scale)}
    \label{figerrE1}
\end{figure}

\pagebreak
\begin{figure}[p]
    \centering
    \begin{subfigure}{0.495\textwidth}
        \centering
        \includegraphics[width=\textwidth]{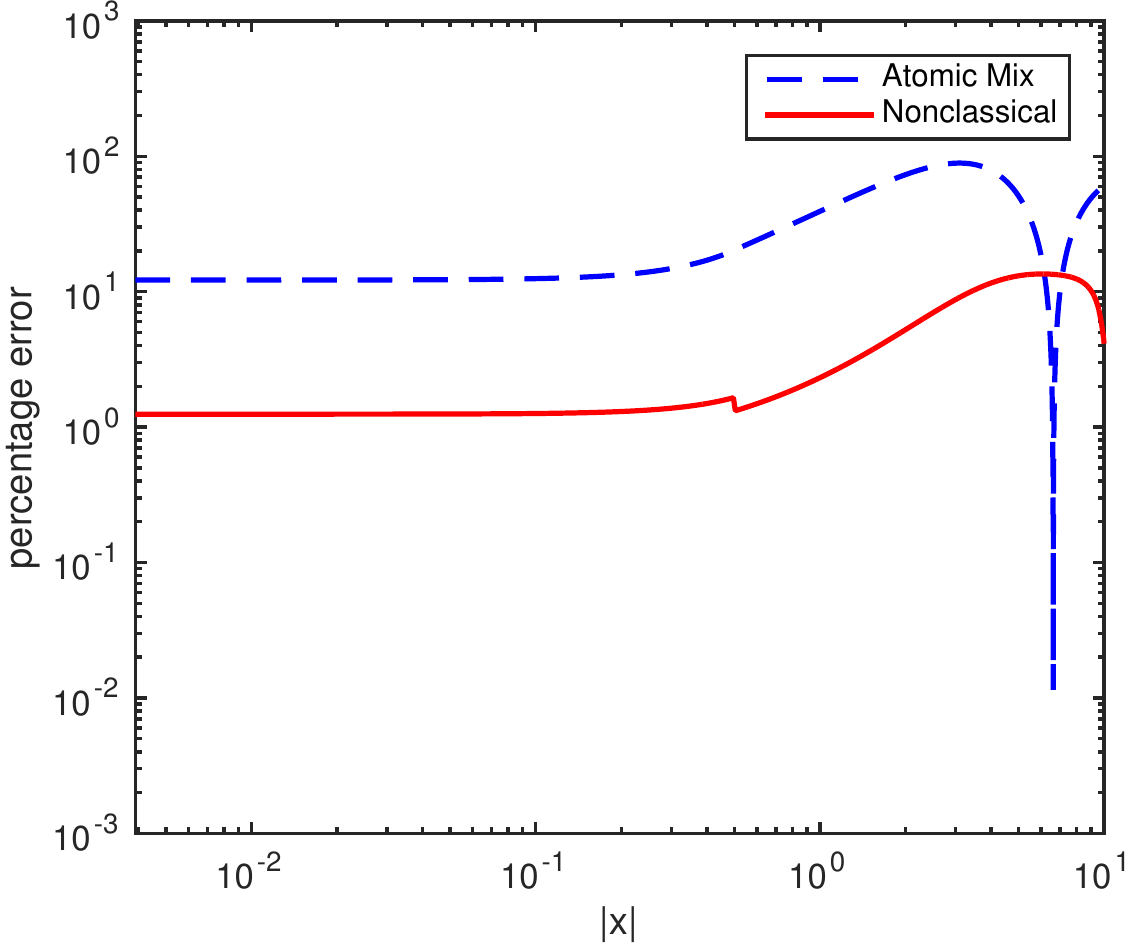}
        \caption{$c_1 = 0.6$}
        \label{figerrE60}
    \end{subfigure}
    \hfill
    \begin{subfigure}{0.495\textwidth}
        \centering
        \includegraphics[width=\textwidth]{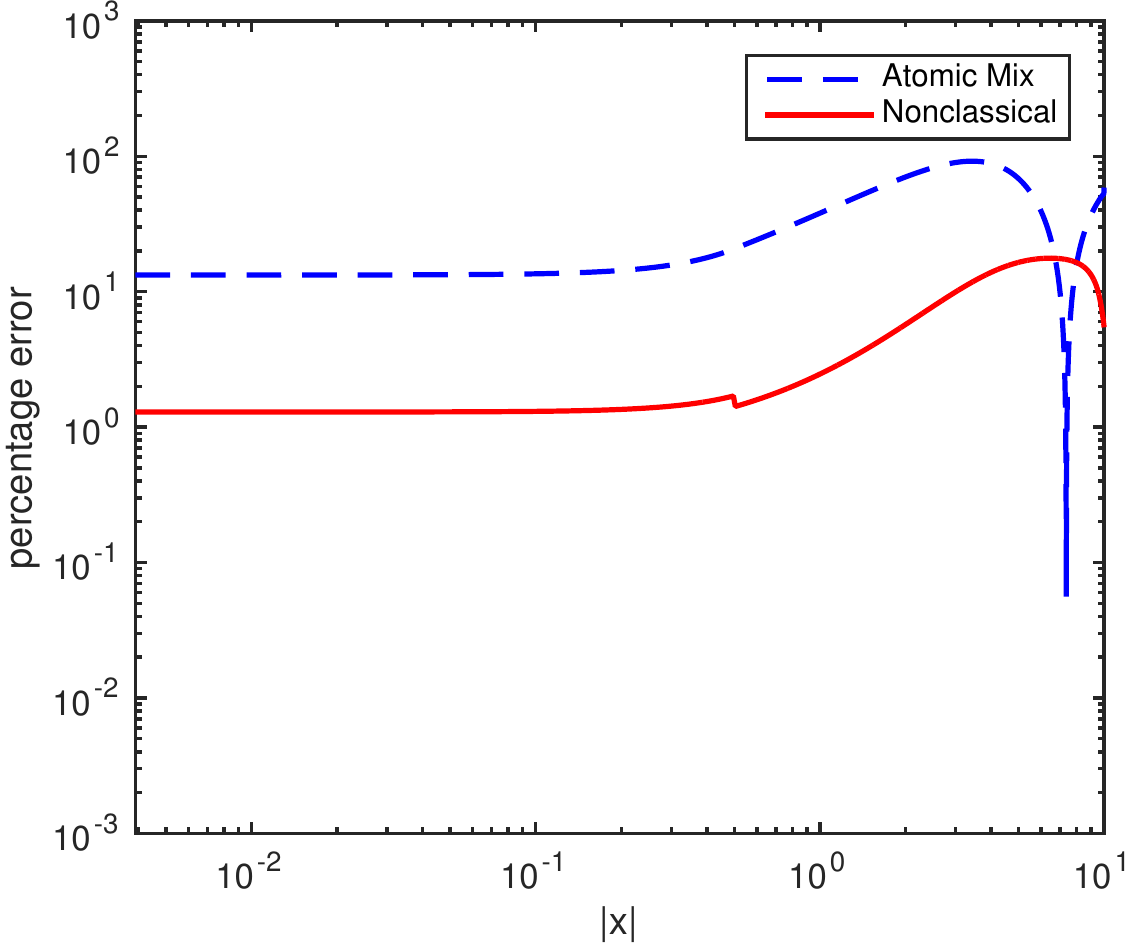}
        \caption{$c_1 = 0.7$}
        \label{figerrE70}
    \end{subfigure}
    \\
    \centering
    \begin{subfigure}{0.495\textwidth}
        \centering
        \includegraphics[width=\textwidth]{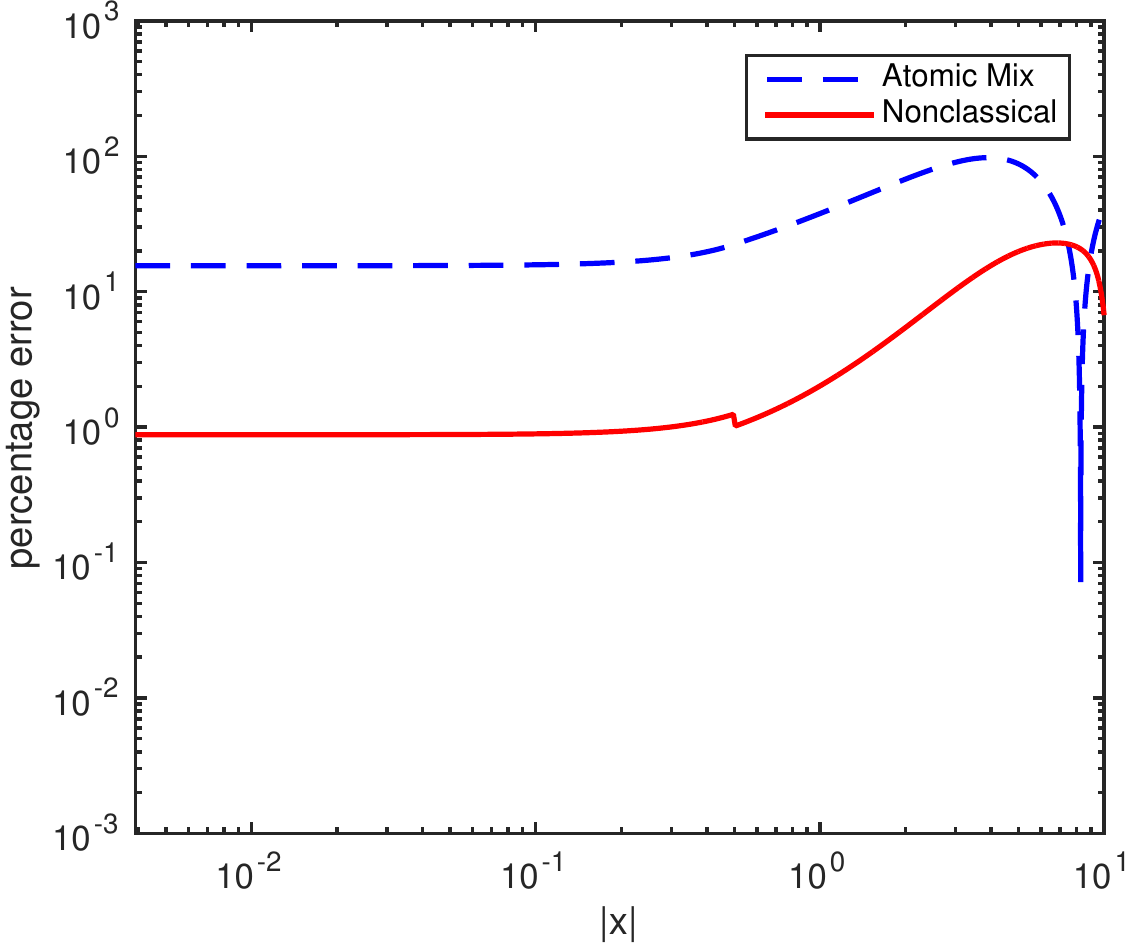}
        \caption{$c_1 = 0.8$}
        \label{figerrE80}
    \end{subfigure}
    \hfill
    \begin{subfigure}{0.495\textwidth}
        \centering
        \includegraphics[width=\textwidth]{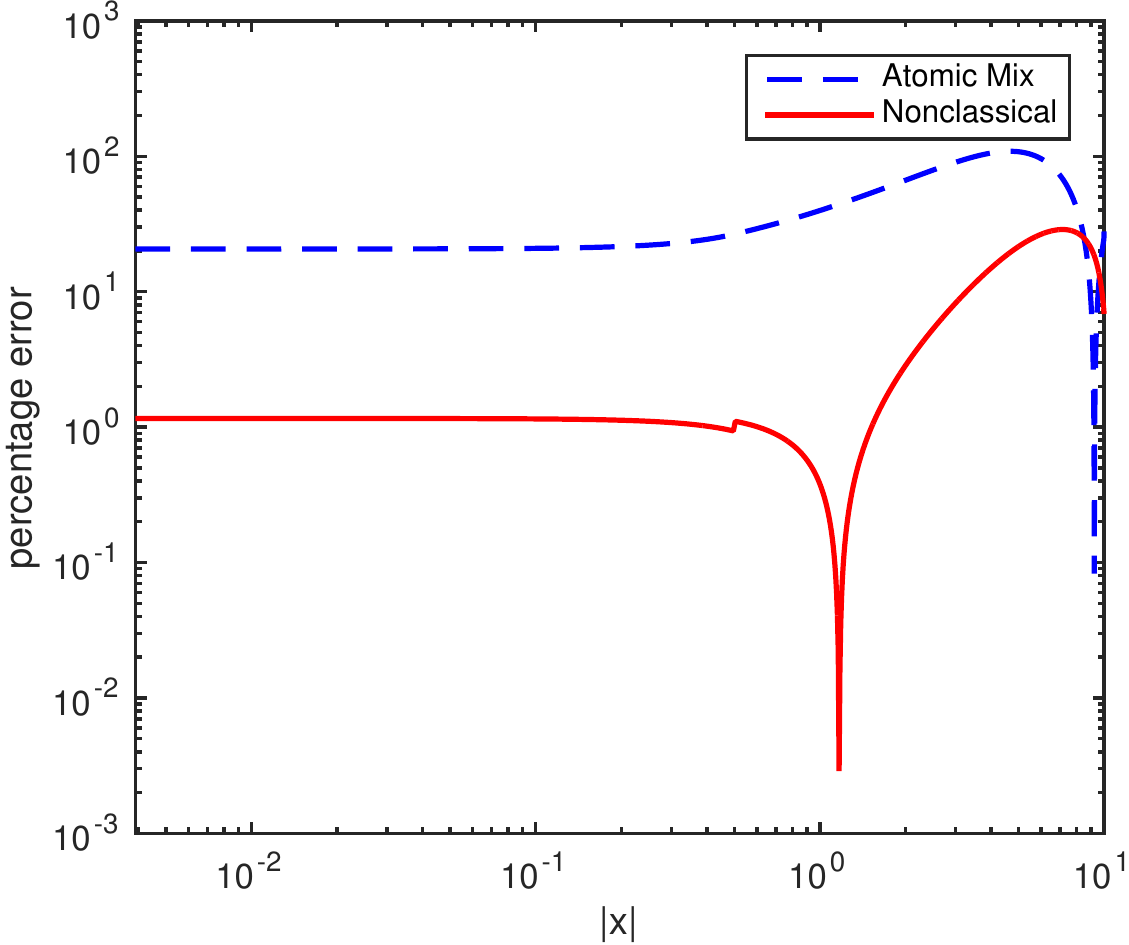}
        \caption{$c_1 = 0.9$}
        \label{figerrE90}
    \end{subfigure}
    \\
    \centering
    \begin{subfigure}{0.495\textwidth}
        \centering
        \includegraphics[width=\textwidth]{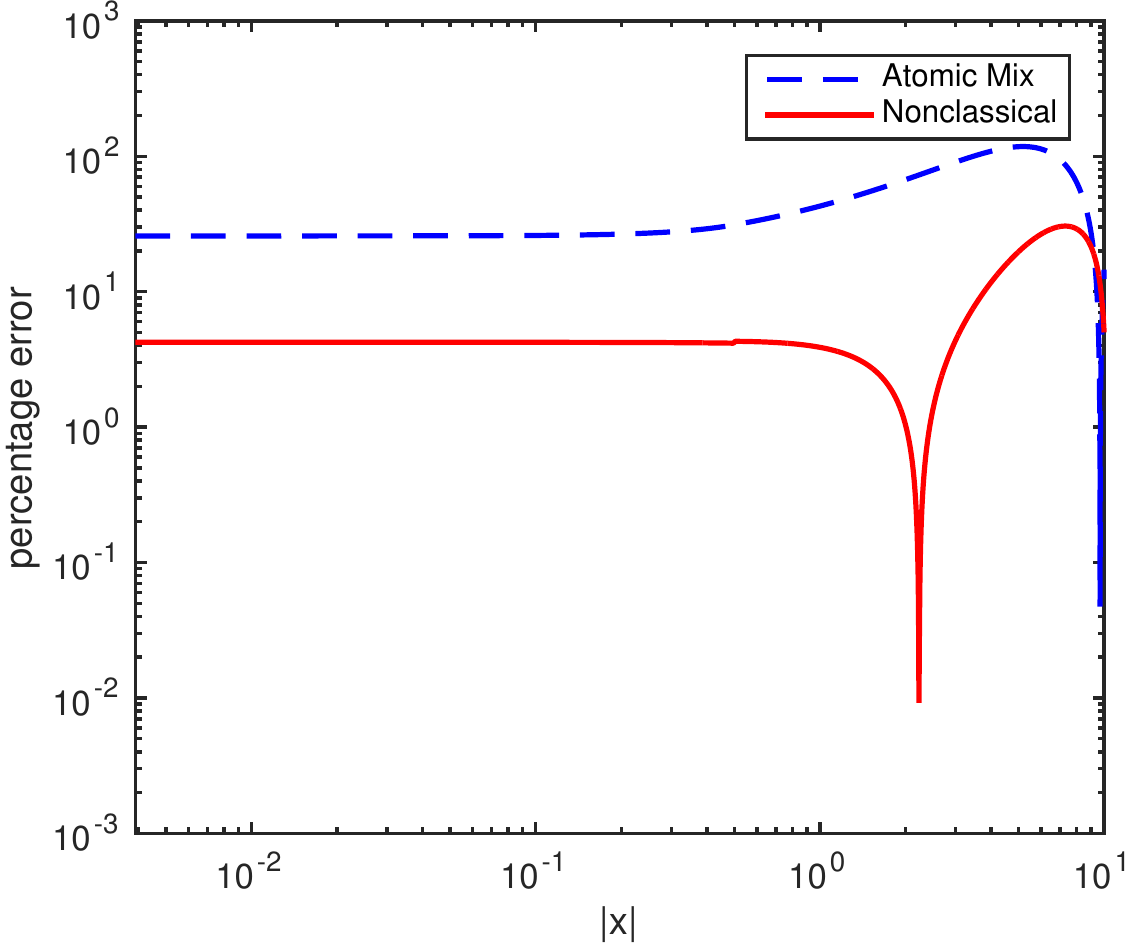}
        \caption{$c_1 = 0.95$}
        \label{figerrE95}
    \end{subfigure}
    \hfill
    \begin{subfigure}{0.495\textwidth}
        \centering
        \includegraphics[width=\textwidth]{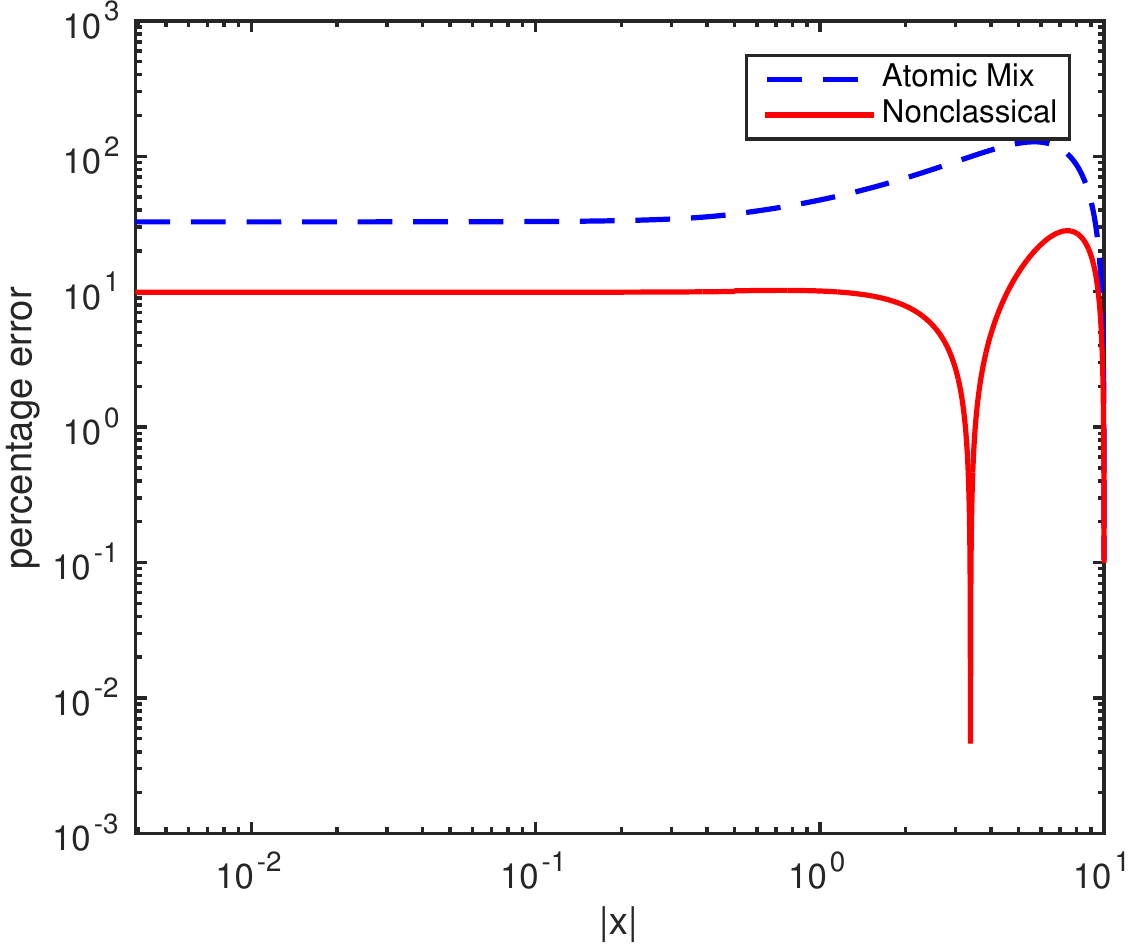}
        \caption{$c_1 = 0.99$}
        \label{figerrE99}
    \end{subfigure}
    \caption{Atomic mix and nonclassical percentage errors with respect to the benchmark solutions for problem set $\setb_2$ (log scale)}
    \label{figerrE2}
\end{figure}

\pagebreak
\begin{figure}[p]
    \centering
    \begin{subfigure}{0.495\textwidth}
        \centering
        \includegraphics[width=\textwidth]{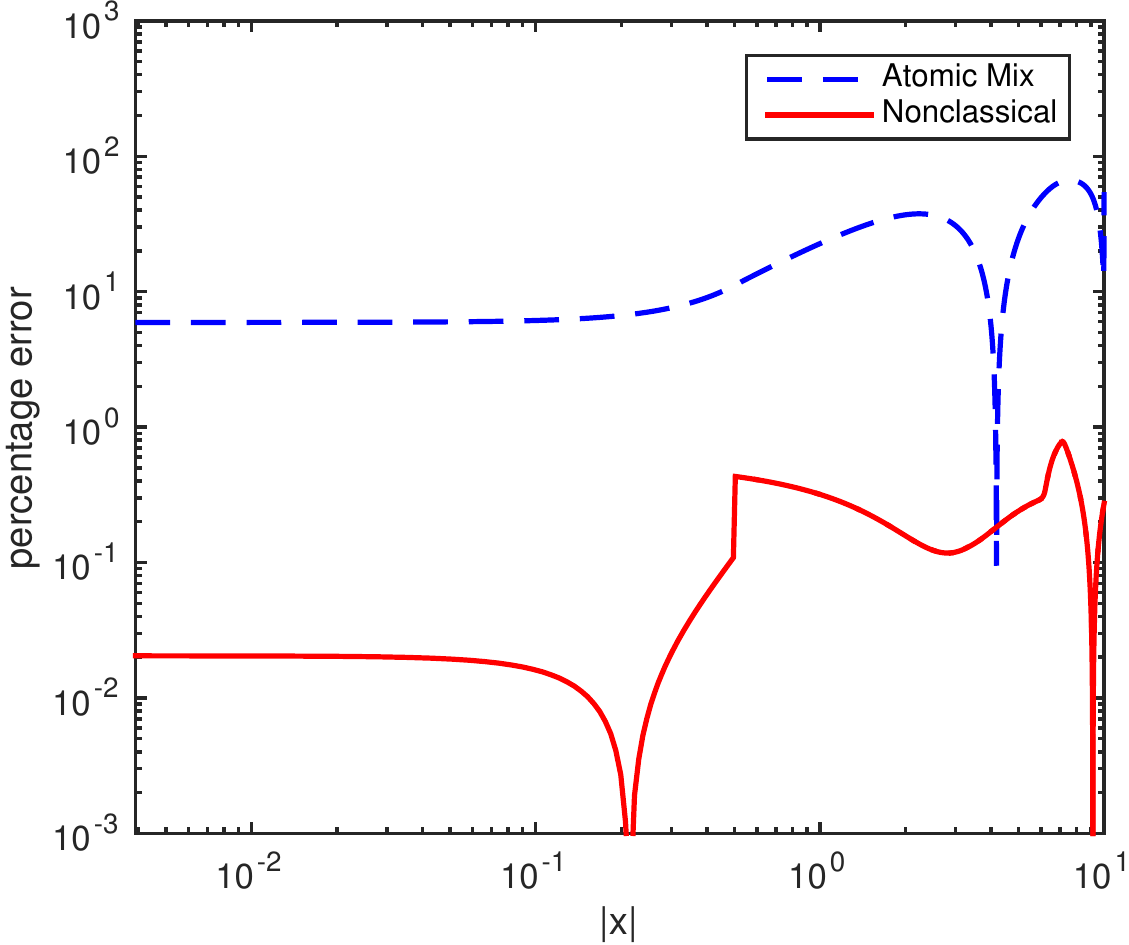}
        \caption{$c_1 = 0.0$}
        \label{figerrF00}
    \end{subfigure}
    \hfill
    \begin{subfigure}{0.495\textwidth}
        \centering
        \includegraphics[width=\textwidth]{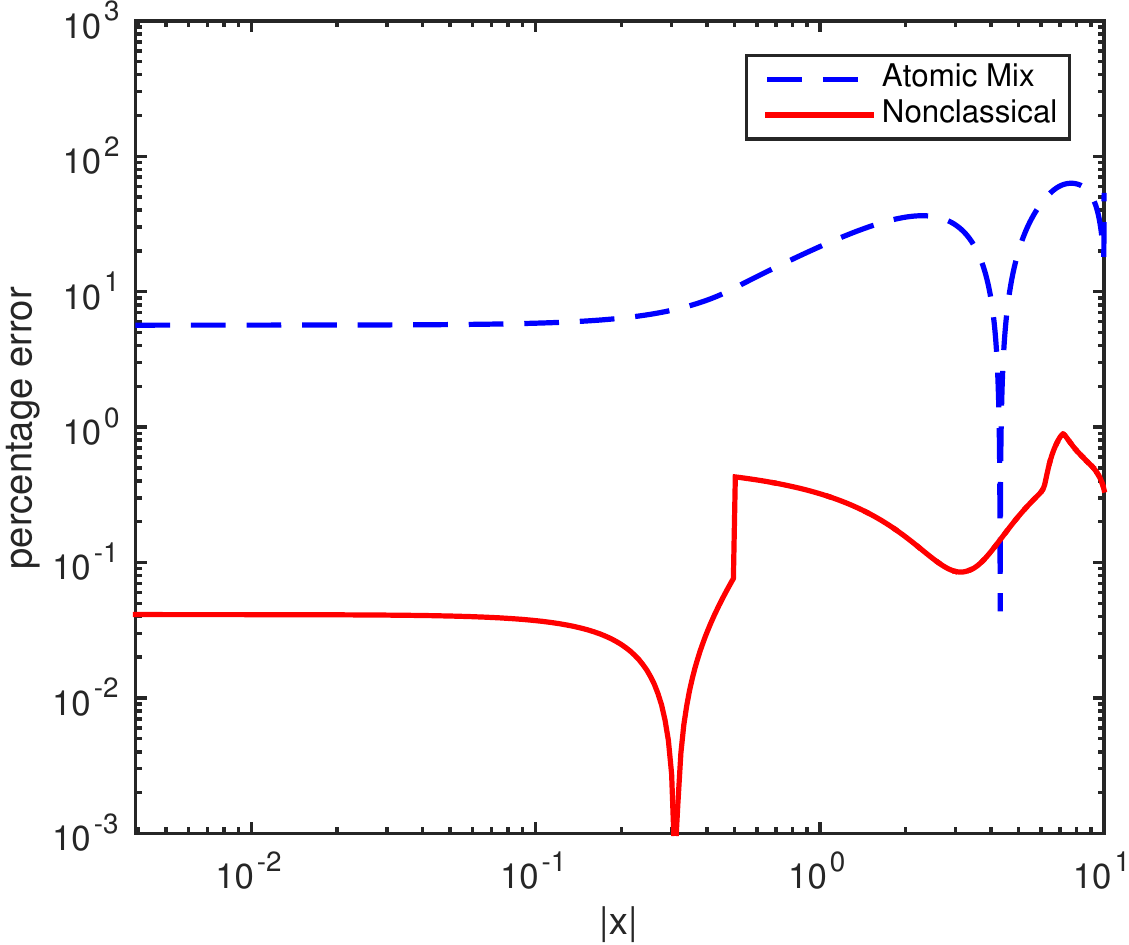}
        \caption{$c_1 = 0.1$}
        \label{figerrF10}
    \end{subfigure}
    \\
    \centering
    \begin{subfigure}{0.495\textwidth}
        \centering
        \includegraphics[width=\textwidth]{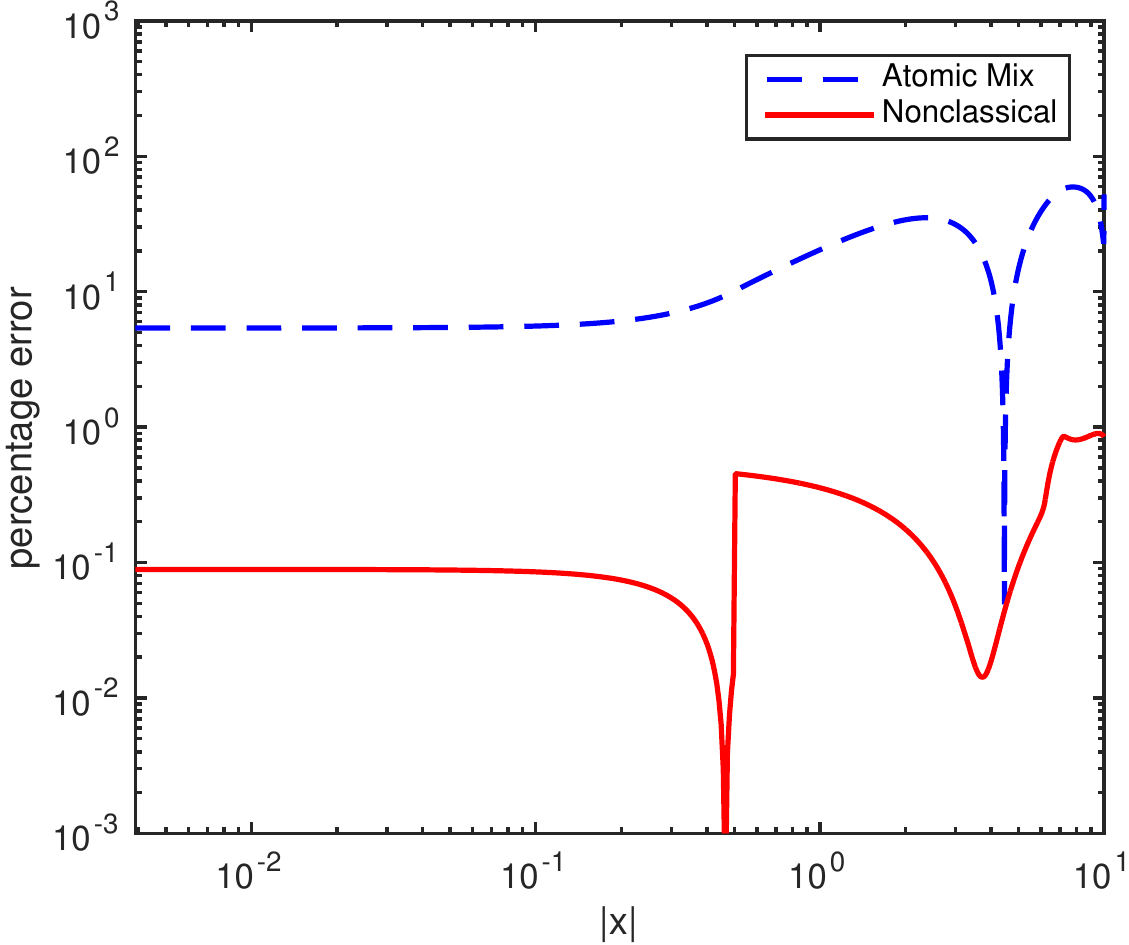}
        \caption{$c_1 = 0.2$}
        \label{figerrF20}
    \end{subfigure}
    \hfill
    \begin{subfigure}{0.495\textwidth}
        \centering
        \includegraphics[width=\textwidth]{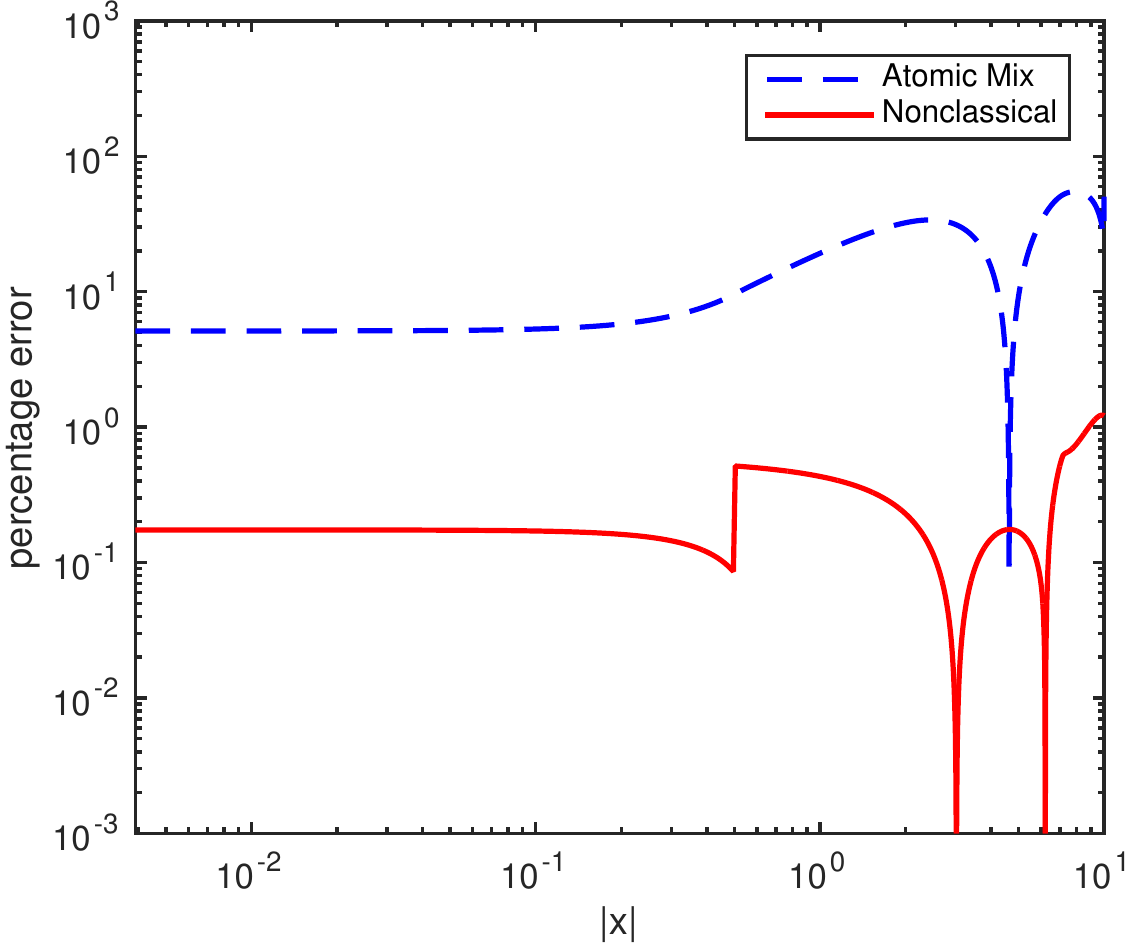}
        \caption{$c_1 = 0.3$}
        \label{figerrF30}
    \end{subfigure}
    \\
    \centering
    \begin{subfigure}{0.495\textwidth}
        \centering
        \includegraphics[width=\textwidth]{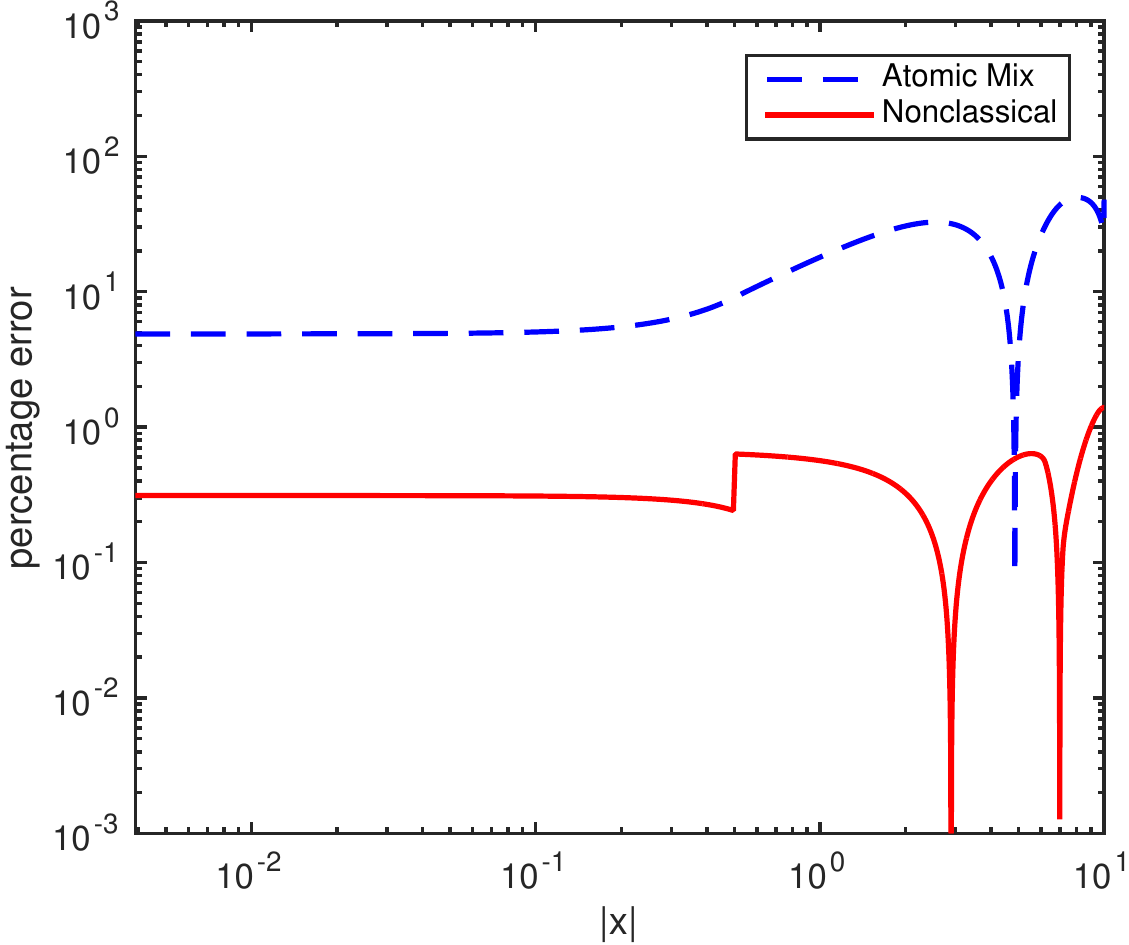}
        \caption{$c_1 = 0.4$}
        \label{figerrF40}
    \end{subfigure}
    \hfill
    \begin{subfigure}{0.495\textwidth}
        \centering
        \includegraphics[width=\textwidth]{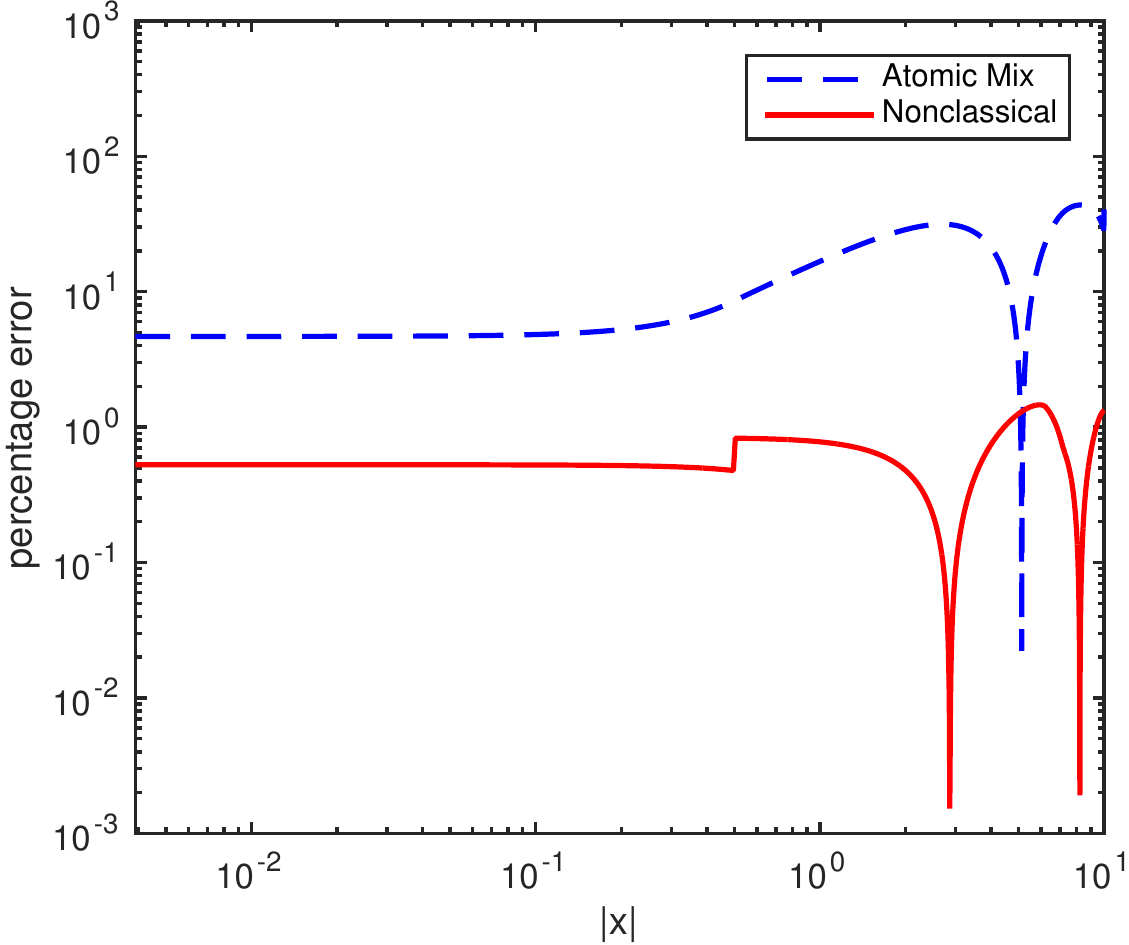}
        \caption{$c_1 = 0.5$}
        \label{figerrF50}
    \end{subfigure}
    \caption{Atomic mix and nonclassical percentage errors with respect to the benchmark solutions for problem set $\setb_3$ (log scale)}
    \label{figerrF1}
\end{figure}

\pagebreak
\begin{figure}[p]
    \centering
    \begin{subfigure}{0.495\textwidth}
        \centering
        \includegraphics[width=\textwidth]{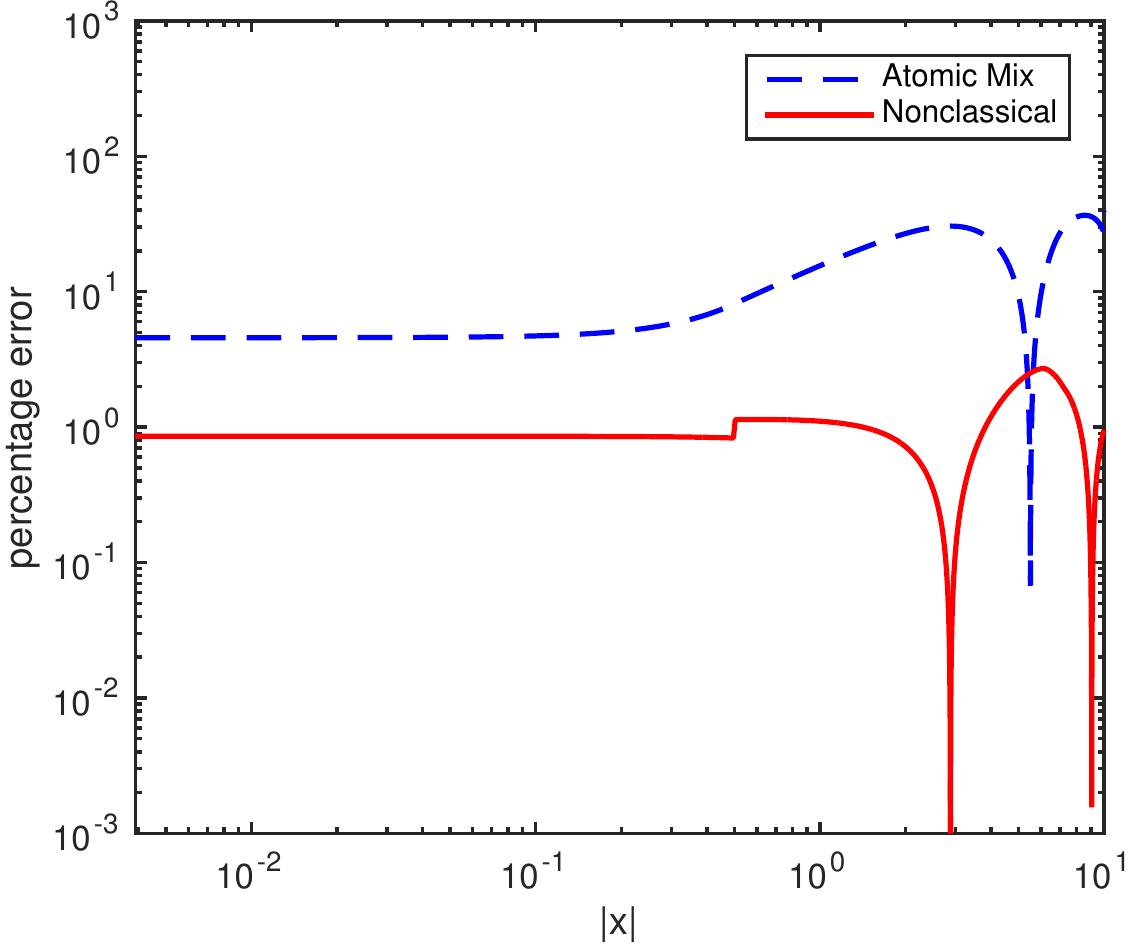}
        \caption{$c_1 = 0.6$}
        \label{figerrF60}
    \end{subfigure}
    \hfill
    \begin{subfigure}{0.495\textwidth}
        \centering
        \includegraphics[width=\textwidth]{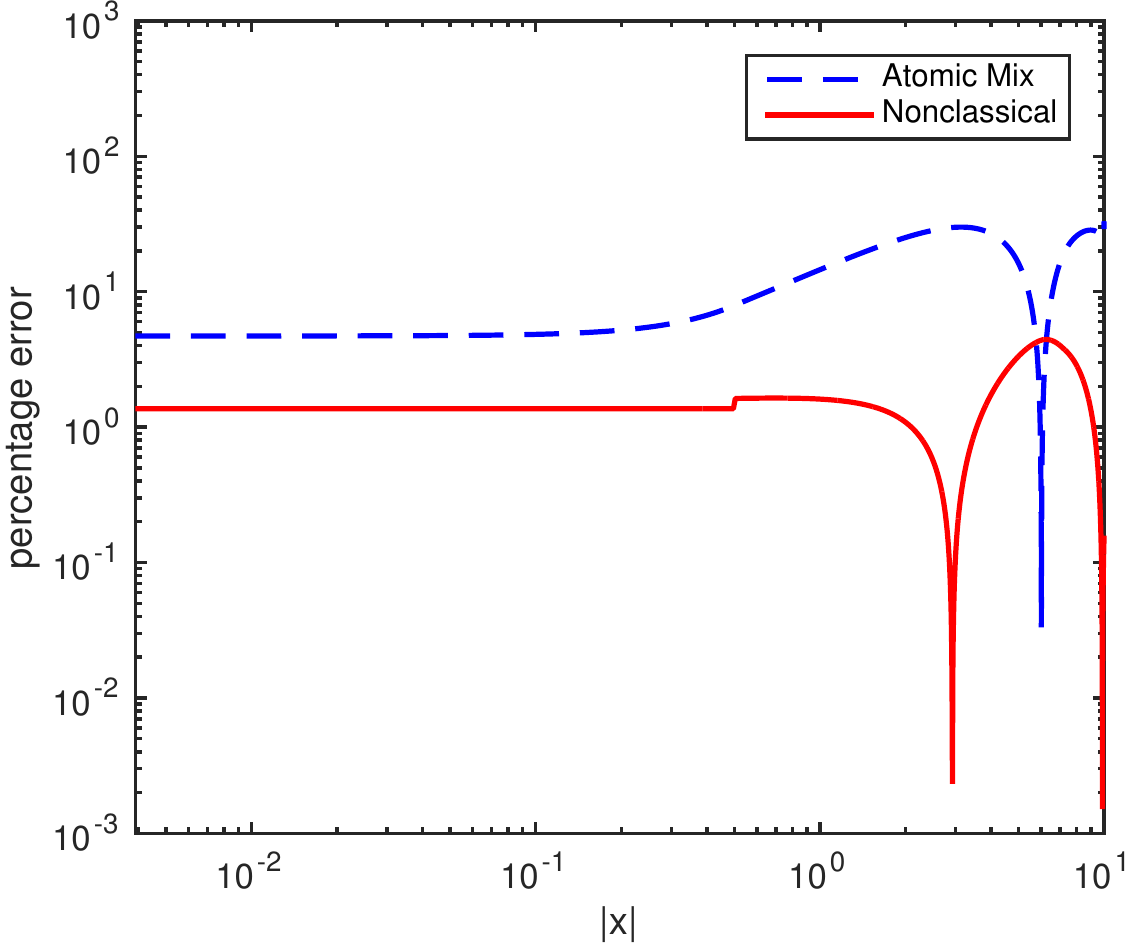}
        \caption{$c_1 = 0.7$}
        \label{figerrF70}
    \end{subfigure}
    \\
    \centering
    \begin{subfigure}{0.495\textwidth}
        \centering
        \includegraphics[width=\textwidth]{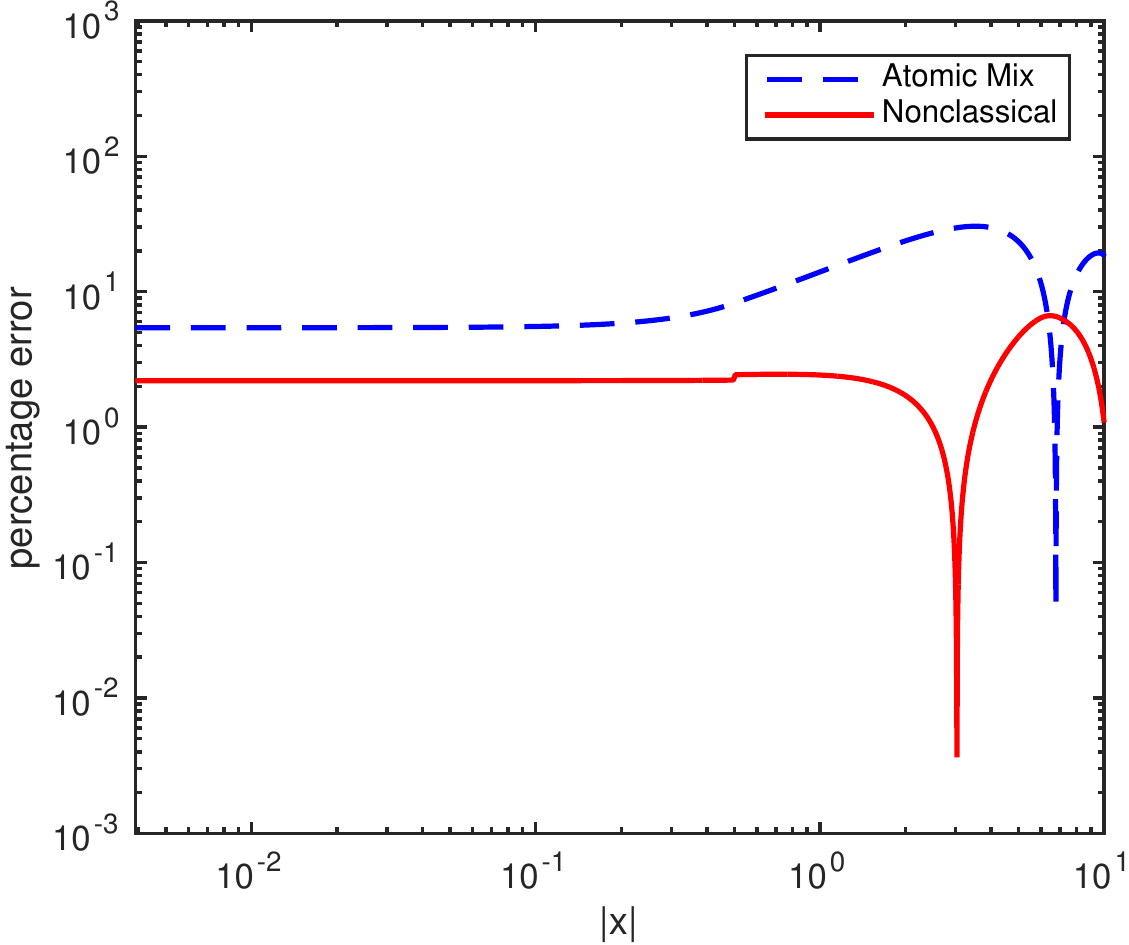}
        \caption{$c_1 = 0.8$}
        \label{figerrF80}
    \end{subfigure}
    \hfill
    \begin{subfigure}{0.495\textwidth}
        \centering
        \includegraphics[width=\textwidth]{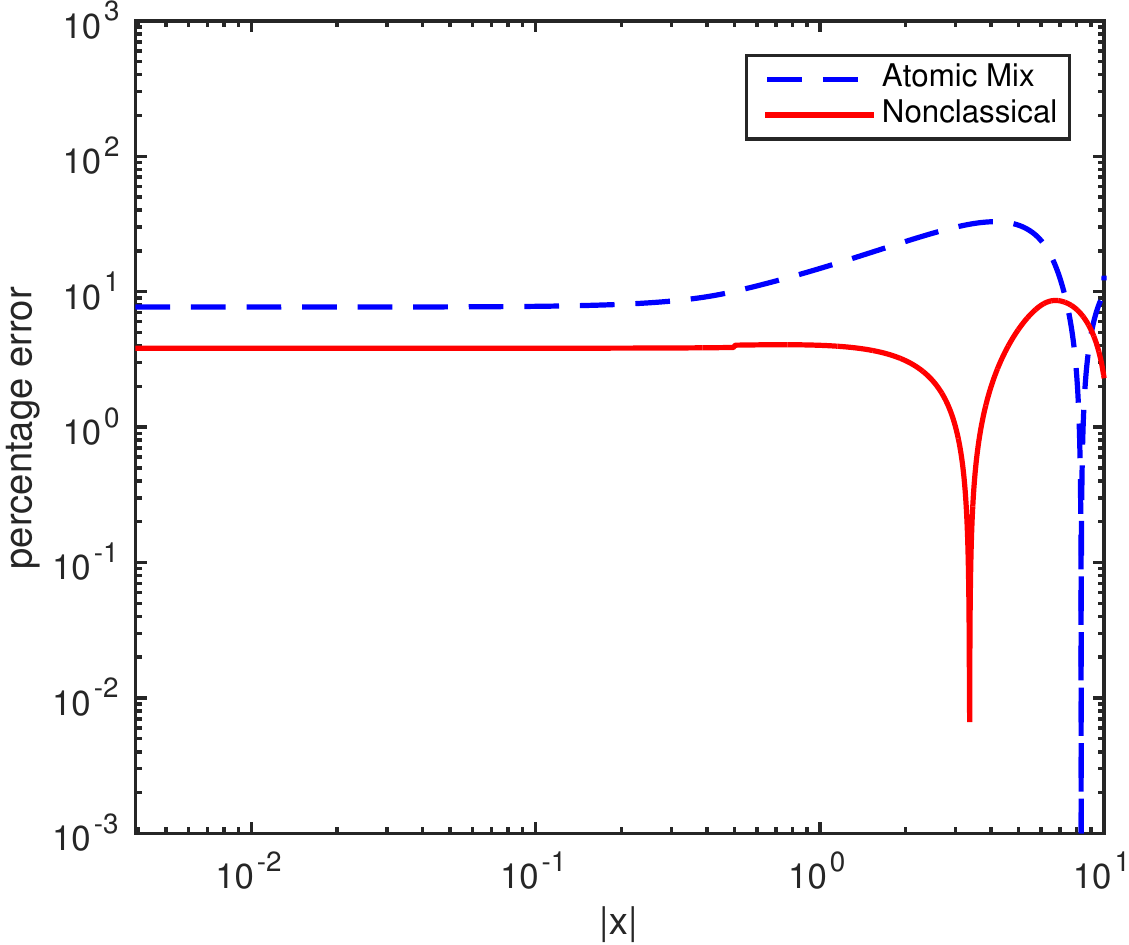}
        \caption{$c_1 = 0.9$}
        \label{figerrF90}
    \end{subfigure}
    \\
    \centering
    \begin{subfigure}{0.495\textwidth}
        \centering
        \includegraphics[width=\textwidth]{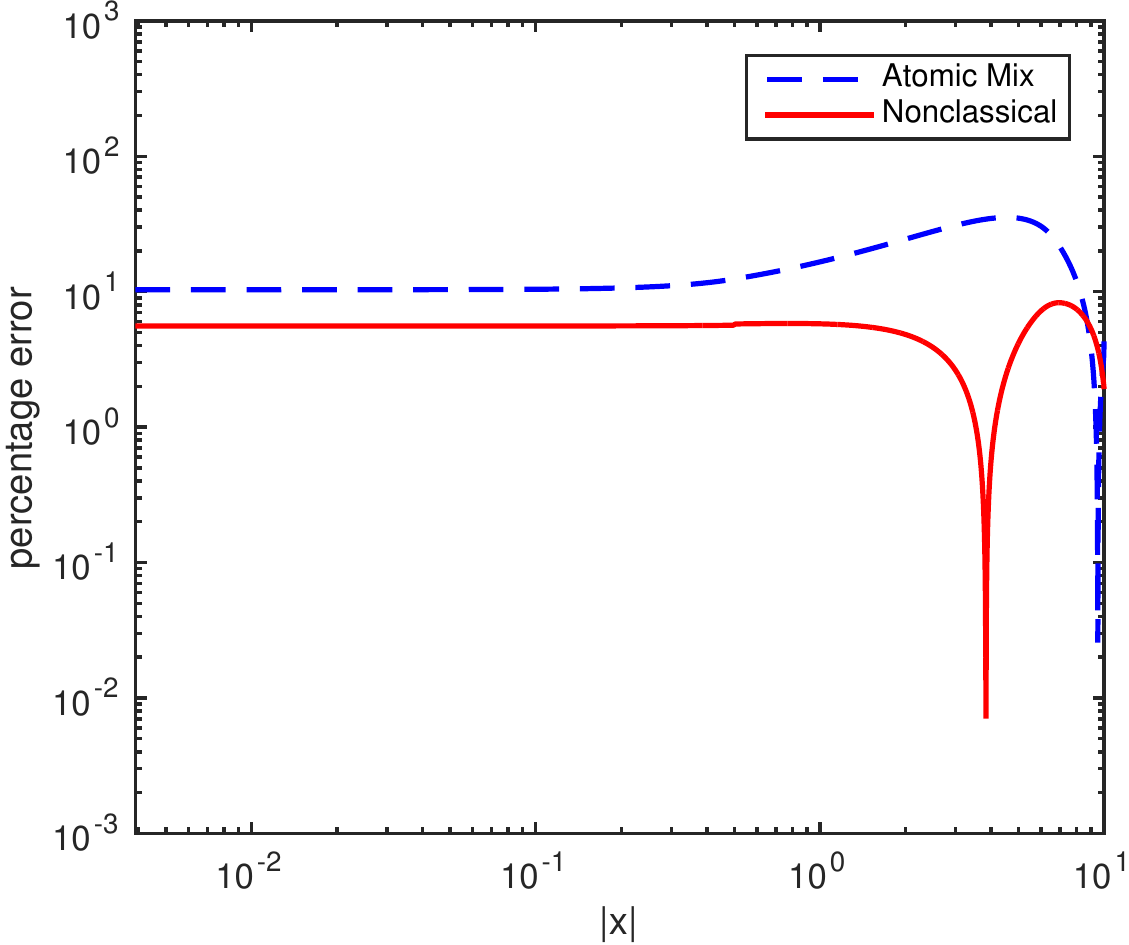}
        \caption{$c_1 = 0.95$}
        \label{figerrF95}
    \end{subfigure}
    \hfill
    \begin{subfigure}{0.495\textwidth}
        \centering
        \includegraphics[width=\textwidth]{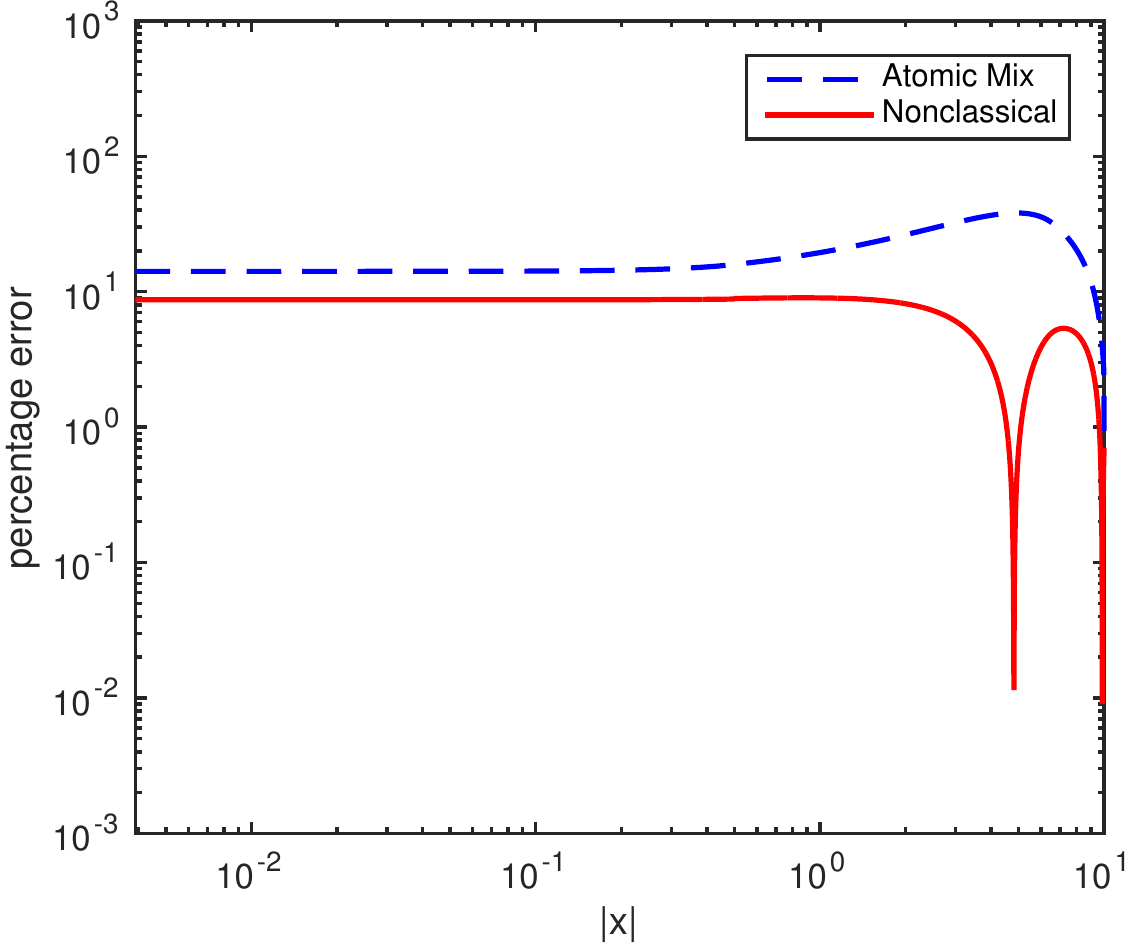}
        \caption{$c_1 = 0.99$}
        \label{figerrF99}
    \end{subfigure}
    \caption{Atomic mix and nonclassical percentage errors with respect to the benchmark solutions for problem set $\setb_3$ (log scale)}
    \label{figerrF2}
\end{figure}

\end{document}